\documentclass[a4paper,11pt]{article}

\usepackage[normalem]{ulem}
\usepackage{jheppub}
\usepackage{color}
\usepackage[table]{xcolor}
\usepackage{epstopdf}
\usepackage{cleveref}
\usepackage{epstopdf}
\usepackage{caption}
\usepackage{subcaption}
\usepackage{amsthm}
\usepackage{slashed}
\usepackage{orcidlink}

\DeclareMathOperator*{\res}{Res}
\DeclareMathOperator*{\sgn}{sgn}
\DeclareMathOperator{\cut}{Cut}
\DeclareMathOperator*{\im}{Im}

\DeclareMathOperator*{\tr}{tr}

\newcommand{\pol}{\ensuremath{\varepsilon}}
\newcommand{\ctpeps}{\ensuremath{i0}}
\newcommand{\dimreg}{\ensuremath{\epsilon}}
\newcommand{\odimreg}[1]{\ensuremath{\mathcal{O}(\dimreg^{#1})}}

\newcommand{\mtree}{\ensuremath{\mathcal{M}_{\text{tree}}}}

\newcommand{\msggs}{\ensuremath{\mathcal{M}_{sggs}}}

\newcommand{\graph}[1]{G_{\text{#1}}}
\newcommand{\rcut}[1]{\ensuremath{\overline{\cut}_{#1}}}

\newcommand{\ctpdof}[1]{\{\dots\}_{#1}}
\newcommand{\ctpdofb}[1]{\overline{\{\dots\}}_{#1}}

\newcommand*\diff{\mathop{}\!\mathrm{d}}
\newcommand*\Diff[1]{\mathop{}\!\mathrm{d}^#1}

\newcommand{\ellint}[1]{\ensuremath{ \int \frac{\Diff{d} \ell_{#1}}{(2 \pi)^d}}}
\newcommand{\elllint}[1]{\ensuremath{\int \prod_{i=1}^{#1}\left(\frac{\Diff{d} \ell_{i}}{(2 \pi)^d}}\right)}

\newcommand{\intw}{\ensuremath{\int \frac{\diff \omega}{2 \pi}}}
\newcommand{\inte}[1]{\ensuremath{\int_{-\infty}^{\infty} \diff \omega \kappa(\omega) \omega^{#1}}}
\newcommand{\intem}[1]{\ensuremath{\int_{-\infty}^{\infty} \diff \omega \kappa(\omega,\mu) \omega^{#1}}}

\newcommand{\ftnorm}{\ensuremath{}}
\newcommand{\iftnorm}{\ensuremath{\frac{1}{2\pi}}}

\newcommand{\deltaE}[2]{\ensuremath{(\Delta E)^{\text{#1}}_{\text{#2}}}}
\newcommand{\delEInc}[1]{\ensuremath{(\overline{\Delta E})_{\text{#1}}}}

\newcommand{\ttcoup}{\ensuremath{(512  G_N^3 E^2 \pi^3)}}
\newcommand{\tttcoup}{\ensuremath{(8192 G_N^4  E^3 \pi^4)}}
\newcommand{\ttttcoup}{\ensuremath{(131072  G_N^5 E^4\pi^5)}}

\theoremstyle{definition}

\newcommand{\eg}{{e.g.~}}
\newcommand{\ie}{{i.e.~}}

\colorlet{cutcolor}{blue!60!white}
\colorlet{blobcolor}{gray!40}

\colorlet{ref-link}{green!70!black!70!blue}
\colorlet{cite-link}{red!60!blue}

\colorlet{fc}{green!50!black}
\colorlet{alex}{blue!50!green}

\hypersetup{
  citecolor = cite-link,
  linkcolor = ref-link,
}

\newcommand{\ml}[1]{\textcolor{blue}{\textbf{ML:}#1}}

\usepackage{tikz}
\usetikzlibrary{calc}
\usetikzlibrary{patterns}
\usetikzlibrary{decorations.pathreplacing}
\usetikzlibrary{decorations.markings}
\usetikzlibrary{decorations.pathmorphing}
\usetikzlibrary{positioning}
\usetikzlibrary{bending}

\pgfdeclarelayer{background}
\pgfsetlayers{background,main}

\usetikzlibrary{shapes,shapes.geometric,arrows,arrows.meta,decorations.pathmorphing,decorations.markings,decorations.pathreplacing,patterns}
\usetikzlibrary{tikzmark}

\tikzset{every picture/.style={baseline={([yshift=-.7ex]current bounding box.center)}}}

\tikzset{every node/.style={font=\scriptsize}}

\tikzset{dir/.style={decoration={markings, mark=at position \halfway with {\arrow{Latex}}},postaction={decorate}}}
\tikzset{cuts/.style={dash pattern=on 4pt off 2pt,draw=cutcolor,ultra thick}}

\tikzset{photon/.style={decorate,decoration={coil,aspect=0,segment length=2,amplitude=1}}}
\tikzset{photon big/.style={decorate,decoration={coil,aspect=0,segment length=4,amplitude=2}}}

\tikzset{potential/.style={%
   photon big}}

\tikzset{potentialf/.style={%
    dashed,thick,gray}}

\tikzset{quad src/.style n args={2}{%
    fill,anchor=center,inner sep=#1,minimum size=#1,label=left:{#2}}}
\tikzset{mass src/.style n args={1}{%
    circle,fill,anchor=center,minimum size=#1,inner sep=#1,label=left:{\(E\)}}}
\tikzset{blob/.style n args={1}{%
    draw=black,fill=blobcolor,circle,minimum size=#1,inner sep=0}}
  
\tikzset{bulk/.style={%
    circle,fill=blobcolor,draw=black,inner sep=1pt}}

\tikzset{pics/cut hill/.style n args={2}{code={%
      \draw[photon big] (#1.center)
      .. controls ($(#1) + (0.7,0)$) and ($(#2)+(0.7,0)$)..
        node[pos=1/2](v){} (#2.center);
        \draw[cuts] ($(v)+(right:0.3)$) -- ++(left:0.6);
      }}}

\tikzset{pics/cut vhill/.style n args={2}{code={%
      \draw[photon big] (#1.center)
      .. controls ($(#1) + (0,0.7)$) and ($(#2)+(0,0.7)$)..
        node[pos=1/2](v){} (#2.center);
        \draw[cuts] ($(v)+(up:0.3)$) -- ++(down:0.6);
      }}}
\tikzset{pics/cut dhill/.style n args={2}{code={%
      \draw[photon big] (#1.center)
      .. controls ($(#1) + (0,-0.7)$) and ($(#2)+(0,-0.7)$)..
        node[pos=1/2](v){} (#2.center);
        \draw[cuts] ($(v)+(up:0.3)$) -- ++(down:0.6);
        }}}

\tikzset{pics/hills/.style args={#1}{code={%
      \node[quad src={3}{\(I^{ij}\)}] (0) at (0,0){};
      \node[quad src={3}{\(I^{mn}\)}] (b) at (0,#1+1){};
      \foreach \ml[remember=\ml as \lastm] in {1,...,#1}
      {
        \node[mass src={2}](\ml) at (0,\ml) {};
        \draw pic {cut hill={\lastm}{\ml}};
        }
        \draw pic {cut hill={#1}{b}};
    }}}

\pgfmathsetmacro{\mampPushback}{1.2}
\tikzset{
  mampControlsCoords/.style={
    execute at begin picture={
      \coordinate (mampCC) at (\mampPushback,0);
   }
 },
every picture/.append style={mampControlsCoords} 
}

\tikzset{pics/mamp/.style n args={3}{code={%
      \pgfmathsetmacro\n{int(#3+2)}
      \draw[photon big](#1.center)
      .. controls ($(#1) + (mampCC)$) and ($(#2)+(mampCC)$)..
      node[bulk,pos=1/2](v){$\mathcal{M}_{\n}$} (#2.center);
      \foreach \ml in {1,...,#3}
      {
        \node[mass src={2}](\ml) at ($(#1)+(up:\ml)$){};
        \begin{pgfonlayer}{background}
          \draw[potential](\ml) -- (v.center);
        \end{pgfonlayer}
      }
      \draw[cuts] ($(v) +(1,0)+(220-90:1.5)$) arc (220-90:320-90:1.5);
    }}}

\title{Higher-Order Tails and RG Flows due to Scattering of\\
Gravitational Radiation from Binary Inspirals}

\author[a]{Alex Edison \orcidlink{0000-0002-5430-9500},}
\author[b]{Mich\`ele Levi}

\affiliation[a]{Department of Physics and Astronomy, Northwestern
	University, Evanston, Illinois, 60208, USA}
\affiliation[b]{Mathematical
	Institute, University of Oxford, Oxford OX2 6GG, United Kingdom}

\emailAdd{alexander.edison@northwestern.edu}
\emailAdd{levi@maths.ox.ac.uk}

\abstract{ We establish and develop a novel methodology to treat
  higher-order non-linear effects of gravitational radiation that is
  scattered from binary inspirals, which employs modern
  scattering-amplitudes methods on the effective picture of the binary
  as a composite particle.  We spell out our procedure to study such
  effects: assembling tree amplitudes via generalized-unitarity
  methods and employing the closed-time-path formalism to derive the
  causal effective actions, which encompass the full conservative and
  dissipative dynamics.  We push through to a new state of the art for
  these higher-order effects, up to the third subleading tail effect,
  at order $G_N^5$ and the $5$-loop level, which corresponds to the
  $8.5$PN order.  We formulate the consequent dissipated energy for
  these higher-order corrections, and carry out a renormalization
  analysis, where we uncover new subleading RG flow of the quadrupole
  coupling.  For all higher-order tail effects we find perfect
  agreement with partial observable results in PN and self-force
  theories, where available.  }

\begin{document}
\maketitle

\section{Introduction}
The direct observation of gravitation waves (GWs) coming from binary
black hole (BBH) merger events \cite{Abbott:2016blz, Abbott:2016nmj,
  TheLIGOScientific:2017qsa, Abbott:2017vtc, Abbott:2017oio,
  Abbott:2017gyy} has shifted precision predictions of GW and BBH
structure from theoretical curiosity to phenomenological imperative.
With four gravitational-wave observatories active through the
LIGO-VIRGO-KAGRA network \cite{LIGOScientific:2014pky,
  VIRGO:2014yos, KAGRA:2020tym} and new ground- and space-based
detectors on the way 
\cite{2017arXiv170200786A,Reitze:2019iox, Maggiore:2019uih}, 
the increasing scope and depth of incoming gravitational-wave data 
threatens to exceed our currently available predictions.  To meet 
this looming demand, the last few decades have seen an explosive growth 
on the theoretical frontier.  

The longest-running framework for studying GW sources during the 
significant inspiral phase is post-Newtonian (PN) General Relativity, 
which deals with simultaneously weakly interacting and
slowly moving bodies; we refer the interested reader to
Ref.~\cite{Blanchet:2013haa} for a comprehensive Living Review on the
subject.  With both the orbital velocity and the gravitational coupling 
as small parameters, PN computations build on classical two-body
Newtonian dynamics, and compute the perturbative corrections in these 
small parameters induced by GR (hence the ``post-Newtonian'' moniker).  
Note that the PN 
approximation treats the perturbation constants as 
$G_N \sim v^2/c^2 \ll 1$, and thus admits half-PN counting through 
single powers of $v$.
Due to the long-standing unique prominence of inspiraling  binaries 
as GW sources, PN calculations have been the primary basis for the 
generation of theoretical gravitational waveforms. 

The state of the art in PN theory is currently focused on the $5$,
$5.5$, and $6$PN order via multiple approaches, including traditional
GR methods \cite{Bini:2019nra,Bini:2020wpo,Bini:2020uiq}, as well as
particle-physics inspired \cite{Goldberger:2004jt} effective field
theory (EFT) methods using Feynman technology
\cite{Blumlein:2019zku,Blumlein:2021txe,Bluemlein2022}.  
Starting at $2.5$PN, radiative effects become essential. The leading 
dissipative contribution, originally derived by Einstein, and later by 
Burke and Throne \cite{Einstein:1918btx,Burke:1970wx,Thorne:1969rba}, 
has come to be known as the ``radiation-reaction'' term.  As of $4$PN, 
the system dynamics must also account for a collection of phenomena 
known as``tail'' effects, in which radiation from the system scatters 
off of the system's own potential background
\cite{Blanchet:1987wq,Blanchet:1992br}.  The leading tail effect has
also been well-studied in the EFT context \cite{Goldberger:2009qd,
  Galley:2015kus, Foffa:2011np, Foffa:2019eeb}.  On the other hand,
the subleading ``tail-of-tail'' has received limited direct study
\cite{Blanchet:1997jj,Goldberger:2009qd}, and the sub-subleading
``tail-of-tail-of-tail'' (T$^3$) has only been computed via
traditional GR methods \cite{Marchand:2016vox}, without a counterpart
EFT computation.

It is also possible to release the small-velocity approximation used
in the PN expansion, and work instead with Special Relativity as the
base theory on top of which gravitational sources produce small
fluctuations, with $G_N$ serving as the only perturbation parameter.
This aptly named ``post-Minkowskian'' (PM) approximation has seen a
surge of interest in recent years, thanks in part to the close
similarity between gravitational-wave source calculations in PM and
computing effective potentials via scattering amplitudes
\cite{Cheung:2018wkq}.  The state of the art in PM calculations has
recently been pushed to $4$PM \cite{Bern2019, Bern2019a, Bern:2021yeh,
  Mogull2021, Jakobsen2022, Jakobsen2022a, Dlapa:2021vgp, Dlapa2023}, 
and is currently one of the driving forces in
understanding certain classes of Feynman integrals.  Even though 
these computations are carried out in the scattering regime,
there are methods for extracting quantities relevant to the bound
problem from them \cite{Cheung:2018wkq, Kaelin2020, Kaelin2020a,
  Cho:2021arx, DiVecchia:2022piu, DiVecchia:2021bdo}.  However, these
mappings between the bound and scattering regimes are expected to
break down beyond $4$PM as a result of the nonlocal-in-time
contributions coming from the tails and similar higher-order effects
\cite{Blanchet:1987wq, Blanchet:1992br, Cho:2021arx,
  Buonanno:2022pgc}.

Thus, the study of tail effects is critical both for their direct
relevance to the real-world PN sources, as well being a key piece for
possibly understanding the connection between bound and unbound
systems.  This work extends and pushes our study of higher-order tail
effects by building on our previous letter,
Ref.~\cite{Edison:2022cdu}. In that paper, we briefly introduced a
novel approach to calculating higher-order tail effects by exploiting
generalized unitarity methods \cite{Bern:1994zx, Bern:1994cg,
  Britto:2004nc, Anastasiou:2006jv}.  Using this approach, we were
able to compute the quadrupole-sourced tail effects up through T$^3$
at $G^4$, which matched state-of-the-art results from traditional GR
methods \cite{Marchand:2016vox}, and surpassed previous attempts using
EFT techniques \cite{Goldberger:2009qd,Galley:2015kus, Foffa:2011np,
  Foffa:2019eeb}.  In this work, we provide a significantly more
detailed discussion of the new methodology, including laying out the
constituent building-block amplitudes, as well as in-depth
calculations through the tail-of-tail.  As further novel results of
this work, we calculate through to the tail-of-tail-of-tail-of-tail
(T$^4$) contribution at G$^5$ to the quadrupole-quadrupole effective
action, \cref{eq:tttt-raw}, and energy dissipation,
\cref{eq:tttt-e-raw,eq:del-e-t4-inc}.  With the new energy-loss term,
we are able to compute new subleading RG flow of the quadrupole
source, \cref{eq:new-rg}, extending the results of
Refs.~\cite{Goldberger:2009qd, Goldberger:2012kf} to further allow for
prediction of all \emph{subleading} logs in tail-induced energy-loss.

The structure of this paper is as follows.  In the next couple of sections,
we provide a somewhat disjoint review of relevant material of EFT and 
Amplitudes methods for our hybrid approach to the computation of tail 
effects. In \cref{sec:eft}, we focus on the EFT setup for the tails problem.
We discuss the relevant separation of scales and the emergence of
tails as a phenomena that signals the breakdown of this separation.  
In \cref{sec:ctp} we lay out the closed-time-path (CTP) formalism adapted 
by Galley et al \cite{Galley:2012hx, Galley:2014wla}, from QFT to our 
classical context as a method of computing dissipative effective actions. 
Subsequently, in \cref{sec:unit} we present some of the basic results and 
modern methods common in the study of scattering amplitudes, and point out 
how they will be of use for the computation of tails.  

\Cref{sec:all-tails,sec:higher-tails} contain the heart of this paper.
In \cref{sec:all-tails} we elaborate on the elements of our novel
methodology, and demonstrate it on the leading radiation-reaction and
tail effects, which have been well-studied in terms of effective
actions.  In section \cref{sec:higher-tails}, we proceed to present
the computation of new effective actions of higher-order tails up
through T$^4$, the first ever computation of this effect, using
generalized unitarity methods.  We relegate details about integration
to the appendices \cref{sec:tens-red,sec:eval-ints}.  In
\cref{sec:renorm}, we first formulate the energy loss through the use
of the CTP approach for the binary inspiral, and we explicitly extract
the related contributions to the radiated energy. With this collection
of dissipation corrections in hand, we proceed to analyze them to
determine the renormalization and RG flow of the quadrupole source,
finding agreement with previous leading EFT results
\cite{Goldberger:2009qd, Galley:2015kus}, and extending the RG flow to
subleading order. Finally, in \cref{sec:compare} we cross-check our
new higher-order results against partial ones, known from PN and
self-force theory, where they overlap \cite{Marchand:2016vox,
  Fujita:2011zk, Fujita:2012cm}, and find perfect agreement.


\section{EFT of Binary as Composite Particle}
\label{sec:eft}
The effective field theory description of binary inspirals in PN
gravity has been formally defined since Goldberger and Rothstein's
seminal work \cite{Goldberger:2004jt}.  We briefly review it here.  We
direct interested readers to Refs.~\cite{Porto:2016pyg, Levi:2018nxp}
for more recent comprehensive reviews of the subject.

Starting from the binary PN assumptions of small velocity and weak
field, the two constituent massive objects have
\emph{non-relativistic} momenta given by
\begin{equation}
  p_i^\mu \sim (m_iv^2, m_iv)
\end{equation}
governed by two small quantities, approximately equated by the virial theorem,
\begin{equation}
  v^2 \sim \frac{G_N m}{r} \ll 1 ,
  \label{eq:pn-assump}
\end{equation}
with $m$ the characteristic mass of the gravitating particles, $r$
their orbital separation, and $v$ their characteristic orbital velocity.
The gravitational field due to the interaction between the two inspiraling 
bodies can then be split into two graviton modes: 
\begin{equation}
  k^{\mu} \sim
  \begin{cases}
    (v/r,1/r) & \text{potential (near zone)}\\
    (v/r,v/r) & \text{radiation (far zone)}
  \end{cases} \,,
\end{equation}
and the recoil from each of the massive bodies interacting with the
gravitons is assumed to be negligible, which allows to handle the 
components as classical sources on non-dynamical worldlines.
The momentum of the potential modes has a dominant spatial
component, and thus they are treated as space-like instantaneous
mediators.  As their name suggests, these modes are responsible for the
gravitational binding of the two-body system.  From these
considerations, a full effective action at the orbital scale was
defined in Ref.~\cite{Goldberger:2004jt} with manifest
power-counting following \cref{eq:pn-assump}.  This effective action
has been used extensively for computations of the binding energy of
gravitational binaries, and even extended to
include spin-induced effects of the binary \cite{Porto:2005ac, Porto:2008tb,
  Porto:2008jj, Porto:2007qi, Levi:2010zu, Levi:2008nh,Kol:2010ze, 
  Levi:2011eq, Levi:2014sba, Levi:2015msa, Levi:2014gsa, Levi:2016ofk, 
  Levi:2015uxa, Levi:2015ixa, Levi:2017kzq, Maia:2017gxn, Maia:2017yok, Levi:2019kgk,Levi:2020lfn,Levi:2020kvb,Levi:2020uwu,
  Kim:2021rfj,Kim:2022pou,Kim:2022bwv,Levi:2022dqm,Levi:2022rrq}. 

Beyond the orbital-scacle conservative sector, namely when radiation
modes also participate in interactions, it is beneficial to consider
as a starting point the entire binary system as a single point
particle moving on a worldline, with its internal structure modeled by
multipole moments coupled to gravity.  This effective action of the
binary as a composite particle, which is analogous to that of the
single compact object with its spin-induced multipoles at the orbital
scale, is given by \cite{Goldberger:2009qd, Levi:2010zu, Ross:2012fc,
  Levi:2015msa,Levi:2018nxp}:
\begin{align}
	S_{\text{eff(c)}}[g_{\mu\nu},y_c^\mu,e_{c\,A}^{\,\,\mu}]
	&= 
	-\frac{1}{16\pi G}\int d^4x
	\sqrt{g}\,R\left[g_{\mu\nu}\right] +\, S_{\text{pp(c)}}
	[g_{\mu\nu}(y_c),y_c^\mu,e_{c\,A}^{\,\,\mu}](\sigma_c)\,,
\end{align}	
with
\begin{align}	
	S_{\text{pp(c)}} [h_{\mu\nu},y_c^\mu,e_{c\,A}^{\,\,\mu}](t)
	&= -\int dt \sqrt{g_{00}}\, \biggr[ E(t) 
	+ \frac{1}{2}\epsilon_{ijk}L^{k}(t)\left(\Omega_{\text{LF}}^{ij}
	+\omega_\mu^{ij}u^\mu\right) \notag \\
	&\qquad- \sum_{l=2}^{\infty} 
	\biggr(\frac{1}{l!}I^{L}(t)\nabla_{L-2}\mathcal{E}_{i_{l-1}i_{l}}
	- \frac{2l}{(l+1)!}J^{L}(t)\nabla_{L-2}\mathcal{B}_{i_{l-1}i_{l}}\biggr)\biggr]\,
	\label{eq:tail-eft}
\end{align}
in terms of the time coordinate $t$ as the composite-particle
worldline parameter.  The point-particle action includes gravitational
couplings to the particle's total energy $E(t)$, its angular momentum
$L^{k}(t)$, and higher multipoles of charge and current type with
definite parity, $I^{L}(t)$ and $J^{L}(t)$ respectively, bearing
symmetric traceless SO$(3)$ (spatial Euclidean) tensor indices.
$\mathcal{E}$ and $\mathcal{B}$ are the respective even- and
odd-parity components of the gravitational curvature tensor
\footnote{$\Omega_{\text{LF}}$ is the generalized angular velocity in
  the local frame, and $\omega$ the Ricci rotation coefficient or spin
  connection, though they will not be relevant to the present work.}.
In the current work, we limit ourselves to the leading (static)
gravitating energy, $E(t) \to E + \mathcal{O}(G_N)$, ignoring various
subleading corrections due to the gravitational interactions.

Since gravity is self-interacting, integrating out the gravitational
field, starting from this EFT of the composite particle, will involve
fully analyzing interactions that include both potential and radiation
modes, as the separation of scales inevitably breaks down at a
sufficiently high perturbative order of the EFT.  The simplest class
of such effects is the scattering of a radiation-mode graviton with
one or more potential-mode gravitons. The interaction with a single
potential mode is referred to as the ``tail'' effect.  We refer to
interactions with $n$ potential modes as (tail-of-)$^{n-1}$tail or
T$^n$ for brevity \footnote{Note that the nomenclature used by
  Blanchet \cite{Blanchet:1997jj, Blanchet:2013haa} makes a
  distinction between, for instance, tail$^2$ and tail-of-tail. In our
  EFT approach it is not particularly useful to make such a
  distinction.}.

Successive terms in the multipole expansion carry increasing powers of
radiation-mode momenta.  As such, the tails related with each of these
multipole sources enter at staggered orders in perturbation theory. In
this work we analyze effects that are sourced only by quadrupoles,
which yield the leading PN contributions in growing orders of the
gravitational coupling constant, $G_N$.  As we will see below, the
analysis of this EFT of the composite particle at the radiation scale
requires regularizing and renormalizing ultraviolet divergences.  One
can simply follow the standard method to handle renormalization in an
EFT by introducing renormalized couplings, which means in this case
modifying the coefficients of the multipole source terms to absorb the
divergences. Matching with the orbital-scale EFT would align such
ultraviolet divergences with infrared divergences in the small-scale
theory.

Two approaches to addressing the tails have been presented in the
literature. The first we refer to as the ``one-point'' formalism,
which was used by Goldberger and Ross to construct the gravitational
radiation from the tail and tail-of-tail \cite{Goldberger:2009qd}.  In
this setup, one computes the graviton one-point amplitude,
$\mathcal{A}_h(k^\mu) \sim \pol_{ij}(k^\mu)I^{ij}(k^0)$, with the
(Fourier transform of the) quadrupole, $I^{ij}(\omega)$, as a
classical source.  From the one-point amplitude it is simple to
construct a graviton differential emission rate, which can then be
appropriately weighted and integrated to extract radiation effects,
for instance the radiated four-momentum.  This approach is elegant
and well-motivated physically when all one desires is extracting
dissipative features.  However, it is ill-suited for extracting the
effect on the \emph{conservative dynamics} of the binary induced by
the tails, which are of phenomenological import
\cite{Blumlein:2021txe, Levi:2018nxp, Bini:2020wpo, Jaranowski:2012eb, 
	Bluemlein2022, Bini:2019nra,Porto:2016pyg, Galley:2015kus}.

The other approach, in which we base the current work and which does
also account for the conservative dynamics, computes the effective
``two-point function'' of the quadrupole on the worldline that results
from \emph{integrating out} gravitational interactions with the
quadrupole source. In this picture we can consider the gravitational
field integrated out as an inaccessible degree of freedom, from which
the quadrupole can gain or lose energy. The final result will be an
effective action which takes the following form in frequency domain:
\begin{equation}
  S^{\text{eff}}
  = \int \diff \omega\ f(\omega) I^{ij}(\omega)I_{ij}(-\omega)\,,
\end{equation}
where the shorthand
$\kappa(\omega) \equiv I^{ij}(\omega)I_{ij}(-\omega)$ will be useful
\cite{Bini:2021qvf}.  Because the result is an effective action for
the evolution of a quadrupole, we have access to both the conservative
dynamics through a Lagrangian, and to radiative observables, via some
generalized calculus of variations within the Closed-Time-Path (CTP)
formalism, see the following \cref{sec:ctp} and later
\cref{sec:ctp-energy}.  This CTP approach was put forward and
popularized by Galley et al, starting in \cite{Galley:2009px}, and has
since been adopted as the standard approach in EFT computations of
tails \cite{Galley:2012qs, Galley:2015kus,Foffa:2019eeb,
  Blanchet:2019rjs, Almeida:2021xwn}.  Prior to our
recent letter \cite{Edison:2022cdu}, there had been no attempt to use
this approach to tackle higher-order (and unknown) tails.


\subsection{Closed-Time-Path Formalism}
\label{sec:ctp}

The nonconservtaive sector requires a more intricate treatment since
the radiating binary is in fact an open system as its leaking energy
via gravitational waves, so time reversal no longer holds beyond the
conservative sector.  As we noted above, while specific setups can be
used to model the radiative features of the binary
\cite{Goldberger:2009qd}, these still run into difficulties
disentangling the causal radiation effects
\cite{Goldberger:2012kf}. Over the last decade, it has become clear
that care must be taken when handling the nonconservative effects
present in inspiraling binaries \cite{Galley:2009px, Galley:2012hx,
  Galley:2012qs, Galley:2014wla}.  In fact,
the approach detailed in Galley et al \cite{Galley:2012qs,
  Galley:2014wla} successfully dealt with radiation reaction and tails
at the level of the effective action \cite{Galley:2009px,
  Galley:2012hx,Galley:2015kus}. This approach is based on the
closed-time-path (CTP) (or ``in-in'') formalism
\cite{kadanoff1962quantum,Keldysh:1964ud,
  Galley:2012hx,Galley:2014wla}, and also fully accounts for the
conservative effects due to tails.

The CTP approach adopted to our worldline EFT provides a classical method of 
integrating out the gravitational field degrees of freedom while maintaining 
time-asymmetry at the level of the resulting effective action.  
This is achieved by formally \emph{doubling all degrees of freedom} in the 
initial full action of the system, and defining the initial CTP
action as: 
\begin{equation}
  \label{eq:ctp-init}
  S_{\text{CTP}} [\ctpdof{1},\ctpdof{2}] = S[\ctpdof{1}] - S^*[\ctpdof{2}] \,,
\end{equation}
where $\ctpdof{i}$ denotes the full set of degrees of freedom of the
system, including all worldline degrees of freedom, and all the field
modes which we plan on integrating out.  After integrating out the
field degrees of freedom, we will obtain a CTP effective action of the
form:
\begin{equation}
  \label{eq:ctp-eff}
  S_{\text{CTP}}^{\text{eff}} = \int \diff t\ \left[
    L(\ctpdofb{1}, t) - L(\ctpdofb{2}, t)
    + K(\ctpdofb{1}, \ctpdofb{2},t)\right]
\end{equation}
where $\ctpdofb{1/2}$ are the remaining worldline degrees of freedom,
$L$ is identified as the \emph{conservative} Lagrangian, and $K$
represents the \emph{nonconservative} potential.  While the initial
action does not contain such a history-mixing term, the process of
integrating out some of the degrees of freedom will produce an
effective mixing contribution.  Once we have obtained the effective action in
terms of the doubled variables, we extract physical dynamics and
observables by varying the action with respect to the $\ctpdofb{1}$
variables, e.g., and then taking the \emph{physical limit} (PL),
$[\ctpdofb{1}-\ctpdofb{2}]|_{\text{PL}} \equiv 0$.

Prior to integrating out it is more useful to perform a change of
variables to $\ctpdof{+} \equiv [\ctpdof{1} + \ctpdof{2}]/2$, and
$\ctpdof{-} \equiv \ctpdof{1} - \ctpdof{2}$, which leads to a modified
propagator matrix:
\begin{equation}
G_{+-} = G_{\text{adv}}, \quad G_{-+} = G_{\text{ret}} 
; \quad 	G_{++} = G_{--} = 0,
\end{equation}
where the scalar component of the graviton propagator is given by:
\begin{equation}
	\label{eq:ctp-props}
	G_{\text{ret}/\text{adv}}(x-x') = \int \frac{\Diff{D} p}{(2 \pi)^D} 
	\frac{e^{-i p_{\mu} (x-x')^{\mu}}}{(p^0 \pm  \ctpeps)^2 - |\vec{p}|^2}
\end{equation}
in the mostly-minus metric convention, with 
$p^0_\text{ret} \equiv p^0+ \ctpeps$ for retarded, 
$p^0_\text{adv} \equiv p^0-\ctpeps$ for advanced, and where
$D\equiv d+1$ with $d$ the number of spatial dimensions of $\vec{p}$. 
In this basis, the conservative contribution to the resulting effective 
action in \cref{eq:ctp-eff} is identified as the part that is
symmetric under $\ctpdofb{+} \leftrightarrow \ctpdofb{-}$, while the
remaining terms are identified as the nonconservative $K$.
Observables in this basis are extracted by performing the calculus of
variations with respect to the $\ctpdofb{-}$ variables, after which
the physical limit
$\ctpdofb{+} \to \ctpdofb{\text{PL}},\ \ctpdofb{-} \to 0$ is
applied.  In \cref{sec:ctp-energy} below, we will derive the 
dissipated energy in the CTP formalism, also specialized in particular 
to the case of tails in binary inspirals.

As discussed in \cref{sec:eft}, the effective action due to tails, which 
amounts to a two-point function of the mass quadrupole, results from 
integrating out its coupling to the gravitational field. Carrying out this 
task in the CTP framework is rather straightforward 
\cite{Galley:2009px, Galley:2012qs,  Galley:2015kus}.  
First, we endow the quadrupoles with CTP labels,
$I^{ij}(\omega) \to I^{ij}_{a}(\omega)$,
$\kappa(\omega) \to \kappa_{ab}(\omega)$ (we use $a$,$b$ for CTP
indices, reserving Latin letters near $i$,$j$ for space-like indices).
Then we sum over the possible CTP labels for the two quadrupoles,
while making consistent CTP label choices for the internal
radiation-mode gravitons.  In the case of tails, this consistent label
choice amounts to having all radiation-mode propagators,
$G^{\text{rad}}_{ab}$, aligned with the quadrupole labels, \eg
\begin{equation} 
  I^{ij}_{-}(\omega) G^{\text{rad}}_{-+}G^{\text{rad}}_{-+} \dots I^{ij}_{+}(-\omega)
  \,.
\end{equation}
Because the CTP propagators address causal propagation and the
potential-mode gravitons are taken to be instantaneous, we do not dress
them with CTP labels.


\section{Amplitudes and Generalized Unitarity}
\label{sec:unit}
The study of scattering amplitudes via the unitarity paradigm has a
long and storied history, and this work is not intended as a review of
the field.  Instead we mention a few specific points that are relevant
for the work at hand, so that non-experts have a point of reference.
For readers interested in more details, we refer to
Refs.~\cite{Elvang:2013cua,Elvang:2015rqa,
  Dixon:2013uaa,Carrasco:2015iwa} as broad introductions to the
subject.  For discussions particularly centered on multi-loop
unitarity methods, we refer the reader to Refs.~\cite{Britto:2004nc,
  Bern:2015ooa, Bourjaily:2017wjl, JJHenrikReview, FivePointN4BCJ,
  Bern:2012uf, Bern:2019prr, Edison:2022jln,
  Edison:2022smn,Carrasco:2023qgz}.


\subsection{Tree Amplitudes}
Tree amplitudes have a number of basic properties that nonetheless
serve as focus points for their study.  They are a description of
\emph{local} scattering interactions between \emph{on-shell} external
particles that are \emph{gauge invariant} and obey
\emph{factorization} rules.  By on-shell, we mean that all of the
external particles have energy-momentum vectors obeying
$p_i^2 = m_i^2$ where $m_i$ is the particle's rest mass.  

Amplitudes
describing massless spin-1 particles are invariant under linearized
gauge transformations, \ie if we explicitly factor polarization
vectors out of an amplitude, they will obey the Ward identity
\begin{equation}
  \mathcal{A}_n = \mathcal{A}_{n\,\mu_1 \cdots \mu_n} \prod \pol_i^{\mu_i}
  \to p_{i}^{\mu_i} \mathcal{A}_{n\,\mu_1 \cdots \mu_n} \prod_{j \neq i} \pol_j^{\mu_j}= 0 \,.
\end{equation}
Amplitudes of massless spin-2 particles (gravitons) are invariant
under linear diffeomorphisms, realized via
\begin{equation}
  \mathcal{M}_n = \mathcal{M}_{n\,\mu_1 \nu_1 \cdots \mu_n \nu_n} \prod \pol_i^{\mu_i}\pol_i^{\nu_i}
  \to p_{i}^{\mu_i}\pol_i^{\nu_i} \mathcal{M}_{n\,\mu_1 \cdots \mu_n} \prod_{j \neq i} \pol_j^{\mu_j}\pol_j^{\nu_j}
  = 0 \,.
\end{equation}
Here we have also introduced a standard notation for dealing with
spin-2 amplitudes: since we require graviton polarizations to be
symmetric traceless tensors we write them as an outer product of two
identical null polarization vectors
\begin{equation}
  \pol^{\mu \nu} \to \pol^\mu \pol^\nu \ ,\qquad \pol^\mu \pol_\mu = 0 \,.
\end{equation}

Locality is the statement that the amplitude can be expressible in
terms of point-like interactions, which in momentum space translates
to only allowing interactions that are polynomials of momenta and
polarizations.  Similarly, factorization is a statement about the
types of pole structures that appear in momentum space.  Concretely,
factorization requires that the residue of a momentum-space amplitude
on a configuration where a sum of external momenta goes on-shell must
be equal to a product of lower-point amplitudes summed over all
theory-allowed intermediate on-shell states
\begin{equation}
  \res_{(p_1 + \dots + p_j)^2=0} \mathcal{A}(1,\dots,j,j+1,\dots,n) = \sum_{\text{states of }i} \delta(p_i^2) \mathcal{A}(1,\dots,j,i) \mathcal{A}(i,j+1,\dots,n) \,.
\end{equation}
An example of the sum over graviton states is in \cref{eq:state-sum} below.  
Graphically, we often represent factorization as
\begin{equation}
\begin{tikzpicture}[scale=0.5]
          \pgfmathsetmacro{\r}{1.1}
          \pgfmathsetmacro{\l}{0.6}
          \pgfmathsetmacro{\y}{-2.2}
          \foreach \x in {1,2,3,4} {
            \pgfmathsetmacro{\a}{135-90*(\x-2)}
            \pgfmathsetmacro{\b}{90-90*(\x-2)}
            \coordinate (\x) at ( \a : \r);
            \draw[thick] (\x) -- ++ ( \a : \l) node[label={[label distance=-8pt]\a:\x}] {};
          }
          \coordinate (c1) at (0,1.5);
          \coordinate (c2) at (0,-1.5);    
          \filldraw[blobcolor] (0,0) circle (\r);
          \draw[cuts] (c1)--(c2);
          \filldraw[blobcolor] (0,0) circle (\l);
    \node[font=\normalsize] at (0,0) {\(\mathcal{A}_4\)};
    \draw (0,0) circle (\r);
  \end{tikzpicture}
  =\begin{tikzpicture}[scale=0.5]
    \pgfmathsetmacro{\r}{1.1}
    \pgfmathsetmacro{\l}{0.6}
    \node[font=\normalsize,circle,fill=blobcolor,draw=black,inner sep=1pt](al) at (-1.5,0) {\(\mathcal{A}_3\)};
    \node[font=\normalsize,circle,fill=blobcolor,draw=black,inner sep=1pt] (ar) at (1.5,0) {\(\mathcal{A}_3\)};
    \draw[thick] (al) -- ++(135:{\r+\l}) node[label={[label distance=-8pt]135:2}] {};
    \draw[thick] (al) -- ++(135+90:{\r+\l}) node[label={[label distance=-8pt]135+90:1}] {};
    \draw[thick] (al) -- (ar);
    \draw[thick] (ar) -- ++(45:{\r+\l}) node[label={[label distance=-8pt]45:3}] {};
    \draw[thick] (ar) -- ++(45-90:{\r+\l}) node[label={[label distance=-8pt]45-90:4}] {};
    \draw[cuts] (0,1) -- (0,-1);
  \end{tikzpicture} \,,
\end{equation}
where we used \textcolor{cutcolor}{dashed colored lines} in the
diagram to highlight the cut internal legs.
While it may seem like factorization is a direct consequence of
locality, geometric constructions of amplitudes like the Amplituhedron
\cite{Arkani-Hamed:2013jha} show that it is possible to manifest
factorization without manifesting locality.

In addition to these standard properties, amplitudes involving
non-Abelian charges, \eg Yang--Mills theory, exhibit
color-kinematics duality \cite{Bern:2008qj,Bern:2019prr}.  The
amplitudes in these theories can be written in terms of numerators
dressing only cubic diagrams, where the kinematic piece of the
numerators obeys the same algebraic relations as the non-Abelian color
charge factors.  These numerators can then be ``double-copied'' by
replacing the color factors with another set of kinematic numerators,
leading to amplitudes in uncolored theories.  The most relevant
example for the current work is that tree amplitudes for gravitational
interactions, both self-interactions and coupling to matter, can be
generated as the double-copy of Yang-Mills amplitudes, also
potentially involving matter \cite{Bern:2010yg, Bern:2010ue,
  Bern:2019prr}.

Over the years, different tree construction methods have been
developed which manifest different properties of amplitudes.  For
instance, direct calculation via standard QFT Feynman rules manifests
locality, factorization, and connection to path integrals and action
principles, but obscures gauge invariance and relationships between
theories. On the other hand, Britto-Cachazo-Feng-Witten
recursion-based constructions
\cite{Britto:2004ap,Britto:2005fq,Arkani-Hamed:2009pfk,Arkani-Hamed:2012zlh}
manifest gauge invariance, on-shell conditions, and high-energy
behavior at the cost of no longer manifesting locality.  Such
recursion relations have been applied in studying candidates for ``black
hole + gravity'' Compton amplitudes
\cite{Arkani-Hamed:2017jhn,Chung:2018kqs,Damgaard:2019lfh,Aoude:2020onz}.
Of primary relevance to the current work is the Cachazo-He-Yuan
formulation of scattering amplitudes
\cite{Cachazo:2013iea,Cachazo:2013hca}, which manifests the
double-copy relations between gauge theory amplitudes and
gravitational amplitudes at the cost of introducing an auxiliary
space.  One of the current authors and Fei Teng developed the publicly
available package
\href{https://gitlab.com/aedison/increasingtrees}{IncreasingTrees} for
efficiently computing gauge and gravity tree amplitudes, including
minimal matter couplings, in this formalism \cite{Edison:2020ehu}.
Using it, we are able to extract all of the building-block amplitudes
needed for the tail computations, which we discuss below in
\cref{sec:tail-bb}.

\subsection{Generalized Unitarity Cuts}
The central concept of the unitarity program is the combining of tree
amplitudes into loop data via \emph{generalized unitarity cuts}.  The
core of generalized-unitarity methods comes from two connected ideas.
First, through tensor reduction and integration relations
\cite{Passarino:1978jh,Laporta:1996mq,Smirnov:2006ry,Smirnov:2012gma},
loop amplitudes can be written in terms of a basis of scalar integrals 
(and often purely-propagator integrands). When written in this basis, 
the coefficient of each basis integral is a theory-dependent algebraic
function of the external data and spacetime dimension.  Writing the
amplitude in this way often exposes significant simplifications and
patterns.  An important and well-studied example is the four-point
one-loop amplitude for any theory, which can be written as
\cite{vanNeerven:1983vr, Bern:1992em, Bern:1993kr, Brown:1952eu,
  tHooft:1978jhc,Passarino:1978jh}
\begin{equation}
   \mathcal{A}_4^{(1)} = 
  \begin{tikzpicture}[scale=0.5]
    \pgfmathsetmacro{\r}{1.1}
    \pgfmathsetmacro{\l}{0.6}
    \pgfmathsetmacro{\y}{-2.2}
    \foreach \x in {1,2,3,4} {
      \pgfmathsetmacro{\a}{135-90*(\x-1)}
      \pgfmathsetmacro{\b}{90-90*(\x-1)}
      \coordinate (\x) at ( \a : \r);
      \draw[thick] (\x) -- ++ ( \a : \l) node[label={[label distance=-8pt]\a:\x}] {};
    }
    \filldraw[blobcolor] (0,0) circle (\r);
    \draw (0,0) circle (\r);
    \filldraw[white] (0,0) circle (\l);
    \draw (0,0) circle (\l);
  \end{tikzpicture}
= c_{\text{box}} \int
  \begin{tikzpicture}[scale=0.4]
    \pgfmathsetmacro{\r}{1.1}
    \pgfmathsetmacro{\l}{0.6}
    \pgfmathsetmacro{\y}{-2.2}
    \foreach \x in {1,2,3,4} {
      \pgfmathsetmacro{\a}{135-90*(\x-1)}
      \pgfmathsetmacro{\b}{90-90*(\x-1)}
      \coordinate (\x) at ( \a : \r);
      \draw (\x) -- ++ ( \a : \l) node[label={[label distance=-8pt]\a:\x}] {};
    }
    \draw (1) -- (2) --(3) --(4)--(1);
  \end{tikzpicture}
+ c_{\text{s-bub}}\int
  \begin{tikzpicture}[scale=0.4]
    \pgfmathsetmacro{\r}{1.1}
    \pgfmathsetmacro{\l}{0.6}
    \pgfmathsetmacro{\y}{-2.2}
    \coordinate (i1) at (0,\l);
    \coordinate (i2) at (0,-\l);
    \foreach \x in {1,2,3,4} {
      \pgfmathsetmacro{\a}{135-90*(\x-1)}
      \pgfmathsetmacro{\b}{90-90*(\x-1)}
      \node[label={[label distance=-10pt]\a:\x}] (\x) at ( \a : \l+\r) {};
    }
    \draw (0,0) circle (\l);
    
    \draw (1) -- (i1)--(2);
    \draw (3)--(i2) --(4);
  \end{tikzpicture}
+ \text{perms} \ ,
\label{eq:basis-decomp}
\end{equation}
in which $c_{\text{box}}$, $c_{\text{s-bub}}$, and what is covered by
``perms'' are all theory dependent.  

While expressing amplitudes in
terms of scalar integral bases is a useful organizing principle on its
own, its real power comes with the help of the second observation: the
perturbative QFT optical theorem can be used to directly construct the
basis coefficients $c_i$.  The key idea, developed by Bern, Dixon,
Dunbar, and Kosower (BDDK) \cite{Bern:1994zx,Bern:1994cg}, is that
unitarity of the $S$-matrix perturbatively requires
\begin{equation}
  -i (T - T^\dagger) = T^{\dagger} T \Rightarrow
  2 \im \left(
  \begin{tikzpicture}[scale=0.5]
    \pgfmathsetmacro{\r}{1.1}
    \pgfmathsetmacro{\l}{0.6}
    \pgfmathsetmacro{\y}{-2.2}
    \foreach \x in {1,2,3,4} {
      \pgfmathsetmacro{\a}{135-90*(\x-1)}
      \pgfmathsetmacro{\b}{90-90*(\x-1)}
      \coordinate (\x) at ( \a : \r);
      \draw[thick] (\x) -- ++ ( \a : \l) node[label={[label distance=-8pt]\a:\x}] {};
    }
    \filldraw[blobcolor] (0,0) circle (\r);
    \draw (0,0) circle (\r);
    \filldraw[white] (0,0) circle (\l);
    \draw (0,0) circle (\l);
  \end{tikzpicture} \right)
= \int \diff \text{LIPS}
  \begin{tikzpicture}[smooth]
    \node (p1) at (0,0) {\(1\)};
    \node (p2) at (4,0) {\(2\)};
    \node (p3) at (4,-1.5) {\(3\)};
    \node (p4) at (0,-1.5) {\(4\)};

    \node (il1) at (1,-0.4) {};
    \node (il2) at (1,-1.1) {};
    \draw[thick] (p1) --(il1);
    \draw[thick](p4)--(il2);
    \draw[thick] (il1) -- ++(.75,0);
    \draw[thick] (il2) -- ++(.75,0);

    \node (ir1) at (3,-0.4) {};
    \node (ir2) at (3,-1.1) {};
    \draw[thick] (p2) --(ir1);
    \draw[thick](p3)--(ir2);
    \draw[thick] (ir1) -- ++(-.75,0);
    \draw[thick] (ir2) -- ++(-.75,0);
    \node[draw=black,fill=blobcolor,ellipse,minimum height=30pt,inner sep=0,align=center] (lb) at (1,-0.75) {Tree};
    \node[draw=black,fill=blobcolor,ellipse,minimum height=30pt,inner sep=0] (rb) at (3,-0.75) {Tree};
    \node[font=\huge] (ox) at (2,-0.75) {\(\otimes\)};
    \node[font=\tiny] at ($(il1)+(.75,.3)$) {\(\ell\)};
    \node[font=\tiny] at ($(il2)+(.75,-.3)$) {\(\ell+k_2+k_3\)};
  \end{tikzpicture}\,,
\label{eq:unitarity-ex}
\end{equation}
where the two trees on the right hand side are fully on-shell
amplitudes, $\diff \text{LIPS}$ is the Lorentz invariant phase space 
measure encoding the loop integral measure and the on-shell cut 
conditions, 
and the $\otimes$ instructs us to sum over all possible on-shell states 
crossing the cut and integrate over the on-shell phase
space of the internal particles.  
We are limiting the drawing to four
external particles in keeping with the example of
\cref{eq:basis-decomp}.  This relation tells us that the branch cut
structure of the integrated amplitude, probed by the left-hand side of
the equality, is related to pole structures of tree amplitudes.  BDDK
demonstrated that careful repeated application of
\cref{eq:unitarity-ex} completely determines the integrated amplitude
\cite{Bern:1994zx,Bern:1994cg}.  

Combination and refinement of these
two ideas leads to relating the right hand side of
\cref{eq:unitarity-ex} with linear combinations of the integral basis
coefficients \cite{Bern:1994zx, Bern:1994cg, Britto:2004nc,
  Anastasiou:2006jv}.  The relations can be exploited algorithmically
by not explicitly evaluating the $\int \diff \text{LIPS}$ in
\cref{eq:unitarity-ex}, but instead using integration reduction to
re-express the rational function of loop momenta generated by the
product of trees in terms of the basis integrals
\cite{Britto:2007tt,Ossola:2006us,Forde:2007mi}.  The on-shell
conditions $\delta(\ell^2)$ and $\delta((\ell+k_2+k_3)^2)$ inside of
$\diff \text{LIPS}$ are then taken as a restriction to only keep basis
elements in the reduction with at least those propagators.

All of the above can be generalized to higher loop orders via
\emph{generalized-unitarity cuts}.  For a graph $G$ in a particular
theory, its generalized-unitarity cut is defined as
\begin{equation}
  \cut_{\graph{}}
  = \sum_{\substack{\text{states}\\\text{of } E(\graph{})}}
  \prod_{v \in V(\graph{})} \mathcal{A}_{\text{tree}}(v) \ ,
  \label{eq:gen-cut}
\end{equation}
where $E(G)$ are the internal edges of G, all of which are taken
on-shell, $V(G)$ are the vertices of $G$, and $\mathcal{A}_{\text{tree}}$ are
the relevant tree amplitudes for the theory under consideration.  For
instance, we can write an iterated version of \cref{eq:unitarity-ex}
for gravitons as
\begin{equation}
  \label{eq:iter-unit}
  \begin{tikzpicture}
    \pgfmathsetmacro{\er}{1.75}
    \pgfmathsetmacro{\etheta}{20}
    \node (e1) at (180+\etheta:\er) {1};
    \node (e2) at (180-\etheta:\er) {2};
    \node (e3) at (\etheta:\er) {3};
    \node (e4) at (-\etheta:\er) {4};
    \node[blob={10pt}] (a1) at (180:0.5*\er){};
    \node[blob={10pt}] (a2) at (0,0){};
    \node[blob={10pt}] (a3) at (0:0.5*\er){};
    \begin{pgfonlayer}{background}
      \draw[photon big] (e1) -- (a1);
      \draw[photon big] (a1) -- (e2);
      \draw pic {cut vhill={a1}{a2}};
      \draw pic {cut dhill={a1}{a2}};
      \draw[photon big] (e3) -- (a3);
      \draw[photon big] (a3) -- (e4);
      \draw pic {cut vhill={a2}{a3}};
      \draw pic {cut dhill={a2}{a3}};
    \end{pgfonlayer}
  \end{tikzpicture}
 =
  \!\!\!\!\! \sum_{\substack{\text{states of }\\ \{\ell_1, \ell_2, \ell_3, \ell_4\}}} \!\!\!\!\!\!\!
  \mtree(1,2,\ell_1,\ell_2) \mtree(-\ell_1,-\ell_2,\ell_3,\ell_4) \mtree(-\ell_3,-\ell_4,3,4).
\end{equation}
Important to note is that it is often impossible to solve all of the
cut conditions defining the on-shell loop momenta in terms of
real-valued Lorentzian kinematics, so generalized unitarity methods
are understood as complex-analytic tools for determining what rational
functions need to be integrated.  The cuts themselves generally do not
have support on the physical integration contour that produces the
amplitude.

The sum over states in \cref{eq:gen-cut} hides most of the complexity
of cut construction.  Since the focus of the current paper involves
gravitational interactions, we restrict our attention to the
particular case of only internal gravitons.  In this situation,
evaluating the state sum for each of the edges crossing the cut
involves inserting a complete set of graviton states in $D$ spacetime 
dimensions via
\begin{align}
  \sum_{\text{states}} \pol^{\mu \nu}_k \pol^{\alpha \beta }_k 
  \equiv \mathcal{P}^{\mu \nu; \alpha \beta}_k 
  &= \frac{1}{2} \left( P_k^{\mu \alpha}P_k^{\nu \beta} + P_k^{\mu \beta}P_k^{\nu \alpha} - 
    \frac{2}{D-2} P_k^{\mu \nu}P_k^{\alpha \beta} \right) \ , 
    \label{eq:state-sum}\\
  P^{\mu \nu}_k &\equiv \eta^{\mu \nu} - \frac{k^\mu q^\nu + k^\nu
                  q^\mu}{k \cdot q } \ ,
\end{align}
in which $q^\mu$ is a null reference vector.  The presence of $q$
serves to make \cref{eq:state-sum} analogous to a gauge-agnostic
graviton propagator (with $(k^2)^{-1}$ replaced with an implicit
$\delta(k^2)$).  In fact, gauge invariance of a cut with respect to
the internal states manifests as $q$-independence of the cut.
Explicitly removing the $q$ dependence can be computationally
intensive for complicated cuts.  Ref.~\cite{Kosmopoulos:2020pcd}
provides an excellent discussion for effective ways of dealing
with these types of $D$-dimensional state sums.

When a basis of integrals for a particular problem is known,
generalized unitarity cuts can be used to identify the basis
coefficients via
\begin{equation}
  \frac{\text{Cut}_{\graph{}}}{|\graph{}|} = \sum_{\substack{\mathcal{I}_i \text{ has propagators}\\\text{compatible with }E(\graph{})}} c_i \mathcal{I}_{i} \ ,
  \label{eq:cut-matching}
\end{equation}
with $|\graph{}|$ the number of symmetries of the graph and where the
sum is over integral basis elements which have at least the same
propagators as $E(G)$.  Integral basis identification and construction is
often a highly non-trivial task involving many subtleties.  As such,
significant effort has been put into identifying particularly good
choices of bases \cite{Bourjaily:2017wjl} and developing
basis-agnostic methods \cite{Gardi:2022wro,Frellesvig:2020qot}.
Luckily, we will see below that the tails have relatively simple and
easy-to-identify integral bases.  Even better, the matching between
cuts and integral-basis coefficients is nearly a direct equality,
modulo details about ``non-planar'' channels that will be discussed in
situ.

\section{The Tail Effect}
\label{sec:all-tails}
The effective field theory approach has been extremely successful at
using analogies with particle physics to improve the understanding of
gravitational dynamics in the bound two-body problem.  Further bringing 
modern amplitudes insights to bear has pushed the frontier for hyperbolic 
encounters in the scattering problem 
\cite{Bern2019, Bern2019a, Bern:2021yeh, Mogull2021, Jakobsen2022,
  Jakobsen2022a, Dlapa:2021vgp, Dlapa2023}, and led to new
developments in direct observable computations \cite{Kosower:2018adc,
  Herderschee:2023fxh}.  
However, in this work as well as in our preceding letter \cite{Edison:2022cdu}, 
we advocate applying both particle analogies and amplitudes methods 
\emph{directly at the level of the composite binary to the gravitational radiation in the bound 
two-body problem}.  
The long-standing analogy between interactions 
of spin-$l/2$ elementary particles, and of classical $l$-th multipole moments, 
and especially higher spin-induced multipoles $S^l$, \cite{Holstein:2008sx, Levi:2015msa}, applied in \cite{Levi:2014gsa,Vaidya:2014kza},
suggests that we can go further, and model even the multipole moments of 
the binary itself -- in terms of
fundamental particles interacting with gravity.  By working with
scattering amplitudes for the particle interactions, rather than
Feynman rules, we will be able to construct the tail effective actions
by combining gauge-invariant on-shell objects via the method of
generalized unitarity.  Doing so removes the need to care about
graviton gauge choices, allows exploiting developments in amplitudes
construction and integration, and more directly highlights the
patterns that appear throughout the tails.

It is also worthwhile to discuss the link between the two-point and
one-point EFT approaches to tails, which ties in significantly
with a unitarity-based perspective
.  
As discussed in \cref{sec:eft}, the one-point approach
deals with calculating the unpolarized cross-section of a graviton
one-point function in the normal way, $\sum_h |\mathcal{A}_h|^2$.
However, this process is almost exactly equivalent to computing
\emph{generalized unitarity cuts} (see \cref{sec:unit}) in the
two-point approach (up to shuffling of terms related to the
regularization schemes). This is since the latter entails 
\emph{inserting a complete set of graviton states between 
	two on-shell amplitudes} as follows:
\begin{equation}
	\int \diff \text{LIPS} \sum_h |\mathcal{A}_h|^2 =
	\begin{tikzpicture}
		\draw[double] (0,-0.5) -- (0,.5);
		\node[draw=black,fill=blobcolor,circle,minimum size=14pt,inner sep=0] (a) at (0,0) {\(\mathcal{A}\)};
		\draw[double] (2,-0.5) -- (2,.5);
		\node[draw=black,fill=blobcolor,circle,minimum size=14pt,inner sep=0] (b) at (2,0) {\(\mathcal{A}^{\dagger}\)};
		\node[draw=white,fill=white,minimum size=4pt] (c) at (1,0) {};
		\begin{pgfonlayer}{background}
			\draw[photon big] (a.center) -- (c.center);
			\draw[photon big] (c.center) -- (b.center);
		\end{pgfonlayer}
		\draw[cuts] ($(c.center)+(0,0.4)$) -- ($(c.center)-(0,0.4)$);
	\end{tikzpicture}
	\Rightarrow
	\begin{tikzpicture}
		\node[quad src={3}{\(I^{ij}\)}] (a) at (0,0){};
		\node[quad src={3}{\(I^{mn}\)}] (b) at (0,5) {};
		\foreach \ml in {1,...,4}
		\node[mass src={2}] (\ml) at (0,\ml){};
		\node (ci) at (2,2.5) {};
		\node (cf) at (0.5,2.5) {};
		\node at (0,1.6) {\(\vdots\)};
		\node at (0,3.6) {\(\vdots\)};
		\node at (0.3,1.6) {\(\vdots\)};
		\node at (0.3,3.6) {\(\vdots\)};
		\draw[photon big](a) to[out=10,in=-10] node[circle,fill=blobcolor,draw=black,inner sep=0pt,pos=1/3,minimum size=14pt] (v1){} node[circle,fill=blobcolor,draw=black,inner sep=0pt,pos=2/3,minimum size=14pt] (v2){} (b) ;
		\begin{pgfonlayer}{background}
			\draw[potential] (1) -- (v1.center);
			\draw[potential] (2) -- (v1.center);
			\draw[potential] (3) -- (v2.center);
			\draw[potential] (4) -- (v2.center);
		\end{pgfonlayer}
		\draw[cuts] (ci)--(cf);
	\end{tikzpicture}
	\,,
	\label{eq:cross-section}
\end{equation}
where the \textcolor{cutcolor}{dashed colored lines} remain a
shorthand for the on-shell state sum.  With the far right-hand-side
we are schematically conveying the diagrammatic expansion of 
$\mathcal{A}$ and $\mathcal{A}^\dagger$
using the diagram style of our two-point approach (as shall also 
be seen in our evaluations below), including
scattering off of the potential modes sourced by the composite
particle's energy $E$. As usual the formal ``all-orders'' amplitudes
are expanded so that the observable contains the desired order of
contributions, including in this case the number of potential-mode
interactions.  

Thus, we can view our unitarity-based two-point formalism as 
alternatively being a method of directly calculating unpolarized 
graviton emission cross-sections without needing to construct 
the individual $\mathcal{A}$ and $\mathcal{A}^{\dagger}$. 
See also a discussion on how this plays out at leading order in 
the recent Ref.~\cite{Almeida2023}.  Note that the
exact perturbative equivalence between the two approaches depends on
aligning various regularization conventions. For example, the two-point
formalism applies dimensional regularization to all of the graviton
loop integrations, while the standard dLIPS that appears in a
cross-section computation normally uses a fixed integer dimension.
Note that a full quantum-like ``GR+matter'' computation, like that 
used in the PM approach, naturally aligns these conventions.  
Yet in the direct PN EFT approach this should be handled with care due 
to the splitting of graviton modes, as well as the treatment of
matter components as classical sources.

\subsection{Building Blocks}
\label{sec:tail-bb}           

For the calculation of the tails, we will need three types of building
blocks.  First, we need to describe the interaction of a quadrupole
with the gravitational field.  The analogy between the interactions of 
classical $l$-th multipoles, in particular of higher-spin multipoles 
$S^l$, and of quantum spin $l/2$ particles, suggests that we model the 
``one-graviton source'' 
associated to the quadrupole, as a $3$-point amplitude of a spin-$1$ 
massive particle radiating a graviton. Extracting the quadrupole-coupling 
directly from a vector-graviton interaction is somewhat involved 
\cite{Holstein:2008sx}. Thus we shall take a shortcut by using the double 
copy.

First, since the quadrupole is a symmetric traceless $SO(3)$ tensor, it 
can be represented as an outer product between two $SO(3)$ vectors
\begin{equation}
  I_{ij} \equiv I_{i} I_{j}\, , \qquad \delta^{ij}I_{i} I_{j} = 0 \,.
\end{equation}
The $SO(3)$ vectors can be covariantized so that in the rest frame of the 
binary we have $I^{\mu} = (0,I^{i})$. Now we consider the fermion-to-spin 
identification of Ref.~\cite{Holstein:2008sx}.  
Applied to the fermion-vector $3$-point amplitude, and interpreting the 
spin vector $S_i$ as the generic $SO(3)$ vector $I^\mu$, we find
\begin{equation}
\mathcal{M}_{fvf}\propto\bar{u}(k_1) \slashed{\pol}_3 u(k_2)
  \to \left[
    \frac{(k_2 \cdot \pol_3)}{2 m} \bar{u}(k_1)  u(k_2)
    - \frac{i}{m^2} \varepsilon_{\mu \nu \rho \sigma}\pol_3^\mu k_1^\nu k_2^\rho I^{\sigma}
  \right] \,,
\end{equation}
where $u$, $\varepsilon$ are the spinors and vector polarization, 
respectively. The second term is proportional to
$\tr(\gamma_5 \slashed{\pol}_3\slashed{k}_2\slashed{k}_1
\slashed{I})$, and the $I$ dependence of both terms, which is implicit 
here in the first term, has the same structure in the non-relativistic 
limit.  

Thus, we shall focus on how the explicit trace term
\begin{equation}
  \mathcal{M}_{fvf}\Big|_{I}\propto \text{tr}(\gamma_5 \slashed{\pol}_3\slashed{k}_2\slashed{k}_1
  \slashed{I})
  \,,
  \label{eq:fvf-hr}
\end{equation}
eventually generates the quadrupole-graviton coupling tensor structure
via the double copy, and fix the overall coupling constant later.  
We take the double copy of \cref{eq:fvf-hr} with itself to produce a 
vector-graviton $3$-point amplitude, with the explicit identification of 
the quadrupole component already made
\begin{equation}
  \mathcal{M}_{vgv} \propto \mathcal{M}_{fvf}^2 \propto
  \left[\tr(\gamma_5 \slashed{\pol}_3\slashed{k}_2\slashed{k}_1
  \slashed{I})\right]^2 \,.
\end{equation}
The square of the parity-odd trace produces a Gram determinant, which
evaluates to
\begin{align}
  \mathcal{M}_{Ig}
  &\propto  \text{GramDet}(k_2,k_1,I, \pol_g)
    =
    \tilde{\lambda} \left[ (\pol_g\cdot k_1)(I \cdot k_g) - (\pol_g\cdot I)(k_1\cdot k_g)\right]^2 \ ,
\label{eq:cov-ig}
\end{align}
after applying momentum conservation and external state conditions,
and absorbing constants into $\tilde{\lambda}$.  

Evaluating this coupling in the rest-frame of particle $1$ yields
\begin{align}
  \mathcal{M}_{Ig} &\to \tilde{\lambda} m_1^2
  (I^i k_g^i \pol_g^0 - I^i \pol_g^i \omega_g)^2 \notag\\
  &= \lambda_I I^{ij}
  (\omega_g k_g^i \pol_g^0 \pol_g^j
  + \omega_g k_g^j \pol_g^0 \pol_g^i - k_g^i k_g^j \pol_g^0 \pol_g^0 
    - \omega_g^2 \pol_g^i \pol_g^j) \, , 
    \label{eq:frame-ig}
\end{align}
where the coupling constant $\lambda_I$ remains to be fixed.
Up to alignment of coupling constants this exactly agrees with the
one-graviton quadrupole operator used throughout EFT-based approaches (see
Refs.~\cite{Goldberger:2004jt,Goldberger:2005cd,Holstein:2008sx,Goldberger:2009qd,Galley:2009px}
for foundational discussions of these types of operators).  
We will write here
\begin{equation}
  \mathcal{M}_{Ig} \equiv \lambda_I J_I^{\mu \nu} \pol_{\mu \nu}
  = \begin{tikzpicture}
    \node[quad src={3}{\(I^{ij}\)},outer sep=0] (a) at (0,0){};
    \node (e) at (1,0) {\(\pol_{\mu \nu}\)};
    \draw[photon big] (a)--(e);
  \end{tikzpicture} \ ,
  \label{eq:quad-src}
\end{equation}
when discussing the quadrupole coupling amplitude in contexts where
the polarizations may be stripped off.  Since we are only focused on
the tails of the quadrupole in this work, we do not require
higher-graviton quadrupole couplings, or higher-multipole couplings.

Second, in addition to the quadrupole coupling, the tails specifically involve
interactions with the gravitational potential of the system.  This
type of coupling is well understood to be analogous with minimal
scalar-graviton interactions \cite{Goldberger:2004jt,Holstein:2008sx}.
As such, we take the potential-mode graviton source to be
\begin{equation}
  \mathcal{M}_{E g} = \mathcal{M}_{sgs} = \frac{\lambda_E}{m_s^2} p_s^\mu p_s^\nu \pol_{\mu \nu}= \begin{tikzpicture}
    \node[mass src={2}] (a) at (0,0){};
    \node (e) at (1,0) {\(\pol_{\mu \nu}\)};
    \draw[photon big] (a)--(e);
  \end{tikzpicture} \,,
  \label{eq:mass-src}
\end{equation}
which in the rest frame of the scalar evaluates to
$\lambda_E \pol_{00}$, in agreement with the EFT definition of the static
source of potential modes, $E h^{00}$, after appropriate alignment of
coupling constants.  

Third, beyond the three-point amplitude
(analogous to a one-point source) for the potential-mode coupling, we
will also require four-point amplitudes to capture the possible
contact terms (analogous to a two-point source) in a gauge-invariant
manner.  The obvious choice for the desired amplitude is the
two-graviton two-scalar extension of \cref{eq:mass-src}, which is
computable as the double-copy of a two-scalar two-gluon amplitude
using
\href{https://gitlab.com/aedison/increasingtrees}{IncreasingTrees} as
\begin{align}
  \mathcal{M}_{E g^2} \overset{?}{=} \msggs 
  &=\frac{\lambda_E}{m_s^2} \lambda_g \left( (\pol_2\cdot k_1)(\pol_3\cdot k_{12})
    - \frac{1}{2}(\pol_2\cdot \pol_3)(k_1\cdot k_2)\right)\Bigg[ \label{eq:sggs}\\
  & \quad\left( (\pol_2\cdot k_1)(\pol_3\cdot k_{12})
    - \frac{1}{2}(\pol_2\cdot \pol_3)(k_1\cdot k_2)\right)
    \left(\frac{1}{2(k_1\cdot k_2)} + \frac{1}{2(k_2\cdot k_3)}\right) \notag\\
  &\quad + 
    \left(
    (\pol_2\cdot k_{13})(\pol_3\cdot k_1) - \frac{1}{2} (\pol_2 \cdot\pol_3)(k_1\cdot k_3)
    \right)
    \frac{1}{2(k_2\cdot k_3)}
    \Bigg] + (2 \leftrightarrow 3) \notag\,,
\end{align}
in which particles $1$ and $4$ are the scalars, and $2,3$ the gravitons, 
all particles have the same external momentum flow, and $\lambda_g$ is 
the graviton self-coupling constant to be fixed later.

However, the amplitude as-is
contains too much information: it encodes not only the graviton
self-interactions and contact terms, but also propagation of an
off-shell scalar particle via the $p_1 \cdot p_2$ and $p_1 \cdot p_3$
poles.  Including the off-shell propagation is in direct tension with
wanting to interpret the scalar as a classical massive object.  The
tension can be resolved by appealing to the identification of the
scalar with a black hole: we treat the scalar mass as the dominant
scale in the problem, and expand the four-point amplitude in said
limit.  Doing so in the rest frame of massive particle leads to
\begin{align}
  \mathcal{M}_{Eg^2} = \msggs(m_s \to \infty) =
  & \frac{\lambda_g\lambda_E }{\omega_{2}^2} \frac{\delta(\omega_2-\omega_3)}{2 (k_2\cdot k_{3})} 
    \Big[ (k_2\cdot k_{3}) \pol_2^0 \pol_3^0
    + \omega_{2}( (\pol_3\cdot k_{2})\pol_2^0 \notag\\
  &- (\pol_2\cdot k_{3})\pol_3^0) - \omega_{2}^2(\pol_2\cdot \pol_{3})\Big]^2 + \mathcal{O}(m_s^{-1}) \, \nonumber \\
  &= \begin{tikzpicture}
    \node[mass src={2}] (a) at (0,0){};
    \node (e1) at (1,0.6) {\(\pol_1^{\mu \nu}\)};
    \node (e2) at (1,-0.6) {\(\pol_2^{\rho \sigma}\)};
    \draw[photon big] (a)--(e1);
    \draw[photon big] (a)--(e2);
  \end{tikzpicture} \ ,
\label{eq:mass-four}
\end{align}
in which we have put back in an explicit $\delta(\omega_2-\omega_3)$
on the leading-order term that results from evaluating the
normally-implicit momentum conserving $\delta^4(\sum k)$ in the
large-mass limit.  Note that this definition is equivalent, up to
external momentum flow conventions and frame choices, with
the ``heavy-mass'' amplitudes of Ref.~\cite{Brandhuber:2021kpo}.

Finally, we will need amplitudes for graviton self-interactions.  For
the work at hand, we only require tree amplitudes,\footnote{This
  follows from the general philosophy that ``tree level is
  classical''.  While not true when matter loops are involved, the
  heuristic does hold for massless force-carrier loops.
  Ref.~\cite{Kosower:2018adc} provides a thorough exploration of the
  topic via explicit $\hbar$ counting.} which are easily constructed
using the
\href{https://gitlab.com/aedison/increasingtrees}{IncreasingTrees}
package.  The tree amplitudes built by the package are normalized by
setting the coefficients of specific kinematic structures to $1$
rather than against a particular choice of coupling, so a coupling
factor of $\lambda_g^{n-2}$ must be included on all of the graviton
tree amplitudes.

Throughout the above discussion, we introduced coupling constants that
should be matched with appropriate references for comparisons to be
accurate.  We choose to specifically match against the conventions of
Refs.~\cite{Goldberger:2009qd,Galley:2015kus}, making our coupling
constants:
\begin{align}
  \lambda_I &= \sqrt{2 \pi G_N}\,, \notag \\
  \lambda_E &= - E \sqrt{8 \pi G_N}\,, \\
  \lambda_g &= - \sqrt{32 \pi G_N} \,. \notag
\end{align}
Notably we will use the fixed-dimension standard definition for $G_N$,
and will introduce a renormalization scale $\mu$ that accounts for the
scale-dependence of working in dimensional regularization.

\subsection{Radiation-Reaction}
The quadrupole-sourced radiation reaction (no interaction with the
background potential), and leading tail, have been studied extensively
\cite{Almeida:2021xwn,Bini:2020wpo,Foffa:2019eeb,Galley:2015kus,Blanchet:2013haa,
	Foffa:2011np,Galley:2012qs,Goldberger:2009qd,Galley:2009px,Galley:2008nss,
	Blanchet:1992br,Blanchet:1987wq},
so serve as verifications for our proposed methods.  They are also
simple enough that we can present the majority of intermediate steps
in detail.

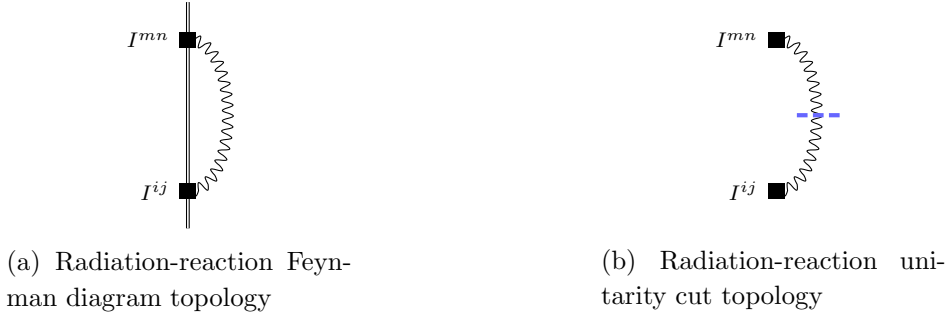
\begin{figure}[h]
  \centering
  \begin{subfigure}[t]{0.3\textwidth}
    \centering
    \begin{tikzpicture}
      \draw[double] (0,-0.5) -- (0,2.5);
      \node[quad src={3}{\(I^{ij}\)},outer sep=0] (a) at (0,0){};
      \node[quad src={3}{\(I^{mn}\)},outer sep=0] (b) at (0,2) {};
      \draw[photon big] (a.center)
      .. controls ($(a) + (0.7,0)$) and ($(b)+(0.7,0)$)..
      (b.center);
    \end{tikzpicture}
    \caption{Radiation-reaction Feynman diagram topology}
    \label{fig:rr-f}
  \end{subfigure}
  \hspace{0.2\textwidth}
  \begin{subfigure}[t]{0.3\textwidth}
    \centering
    \begin{tikzpicture}
      \path (0,-0.5) -- (0,2.5);
      \node[quad src={3}{\(I^{ij}\)}] (a) at (0,0){};
      \node[quad src={3}{\(I^{mn}\)}] (b) at (0,2) {};
      \draw pic {cut hill={a}{b}};
    \end{tikzpicture}
    \caption{Radiation-reaction unitarity cut topology}
    \label{fig:rr-u}
  \end{subfigure}
  \caption{The quadrupole radiation-reaction diagrams.}
  \label{fig:rr}
\end{figure}

Beginning with radiation reaction, there is only one diagram topology
that we need to consider in both the EFT/Feynman and generalized-unitarity
perspectives, namely \cref{fig:rr}.  This is rather straightforward to
see.  When we label the momentum flow of the gravitons in the Feynman
diagram, \cref{fig:rr-f}, and impose momentum conservation 
(in this case non-existent), we find only a single momentum $\ell^\mu$ and
only one possible momentum invariant, $\ell^2 = \omega^2 - \ell_E^2$.
Thus, the relevant integral family consists of
\begin{equation}
  \ellint{E} \frac{ \ell_E^{i_1} \dots \ell_E^{i_n}}{(-\ell_E^2 + \omega^2)^{\lambda}} 
  \ , \qquad \lambda \in \mathbb{Z}_+,\, n \ge 0 \,.
\end{equation}
In fact, symmetry and integral relations always allow us to write any
integral of this type in terms of $\omega^n \delta^{i_1 \dots i_n}$
(see \cref{sec:tens-red} for a brief discussion of our systematic
method for dealing with the tensor reductions appearing in the tails),
and the single basis integral
\begin{equation}
  F^{(1)}(1; \omega^2) =  \ellint{E} \frac{1}{(-\ell_E^2 + \omega^2)} = - \frac{\Gamma(1-d/2)(-\omega^2)^{d/2-1}}{(4 \pi)^{d/2}} \,,
\label{eq:rr-basis-raw}
\end{equation}
where $d$ is the dimensional regularization dimension $d = 3 +\dimreg$
and $\omega$ has an imaginary part set by the $\ctpeps$ prescription for 
the propagator.
In the Feynman prescription, the effective action would then be written as
\begin{equation}
  S_{\text{RR}} = \intw c_{\text{RR}} F^{(1)}(1; \omega^2 + \ctpeps) \,.
\end{equation}
Working in the CTP prescription, we instead need to sum over advanced
and retarded propagators.  In turn, this means that the radiation
reaction contribution to the CTP effective action \emph{must} be
expressible as
\begin{equation}
  S_{\text{RR}} = \intw \left(c^{-+}_{\text{RR}}
    F^{(1)}(1; \omega_R^2)
     + c^{+-}_{\text{RR}} F^{(1)}(1; \omega_A^2) \right)
    \,.
\end{equation}
Our goal is now to determine $c_{\text{RR}}$ using unitarity methods.
It is also worth pointing out that the different $\ctpeps$
prescriptions change the way that the $\sqrt{-\omega^2}$ in
\cref{eq:rr-basis-raw} (and later $\log (- \omega^2)$) will be
analytically continued.  The Feynman prescription tells us to do the
continuation using a \emph{fixed} imaginary part of $\omega^2$, while
the advanced and retarded prescriptions tell us to treat the imaginary
part as \emph{depending on the sign of} $\omega$.

Following the ideas discussed in \cref{sec:unit}, $c_{\text{RR}}$
should be directly related to the generalized unitarity cut of the
diagram divided by the symmetry factor for the diagram.  We calculate
the generalized unitarity cut, following~\cref{eq:gen-cut}, as the product
of two quadrupole amplitudes, \cref{eq:quad-src}, summed over the
$D = d+1$-dimensional states of the internal graviton using
\cref{eq:state-sum} to find
\begin{align}
  \cut^{ab}_{\text{RR}}
  &= \sum_{\text{grav states}} \mathcal{M}_{I_{a}(-\omega)} \mathcal{M}_{I_{b}(\omega)} \notag\\
  &= \lambda_I^2 J_{I_{a}(-\omega)}^{\mu \nu}\  
    \mathcal{P}^{\mu \nu; \alpha \beta} J_{I_{b}(\omega)}^{\alpha \beta} \Big|_{\ell^2 = \omega^2 - \ell_E^2 = 0} \notag\\
  &= (2 \pi G_N) \delta(\omega^2 - \ell_E^2)
    \left( J_{I_{a}(-\omega)}^{\mu \nu}J_{I_{b}(\omega)}^{\mu \nu} 
    - \frac{J_{I_{a}(-\omega)}^{\mu \mu} J_{I_{b}(\omega)}^{\nu \nu}}{D-2} \right) \,,
\end{align}
with CTP labels $a,b$.  Inserting the definitions for $J^{\mu \nu}$
and performing tensor reductions following \cref{sec:tens-red}, we
arrive at
\begin{equation}
  \cut^{ab}_{\text{RR}} = (2 \pi G_N) \kappa_{ab}(\omega) \omega^4
  \frac{(d+1)(d-2) }{(d+2)(d-1)} \delta(\omega^2 - \ell_E^2) \,.
  \label{eq:rr-cut}
\end{equation}
We see that the cut only depends on the CTP labels through
$\kappa^{ab}(\omega)$, and this turns out to be true for the rest of
the calculations we will approach in this paper.  As a result of this
observation, we define the unindexed cut as
\begin{equation}
  \cut^{ab}_{\text{RR}} \equiv \cut_{\text{RR}} \kappa_{ab}(\omega) \ ,
\end{equation}
and slightly restructure the CTP effective action as
\begin{align}
S_{\text{RR}} &= \intw a_{\text{RR}}\left( \kappa_{-+}(\omega)
    F^{(1)}(1; \omega_R^2)
    + \kappa_{+-}(\omega) F^{(1)}(1; \omega_A^2) \right) \notag\\
  &\equiv \intw a_{\text{RR}} F^{(1)}_{\text{CTP}}(1;\omega^2) \,.
\end{align}
This makes it clear that while a fully detailed treatment would
involve separately matching coefficients between the different CTP
branches, the net result of the matching in this case will be the same
as if we had ignored the $\ctpeps$ prescription which only contributes
through the integration contour of the basis integrals.  This
continues to occur for all of the higher-order tails considered below as well.

We proceed with reconstructing $a_{\text{RR}}$ by matching the cut
against the basis integral.  Because the cut is already independent of
the loop momentum, we do not have any reduction to perform.  Thus the
matching process is very simple
\begin{align}& \cut_{\text{RR}} = 
    (2 \pi G_N)\omega^4
    \frac{(d+1)(d-2) }{(d+2)(d-1)} \delta(\omega^2 - \ell_E^2)
    = |\graph{RR}| a_{\text{RR}} \delta(\omega^2 - \ell_E^2) \,.
\end{align}
We now need to determine the symmetry factor $|\graph{RR}|$.  The
purpose of dividing by the symmetry factor is to compensate for
possible over-counting of redundant information by the cut.  Since our
implementation of CTP sums over the advanced and retarded branches, we
need to make sure that the cut is not double-counting contributions
across branches.  The simplest way to do so is to count the up/down
reflection (in our drawing convention) as a symmetry of
the cut.  Thus, in the current case we have $|\graph{RR}|=2$.  Many of
the diagrams for the higher-order tails also include this reflection as
part of the symmetry factor.  Putting everything together leads to our
CTP effective action for radiation reaction
\begin{equation}
  S_{\text{RR}} = \frac{(2 \pi G_N)}{2}\frac{(d+1)(d-2) }{(d+2)(d-1)} 
  \intw \omega^4 F^{(1)}_{\text{CTP}}(1;\omega^2) \,.
\end{equation}
Inserting the master integral definition from \cref{eq:rr-basis-raw},
expanding in $d=3+\dimreg$\footnote{Note that our choice of $d$
  differs from the standard analytic continuation of
  $d-d_{\mathbb{Z}} = -2 \dimreg$.  This choice will of course drop
  out of the final observables.}, and performing the CTP sum, we arrive
at
\begin{equation}
\label{eq:rr-raw}
S_{\text{RR}} = -i \frac{G_N}{5} \intw
\,\omega^5 \kappa_{-+}(\omega)\Bigg[ 1 - \frac{\dimreg}{2}
\left( i\pi \sgn\omega + \left[\frac{9}{10}-\log\left( \frac{\omega^2 e^{\gamma_E}}{\mu^2 \pi} \right)\right] \right) + \odimreg{2} \Bigg] \ ,
\end{equation}
in which we have introduced the standard renormalization scale to the
logarithm, and have retained the $\odimreg{1}$ piece for later use in
counterterm analysis.  Note also the appearance of the
$i \pi \sgn \omega$ term, which is a result of using the modified
$\ctpeps$ prescription.  The standard QFT Feynman prescription would
produce a definite sign rather than the $\sgn \omega$.

\subsection{Tail}
We are now ready to approach the leading tail calculation.  The broad
strokes are similar to the radiation reaction, with just a few new
pieces necessary for the higher tails.  From now on, we drop the
explicit $E$ label on the loop momenta as we will always be working
with integrated Euclidean loop momenta and an explicit frequency as
the scale.

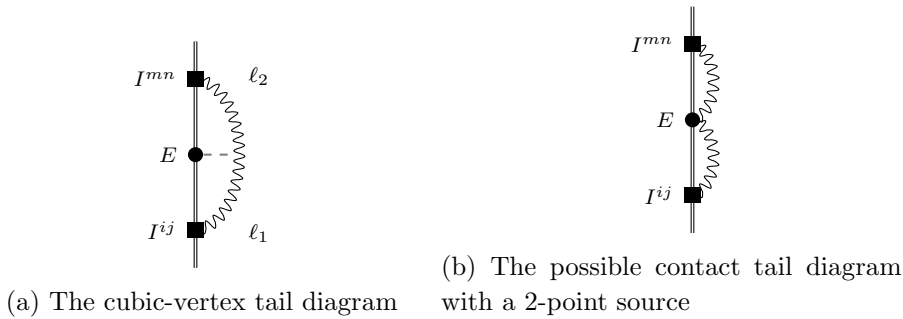
\begin{figure}[h]
  \centering
   \begin{subfigure}[b]{0.4\textwidth}
    \centering
    \begin{tikzpicture}[smooth]
      \draw[double] (0,-0.5) -- (0,2.5);
      \node[quad src={3}{\(I^{ij}\)}] (a) at (0,0){};
      \node[mass src={2}] (m) at (0,1){};
      \node[quad src={3}{\(I^{mn}\)}] (b) at (0,2) {};
      \draw[photon big](a.center) to[out=10,in=-10]
      node[pos=1/4,label=below right:{\(\ell_1\)}]{}
      node[pos=1/2] (v){}
      node[pos=3/4,label=above right:{\(\ell_2\)}] {}(b.center) ;
      \draw[potentialf] (m) -- (v);
    \end{tikzpicture}
    \caption{The cubic-vertex tail diagram}
    \label{fig:tail-cubic}
  \end{subfigure}
  \begin{subfigure}[b]{0.4\textwidth}
    \centering
    \begin{tikzpicture}
      \draw[double] (0,-0.5) -- (0,2.5);
      \node[quad src={3}{\(I^{ij}\)}] (a) at (0,0){};
      \node[mass src={2}] (m) at (0,1){};
      \node[quad src={3}{\(I^{mn}\)}] (b) at (0,2) {};
      \draw[photon big] (a.center) to[out=20,in=-20] (m.center);
      \draw[photon big](m.center) to[out=20,in=-20] (b.center);
    \end{tikzpicture}
    \caption{The possible contact tail diagram with a 2-point source}
    \label{fig:tail-cont}
  \end{subfigure}
 
  \caption{The possible Feynman diagrams needed to evaluate the
    leading tail contribution.}
  \label{fig:tail-diags}
\end{figure}

The first important development is that there is more than one Feynman
diagram, with a single energy coupling (analogous to the so-called 
``self-energy'' graphs), that 
contributes to the process (depending on gauge choices),
shown in \cref{fig:tail-diags}. Therefore we should take some care with
defining our basis of momentum invariants and integral family.  A
maximally convenient basis of momentum invariants to use is one which
contains all possible inverse propagators of the diagrams under
consideration.  For the tail diagrams, we can use the labelings of
$\ell_1$ and $\ell_2$ as shown in \cref{fig:tail-cubic} to define the
inverse propagator basis
\begin{equation}
  Q_1 = \omega^2 - \ell_1^2 \ , \qquad Q_2 = \omega^2 - \ell_2^2 
  \ , \qquad Q_3 = - (\ell_1 + \ell_2)^2 \ , 
\end{equation}
with signs set by a mostly-minus metric to best align with
the standard integration convention for propagator signs, see
\cref{sec:ints-analytic} for more details, and where we have now made the
fact that the $\ell_i$ are the Euclidean spatial part of the momenta
implicit.  Important to note is that $Q_3$ is chosen as a purely
spatial momentum: it is the ``potential-mode'' propagator expected from
the interaction of the quadrupole with the static background
potential.  From this basis, it is obvious that the integral family we
need to consider is
\begin{equation}
  F^{(2)}(\lambda_1, \lambda_2, \lambda_3) = \elllint{2} Q_1^{-\lambda_1}Q_2^{-\lambda_2}Q_3^{-\lambda_3} \ , 
\end{equation}
as it will cover any possible contributions from both diagram
topologies.  Analyzing the integral family, we again find a single
basis integral, $F^{(2)}(1,1,0)$, implying that the effective action
can be written as
\begin{align}
  S_{\text{T}} &= \intw \left(c^{-+}_{\text{T}}
    F^{(2)}(1,1,0; \omega_R^2)
    + c^{+-}_{\text{T}} F^{(2)}(1,1,0; \omega_A^2) \right) \notag \\
  &\equiv
    \intw a_{\text{T}} F^{(2)}_{\text{CTP}}(1,1,0)\,.
    \label{eq:tail-raw-s}
 \end{align}
 
There are two interesting observations about the integral basis.
 First, the basis integral actually factorizes!  $Q_1$ and $Q_2$
 depend separately on the two loop momenta so we have
\begin{equation}
  F^{(2)}(1,1,0) = \left(\ellint{1} \frac{1}{Q_1}\right)
  \left( \ellint{2} \frac{1}{Q_2} \right)=F^{(1)}(1;\omega^2)^2 \,.
  \label{eq:tail-basis}
\end{equation}
Since $F^{(1)}(1;\omega^2)$ is finite in dimensional regularization,
so is $F^{(2)}(1,1,0)$.  Second, we have the important integral
relation
\begin{equation}
  F^{(2)}(1,1,1) = - \frac{(d-2)}{2(d-3) \omega^2} F^{(2)}(1,1,0) \,.
\end{equation}
This relation highlights that the only source of divergences in
the tail comes from terms in which all three propagators are present.

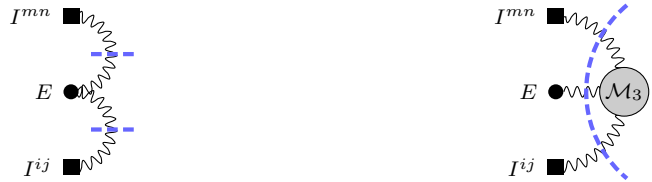
\begin{figure}[h]
  \centering
  \begin{subfigure}[t]{0.4\textwidth}
    \centering
  \begin{tikzpicture}
    \draw pic {hills={1}};
  \end{tikzpicture}
  \caption{The unitarity cut diagram needed to evaluate the leading
    tail contribution}
  \label{fig:tail-cut}
\end{subfigure}
\quad
\begin{subfigure}[t]{0.4\textwidth}
  \centering
  \begin{tikzpicture}
    \node[quad src={3}{\(I^{ij}\)}] (a) at (0,0){};
    \node[quad src={3}{\(I^{mn}\)}] (b) at (0,2) {};
    \draw pic {mamp={a}{b}{1}};
  \end{tikzpicture}
  \caption{The unitarity cut used to check the pole of the basis
    cut.}
  \label{fig:tail-recut}
\end{subfigure}
\caption{The unitarity cut diagrams used for analysis of the tail.}
\label{fig:all-tail}

\end{figure}

Now that we have identified the basis integral, we set out to evaluate
the corresponding cut.  From the basis integral, we know that the
needed cut must contain two radiation mode propagators.  Within the
tail framework, there is only one possible configuration of tree
amplitudes with two radiation propagators: the
quadrupole-mass-quadrupole contraction with a two-point mass
amplitude, as shown in \cref{fig:tail-cut}.  Assembling the cut as the
product of tree amplitudes from \cref{eq:quad-src,eq:mass-four} and
inserting the sum over graviton states, we have
\begin{align}
  \cut_{\text{T}}^{ab} &= \sum_{\text{states}} \mathcal{M}_{I(-\omega)} \mathcal{M}_{Eg^2} 
  \mathcal{M}_{I(\omega)} \Big|_{Q_1=0,Q_2=0} \notag \\
           &= \lambda_I^2 J_{I_a(-\omega)}^{\mu \nu} P^{\mu \nu; \alpha \beta}
             \mathcal{M}_{Eg^2}^{\alpha \beta; \gamma \sigma} P^{\gamma \sigma; \rho \tau} 
             J_{I_b(\omega)}^{\rho \tau} \delta(Q_1) \delta(Q_2) \,.
\end{align}
Evaluating the contractions and performing the tensor reduction to
$\kappa_{ab}(\omega)$ (but not employing IBP reductions yet), we arrive at
\begin{align}
  \cut_{\text{T}} = \frac{4 \pi^2 G_N^2E (d-2)}{ (d-1)^3(d+2) \omega^2}
  \delta(Q_1) \delta(Q_2) \Bigg(
  &(d-2)Q_3^3 + 8(d-2) Q_3^2 \omega^2 + 4(d^2 + 4 d - 9)Q_3 \omega^4 \notag\\
  &+16 (d^2-1)\omega^6 + \frac{8(d^2-1)(d+1) \omega^8}{Q_3}  \Bigg)\,. \label{eq:cut-t}
\end{align}

Importantly, if we had used the full two-scalar two-graviton
amplitude, \cref{eq:sggs}, instead of \cref{eq:mass-four}, $\mathcal{M}_{Eg^2}$, 
then the ``virtual black hole'' poles would have entered the cut carrying
dependence on the reference vector $q$ from the physical state
projector, \cref{eq:state-sum}.  The dependence on $q$ drops exactly
for the leading-order term in the large-$m_s$ expansion.  We have
written the cut as a Laurent series in the uncut momentum invariant
$Q_3$ to manifest the factorization property.  This allows us to
verify the $Q_3^{-1}$ part of the cut by evaluating a different cut,
the one shown in \cref{fig:tail-recut}.  This new cut is built from
the one-point mass term, \cref{eq:mass-src}, as well as a three-point
all-graviton amplitude.  Evaluating it, we find
\begin{align}
  \cut_{\text{\cref{fig:tail-recut}}}
  &=\lambda_I^2 \left(J_{I_a(-\omega)}^{\mu_1 \nu_1} P^{\mu_1 \nu_1; \mu_2 \nu_2}\right)
    \mathcal{M}_{ggg}^{\mu_2 \nu_2;\alpha_2 \beta_2;\rho_2 \tau_2} \notag \\
  &\times \left(P^{\alpha_2 \beta_2; \alpha_1 \beta_1}\mathcal{M}_{sgs}^{\alpha_1 \beta_1}\right)\left( P^{\rho_2 \tau_2; \rho_1 \tau_1} 
    J_{I_b(\omega)}^{\rho_1 \tau_1}\right) \delta(Q_1) \delta(Q_2)\delta(Q_3) \notag \\
  &= \frac{32 \pi^2 G_N^2E (d-2)(d^2-1)(d+1)}{ (d-1)^3(d+2)}\omega^6
  \delta(Q_1) \delta(Q_2) \delta(Q_3) \ ,
\end{align}
in exact agreement with the $Q_3$ residue of \cref{eq:cut-t}.  The
polynomial-in-$Q$ terms are known in the language of generalized
unitarity as ``contact'' terms because they correspond to diagrams in
which some propagators have been collapsed into a contact-like
interaction.

The basis cut in \cref{eq:cut-t} is still a function of the loop
momentum via $Q_3$.  Thus in order to match it with $a_{\text{T}}$, we
need to reduce it using integration-by-parts relations
\cite{Laporta:1996mq,Smirnov:2006ry,Smirnov:2012gma}. Since there is
only one basis integral, we do not need to worry about the support of
the $\delta$s.  We use FIRE6 \cite{Smirnov:2019qkx} to automate the
reduction process, after which we find
\begin{equation}
  \rcut{\text{T}} = -16 G_N^2 \pi^2 \omega^4 
  \frac{(d-2)(12-2d+5d^2 - 4d^3 + d^4)}{(d-3)(d-1)^2d(d+2)}
  \delta(Q_1) \delta(Q_2) \,.
\end{equation}
For later use, we name the polynomial of $d$ occurring in the
numerator as
\begin{equation}
  \mathcal{P}_4 = 12-2d+5d^2 - 4d^3 + d^4 \,.
  \label{eq:p4}
\end{equation}
Constructing the coefficient in the effective action,
\cref{eq:tail-raw-s}, now only requires normalizing by the symmetry
factor of the graph, $|\graph{T}| = 2$.  Thus we have
\begin{equation}
  S_{\text{T}} = -8 G_N^2 \pi^2
  \frac{(d-2)(12-2d+5d^2 - 4d^3 + d^4)}{(d-3)(d-1)^2d(d+2)}
  \intw   \omega^4 F^{(2)}_{\text{CTP}}(1,1,0) \,.
\end{equation}
The basis integral is straightforward to evaluate using
\cref{eq:tail-basis}.  Even though the basis integral is finite, the
coefficient now contains an explicit pole in $d_{\text{crit}}=3$.
Thus we need to expand the entire action using the dimension
regularization $d=3+\dimreg$, yielding
\begin{align}
\label{eq:t-raw}	
  S_{\text{T}} = \frac{2}{5} G_N^2 E
  & \intw \,\omega^6 \kappa_{-+}(\omega)\Bigg\{
    \frac{1}{\dimreg}+\left(\log\left(\frac{\omega^2 e^{\gamma_E}}{\mu^2 \pi}\right)
    -\frac{41}{30}- i \pi  \text{sgn}(\omega )\right) \notag \\
  & \quad +\dimreg\Bigg[
    - i \pi\sgn(\omega )  \left(\log\left(\frac{\omega^2 e^{\gamma_E}}{\mu^2 \pi}\right)-\frac{41}{30}\right)  \\
  & \quad  + \log\left(\frac{\omega^2 e^{\gamma_E}}{\mu^2 \pi}\right)
    \left(\frac{1}{2}\log\left(\frac{\omega^2 e^{\gamma_E}}{\mu^2 \pi}\right)-\frac{41}{30}\right)
    -\frac{9}{4} \zeta_{2}+\frac{3667}{1800} \Bigg] + \odimreg{2}
    \Bigg\} \,. \notag
\end{align}
As already noted in \cite{Edison:2022cdu} this effective action agrees with the one derived by 
Galley et al in Ref.~\cite{Galley:2015kus} up to the overall sign.


\section{Higher-Order Tails}
\label{sec:higher-tails}
With the building blocks at our disposal, and well-known leading tail effects 
demonstrated in our methodology, we are ready to address higher-order tails.  
In this section we discuss the computation of the CTP effective action contributions
for T$^2$ through T$^4$. In our preceding letter \cite{Edison:2022cdu}
the T$^2$ and T$^3$ effective actions have been computed for the first time, 
which we present here in detail. We additionally provide in this paper 
the first newly-computed T$^4$ effective action.

\subsection{Tail-of-Tail}
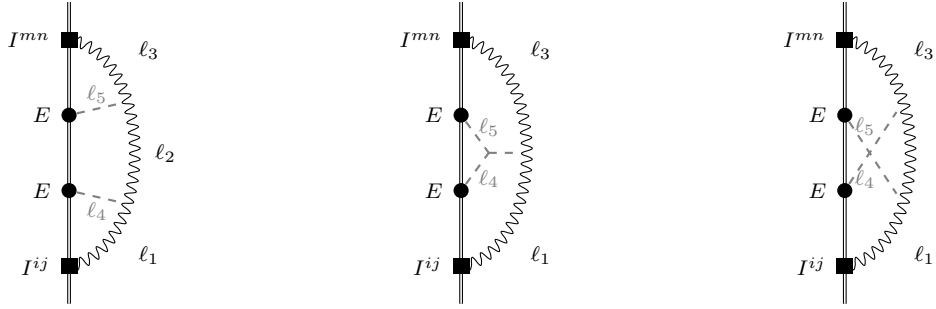
\begin{figure}[h]
  \centering
  \begin{subfigure}[t]{0.3\textwidth}
    \centering
   \begin{tikzpicture}[smooth]
      \draw[double] (0,-0.5) -- (0,3.5);
      \node[quad src={3}{\(I^{ij}\)}] (a) at (0,0){};
      \node[mass src={2}] (m1) at (0,1){};
      \node[mass src={2}] (m2) at (0,2){};
      \node[quad src={3}{\(I^{mn}\)}] (b) at (0,3) {};
      \draw[photon big](a.center) to[out=10,in=-10]
      node[pos=1/4,label=below right:{\(\ell_1\)}]{}
      node[pos=1/3] (v1){}
      node[pos=1/2,label=right:{\(\ell_2\)}]{}
      node[pos=2/3] (v2){}
      node[pos=3/4,label=above right:{\(\ell_3\)}] {}(b.center) ;
      \draw[potentialf] (m1) --node[pos=1/2,below=-2pt]{\(\ell_4\)} (v1);
      \draw[potentialf] (m2) -- node[pos=1/2,above=-2pt]{\(\ell_5\)} (v2);
    \end{tikzpicture}
    \caption{The cubic tail-of-tail diagram with three radiation
      propagators and two potential propagators.}
    \label{fig:tt-rad}
  \end{subfigure}
  \quad
    \begin{subfigure}[t]{0.3\textwidth}
    \centering
   \begin{tikzpicture}[smooth]
      \draw[double] (0,-0.5) -- (0,3.5);
      \node[quad src={3}{\(I^{ij}\)}] (a) at (0,0){};
      \node[mass src={2}] (m1) at (0,1){};
      \node[mass src={2}] (m2) at (0,2){};
      \node[quad src={3}{\(I^{mn}\)}] (b) at (0,3) {};
      \draw[photon big](a.center) to[out=10,in=-10]
      node[pos=1/4,label=below right:{\(\ell_1\)}]{}
      node[pos=1/2](v1){}
      node[pos=3/4,label=above right:{\(\ell_3\)}] {}(b.center) ;
      \coordinate (vc) at ($(v1) + (180:0.5)$);
      \draw[potentialf] (m1) --node[pos=1/2,below right=-5pt]{\(\ell_4\)} (vc);
      \draw[potentialf] (m2) -- node[pos=1/2,above right=-5pt]{\(\ell_5\)}(vc);
      \draw[potentialf](vc)--(v1);
    \end{tikzpicture}
    \caption{The cubic tail-of-tail diagram with two radiation
      propagators and three potential propagators.}
    \label{fig:tt-pot}
  \end{subfigure}
  \quad
    \begin{subfigure}[t]{0.3\textwidth}
    \centering
   \begin{tikzpicture}[smooth]
      \draw[double] (0,-0.5) -- (0,3.5);
      \node[quad src={3}{\(I^{ij}\)}] (a) at (0,0){};
      \node[mass src={2}] (m1) at (0,1){};
      \node[mass src={2}] (m2) at (0,2){};
      \node[quad src={3}{\(I^{mn}\)}] (b) at (0,3) {};
      \draw[photon big](a.center) to[out=10,in=-10]
      node[pos=1/4,label=below right:{\(\ell_1\)}]{}
      node[pos=1/3] (v1){}
      node[pos=1/2]{}
      node[pos=2/3] (v2){}
      node[pos=3/4,label=above right:{\(\ell_3\)}] {}(b.center) ;
      \draw[potentialf] (m1) --node[pos=1/3,below]{\(\ell_4\)} (v2);
      \draw[potentialf] (m2) -- node[pos=1/3,above]{\(\ell_5\)} (v1);
    \end{tikzpicture}
    \caption{The cubic tail-of-tail diagram with non-planar propagators.}
    \label{fig:tt-np}
  \end{subfigure}
  \caption{The tail-of-tail diagrams with cubic vertices that are used 
  	to select an advantageous basis of momentum invariants.}
  \label{fig:tt-cubic}
\end{figure}

We now proceed to the tail-of-tail calculation.  As in the previous
cases, our first task is to identify our basis of momentum invariants
and the associated integral family.  With two energy couplings, we
will have 3 free loop momenta from which to build momentum invariants.
A quick counting shows us that whatever basis of invariants we choose
needs to cover 3 $\ell_i^2$ plus $\binom{3}{2} = 3$ independent choices
of $\ell_i \cdot \ell_j$.  In line with what we did for the tail, it
is useful to analyze the cubic tail-of-tail diagrams, shown in
\cref{fig:tt-cubic}, to cover as much of the momentum basis as
possible using inverse propagators.  All diagrams have 5 propagators,
but only 4 can be chosen in common between \cref{fig:tt-rad} and
\cref{fig:tt-pot}: $\omega^2 - \ell_1^2$, $\omega^2 - \ell_3^2$,
$-\ell_4^2$, $-\ell_5^2$ are propagators in both.  However, this means
that the unique propagators in each diagram, $\omega^2 - \ell_2^2$ in
\cref{fig:tt-rad} and $-(\ell_1 + \ell_3)^2$ in \cref{fig:tt-pot} can
make up the fifth and sixth needed invariants.  We make the following 
particular labeling choice:
\begin{align}
  Q_1 &= \omega^2 - \ell_1^2 \ ,
  &\qquad
  Q_2  &= \omega^2 - \ell_3^2 \ ,
  &\qquad  \notag\\
  Q_3  &= -\ell_4^2 \ ,
  & \qquad 
  Q_4 &= -\ell_5^2= -(\ell_1 + \ell_4 +\ell_3)^2 \ , \\
  Q_5 &= \omega^2 - \ell_2^2 = \omega^2-(\ell_1 + \ell_4)^2 \ ,
  &\qquad
  Q_6 &= -(\ell_1 + \ell_3)^2 \notag\,.
\end{align}
At this point, it is also worth pointing out that there is an
additional ``non-planar'' propagator
$\omega^2 - (\ell_1 + \ell_5)^2 =Q_1 + Q_2 +Q_3 +Q_4 - Q_5 - Q_6 $
that could appear through a diagram like \cref{fig:tt-np}, which
appears to spoil the ability to define a single integral family to
cover all possible propagator structures.  In particular it will show
up later through the $u$-channel pole of a four-graviton amplitude.
However, since we are working to leading order in $G_N$, we expect the
energy couplings to be time-independent and thus indistinguishable.
This means that we can ``uncross'' the legs in \cref{fig:tt-np} (and
similar diagrams) which results in \cref{fig:tt-rad} (or similar)
\emph{except with $\ell_4 \leftrightarrow \ell_5$}, allowing us to
rewrite any integrals that appear with a
$\omega^2 - (\ell_1 + \ell_5)^{2}$ pole back in terms of the
propagator basis using the relabeling
$Q_5 \to Q_1 + Q_2 +Q_3 +Q_4 - Q_5 - Q_6$.

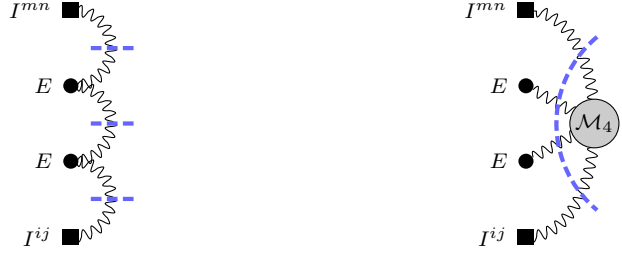
\begin{figure}[h]
  \centering
    \begin{subfigure}[b]{0.4\textwidth}
    \centering
    \begin{tikzpicture}
      \draw pic {hills={2}};
    \end{tikzpicture}
    \caption{The double-contact TT diagram}
    \label{fig:tt-hills}
  \end{subfigure}
  \begin{subfigure}[b]{0.4\textwidth}
    \centering
    \begin{tikzpicture}[smooth]
      \node[quad src={3}{\(I^{ij}\)}] (a) at (0,0){};
      \node[quad src={3}{\(I^{mn}\)}] (b) at (0,3) {};
      \draw pic {mamp={a}{b}{2}}{};
    \end{tikzpicture}
    \caption{The bulk contact TT diagram}
    \label{fig:tt-bulk}
  \end{subfigure}
  \caption{The unitarity cut diagrams needed to evaluate the tail-of-tail.}
  \label{fig:tt-diags}
\end{figure}

With the non-planar consideration dealt with, we have successfully
identified all propagators that can appear and used them to span the
set of momentum invariants.  Thus, we capture all possible
contributions coming from \cref{fig:tt-cubic} or their contact
diagrams using the single integral family
\begin{equation}
  F^{(3)}(\lambda_1,\lambda_2,\lambda_3,\lambda_4,\lambda_5,\lambda_6) 
  = \int \left(\frac{\Diff{d}\ell}{(2 \pi)^d}\right)^3 
  \frac{1}{Q_1^{\lambda_1}Q_2^{\lambda_2}Q_3^{\lambda_3} 
  Q_4^{\lambda_4}Q_5^{\lambda_5} Q_6^{\lambda_6}} \ ,
\end{equation}
with integer values of $\lambda_i$.  This integral family can be
reduced in terms of two relevant basis integrals \footnote{The basis
  technically has additional integrals in it, for instance
  $F^{(3)}(1,0,0,1,1,1)$, due to a rotational symmetry of the
  graphical representation of the family.  This rotational symmetry is
  broken in the actual computation by the identification of which
  radiation propagator is sourced from which quadrupole.}
\begin{equation}
  \mathcal{I}^{(3)} = \{F^{(3)}(1,1,0,0,1,0),F^{(3)}(1,1,1,1,0,0)\} \ ,
\end{equation}
with topologies corresponding to the unitarity cut diagrams shown in
\cref{fig:tt-diags}.  Thus the effective action for the tail-of-tail
will be
\begin{align}
  S_{\text{TT}}
  &= \intw\left[ a_{\text{TT},1}F_{\text{CTP}}^{(3)}(1,1,0,0,1,0)
    +a_{\text{TT},2} F_{\text{CTP}}^{(3)}(1,1,1,1,0,0) \right] \,.
    \label{eq:tt-raw-s}
 \end{align}
 We now proceed to evaluate the cuts in \cref{fig:tt-diags} to
 determine $a_{\text{TT},1}$ and $a_{\text{TT},2}$.

The process for evaluating the first cut, \cref{fig:tt-hills},
proceeds almost identically to evaluating the tail cut.  We have
\begin{align}
  \cut_{\text{\cref{fig:tt-hills}}}
  &= \lambda_I^2 \delta(Q_1)\delta(Q_2)\delta(Q_5)J_{I(-\omega)}^{\mu_1 \nu_1} 
  P^{\mu_1 \nu_1; \mu_2 \nu_2} \mathcal{M}_{Eg^2,1}^{\mu_2 \nu_2; \mu_3 \nu_3} \notag \\
  & \qquad \times 
  P^{\mu_3 \nu_3; \mu_4 \nu_4}\mathcal{M}_{Eg^2,2}^{\mu_4 \nu_4; \mu_5 \nu_5}
  P^{\mu_5 \nu_5; \mu_6 \nu_6} J_{I(\omega)}^{\mu_6 \nu_6} \,,
\end{align}
where we have added subscripts to the $\mathcal{M}_{Eg^2}$ amplitudes to
distinguish the two momentum labelings.  Inserting the relevant
definitions and evaluating all of the index contractions results in an
expression that is hundreds of terms long, but has the general structure
\begin{align}
  \cut_{\text{\cref{fig:tt-hills}}} = \ttcoup
  & \delta(Q_1) \delta(Q_2) \delta(Q_5)
    \Big( \frac{1}{Q_3 Q_4} g_{1}(Q_6,\omega,d) 
    + \frac{1}{Q_3} g_{2}(Q_4,Q_6,\omega,d) \notag \\
  &+ \frac{1}{Q_4} g_{3}(Q_3,Q_6,\omega,d)
    + g_{4}(Q_3,Q_4,Q_6,\omega,d) \Big) \,,
    \label{eq:tt-hills-cut}
\end{align}
where each of the $g_i$ are polynomials in the $Q$s, but rational in
$d$ and $\omega$.  This specific form of the cut highlights the
available factorization channels that can be cross-checked: $g_2$ and
$g_3$ are partial contacts that can be checked by calculating the
relevant cuts; $g_1$ is a channel that \emph{overlaps} with
\cref{fig:tt-bulk} and thus serves both as a check and as
demonstration for why symmetry factors are necessary in the cut
matching process, which we will show after constructing the other cut.

To constrain the integral basis coefficients using the cut, we must
reduce it to the basis.  As before, we do so by treating the
$\delta(Q_i)$s as propagators and reducing using standard methods, but
since we have more than one basis element, the $\delta(Q_i)$s instruct
us to only keep the parts of the reduction which have at least $Q_1$,
$Q_2$, and $Q_5$ as propagators.  In particular, the $g_2$, $g_3$,
and $g_4$ terms produce only the needed contributions, but the
reduction of the $g_1$ term will produce both basis elements.  We only
keep the one involving $F^{(3)}(1,1,0,0,1,0)$.  Performing the
reduction in this manner, we arrive at
\begin{align}
  \rcut{\text{\cref{fig:tt-hills}}}
  &=
  \ttcoup
  \frac{(d-2)(12-2d + 5d^2 - 4 d^3 + d^4)^2  \omega^4}
  {4(d-3)^2 (d-1)^3 d^2 (d+1)(d+2)} 
    \delta(Q_1) \delta(Q_2) \delta(Q_5) \notag \\
   &=
  \ttcoup
  \frac{(d-2)\mathcal{P}_4^2  \omega^4}
  {4(d-3)^2 (d-1)^3 d^2 (d+1)(d+2)} 
  \delta(Q_1) \delta(Q_2) \delta(Q_5)\,.
  \label{eq:tt-hills-rcut}
\end{align}

We now turn our attention to the second cut, \cref{fig:tt-bulk}, which
is constructed as
\begin{align}
  \cut_{\text{\cref{fig:tt-bulk}}} = \lambda_I^2
  (J_{I(-\omega)} P)^{\mu_1 \nu_1}
  (\mathcal{M}_{sgs,1} P)^{\mu_2 \nu_2}
  \mathcal{M}_4^{\mu_1 \nu_1 \dots \mu_4 \nu_4}
  (P \mathcal{M}_{sgs,2})^{\mu_3 \nu_3}
  (P J_{I(\omega)})^{\mu_4 \nu_4} \,.
  \label{eq:tt-bc-ms}
\end{align}
This cut is also almost one hundred terms long, but has the schematic form
\begin{align}
  \cut_{\text{\cref{fig:tt-bulk}}} =\ttcoup
  & \delta(Q_1) \delta(Q_2) \delta(Q_3)\delta(Q_4) \Big( \frac{1}{Q_5} h_{1}(Q_6,\omega,d) 
    + \frac{1}{Q_6} h_{2}(Q_5, \omega, d) \notag \\
  &+ \frac{1}{Q_5 {+} Q_6} h_{3}(Q_5-Q_6,\omega,d)
    + h_{4}(Q_5,Q_6,\omega,d)\Big) \,,
    \label{eq:tt-bulk-cut}
\end{align}
with the $h_i$ having similar properties as the $g_i$ above.  The
arrangement is again chosen to manifest pole structures and
factorization: $h_2$ contains all contributions with the pole
structures of \cref{fig:tt-pot}, $h_1$ all those with the poles of
\cref{fig:tt-rad}, $h_3$ from \cref{fig:tt-np}, and $h_4$ the contact
contribution.

From here we can delve into the overlapping channels to highlight the
internal consistency checks and the need for relative symmetry
factors.  The objects of interest for the discussion are the
overlapping channels from \cref{eq:tt-hills-cut} and
\cref{eq:tt-bulk-cut}
\begin{align}
  g_1(Q_6,\omega,d)
  &= h_1(Q_6,\omega,d) = -h_3(Q_5-Q_6,\omega,d) \notag\\
  &= \frac{1}{8(d-1)^3(d+2)}\Big[
    (-2 + d)^2 Q_6^4 + 4 (-2 + d)^2 Q_6^3 \omega^2 \notag \\
  & +  4 (d-1)(-9 + d + d^2) Q_6^2 \omega^4 + 
    8 (-2 + d) (-1 + d) (1 + d) Q_6 \omega^6  \notag \\
  &+  8 (-2 + d) (-1 + d)^2 (1 + d) \omega^8 \Big]\,.
\end{align}
We immediately see that the overlapping channels agree: when we
evaluate the $Q_3 = Q_4 = 0$ residue of \cref{eq:tt-hills-cut},
essentially picking out the $g_1$ contribution, we get exactly the
same expression as when evaluating the $Q_5=0$ residue of
\cref{eq:tt-bulk-cut}.  However, when we ``uncross'' the $Q_5+Q_6$
pole using $Q_5 \to - Q_5 - Q_6$ (the consequence of the
$\ell_4 \leftrightarrow \ell_5$ relabeling, with the help of the
on-shell conditions), the cut becomes
\begin{align}
  \cut_{\text{\cref{fig:tt-bulk}}} =\ttcoup
  & \delta(Q_1) \delta(Q_2) \delta(Q_3)\delta(Q_4) \Big( \frac{1}{Q_5} 2 h_{1}(Q_6,\omega,d) 
    + \frac{1}{Q_6} h_{2}(Q_5, \omega, d) \notag \\
  &
    + h_{4}(Q_5,Q_6,\omega,d)\Big) \,.
    \label{eq:tt-bulk-cut-uc}
\end{align}
Through the $Q_5^{-1} 2 h_1(Q_6,\omega,d)$ term, we see that
$\cut_{\text{\cref{fig:tt-bulk}}}$ is actually double-counting its
contribution with respect to $\cut_{\text{\cref{fig:tt-hills}}}$.
This misalignment is exactly what is compensated for by the $|G|$
normalization of cut matching, \cref{eq:cut-matching}.  In the current
case, the diagram of \cref{fig:tt-bulk} has an extra symmetry of
swapping the energy sources (or $\ell_4 \leftrightarrow \ell_5$ as we
actually use it) that \cref{fig:tt-hills} does not have.

After the uncrossing, it is now also straightforward to reduce the
cut.  Again we only keep the reductions with propagators aligning with
the $\delta(Q_i)$, namely those producing $F^{(3)}(1,1,1,1,0,0)$.
Doing so yields
\begin{align}
  \rcut{\text{\cref{fig:tt-bulk}}}
  &= \ttcoup \omega^6  \delta(Q_1) \delta(Q_2) \delta(Q_3)\delta(Q_4) \label{eq:tt-bulk-rcut}\\
  & \times \frac{(2d-3)
  \left(960 - 1696 d + 424 d^2 - 476 d^3 + 330 d^4
  - 39 d^5 + 53 d^6 - 45 d^7 + 9 d^8\right)
  }{3(d-3)(d-1)^3 d (d+1)(d+2) (3d-4)(3d-2)} \notag
    \,.
\end{align}
We similarly name the $d$ polynomial in the numerator
\begin{equation}
  \mathcal{P}_8 = 960 - 1696 d + 424 d^2 - 476 d^3 + 330 d^4
  - 39 d^5 + 53 d^6 - 45 d^7 + 9 d^8 \,.
  \label{eq:p8}
\end{equation}

With both cuts in hand, we can now construct the effective action.
The cuts are again in one-to-one correspondence with the coefficients,
\begin{subequations}
\begin{align}
  a_{TT,1}\, \delta(Q_1) \delta(Q_2) \delta(Q_5)
  &= \frac{\rcut{\text{\cref{fig:tt-hills}}}}{|\graph{\cref{fig:tt-hills}}|}
  = \frac{\rcut{\text{\cref{fig:tt-hills}}}}{2} \ ,\\
  a_{TT,2}\, \delta(Q_1) \delta(Q_2) \delta(Q_4)\delta(Q_5)
  &= \frac{\rcut{\text{\cref{fig:tt-bulk}}}}{|\graph{\cref{fig:tt-bulk}}|}
  = \frac{\rcut{\text{\cref{fig:tt-bulk}}}}{4} \,.
\end{align}
\end{subequations}
The integrals themselves are straightforward to evaluate.
$F^{(3)}(1,1,0,0,1,0)$ is just the cube of a one-propagator integral,
while $F^{(3)}(1,1,1,1,0,0)$ is evaluable via bubble iteration.
More details on the evaluations are provided in \cref{sec:ints-analytic}.
We thus have all of the information required to construct the
tail-of-tail effective action via \cref{eq:tt-raw-s}.  Expanding in
$d=3+\dimreg$ and performing the CTP sum yields
\begin{align}
  \label{eq:tt-raw}
  S_{\text{TT}} = \frac{214}{525} G_N^3 E^2 \intw
  &  \omega^7 \kappa_{-+}(\omega)\Bigg\{
    \frac{i}{\dimreg}
    +\left[\frac{3}{2} i \log\left(\frac{\omega^2 e^{\gamma_E}}{\mu^2 \pi}\right)
    +\frac{3 \pi  \sgn(\omega )}{2}-
    \frac{420}{107} i\zeta_{2}-\frac{675359 }{89880}i\right] \notag \\
  &+\dimreg \Bigg[ \pi\sgn(\omega ) \left(
    \frac{9}{4}  \log\left(\frac{\omega^2 e^{\gamma_E}}{\mu^2 \pi}\right)
    -\frac{ (352800 \zeta_{2}+675359)}{59920}\right)    \notag \\
  &\quad  + i \log\left(\frac{\omega^2 e^{\gamma_E}}{\mu^2 \pi}\right)
    \left(\frac{9}{8} \log\left(\frac{\omega^2 e^{\gamma_E}}{\mu^2 \pi}\right)
    -\frac{ (352800\zeta_{2}+675359)}{59920}\right) \notag \\
 &\quad   +\frac{4569}{856} i \zeta_{2}-\frac{1050 }{107}i \zeta_3+\frac{1259125247 }{37749600}i\Bigg] + \odimreg{2} \Bigg\} \ ,
\end{align}
with $\zeta_n$ the Riemann zeta values: $\zeta_2= \frac{\pi^2}{6}$,
$\zeta_3 = 1.20206\dots$, $\zeta_4 = \frac{\pi^4}{90}$, etc.  This
result, first reported in Ref.~\cite{Edison:2022cdu}, is the first time
the tail-of-tail effective action as been computed.

\subsection{Tail-of-Tail-of-Tail}
\label{sec:ttt}
\begin{figure}[h]
  \centering
  \begin{subfigure}[t]{0.22\textwidth}
    \centering
   \begin{tikzpicture}[smooth]
      \draw[double] (0,-0.5) -- (0,4.5);
      \node[quad src={3}{\(I^{ij}\)}] (a) at (0,0){};
      \node[mass src={2}] (m1) at (0,1){};
      \node[mass src={2}] (m2) at (0,2){};
      \node[mass src={2}] (m3) at (0,3){};
      \node[quad src={3}{\(I^{mn}\)}] (b) at (0,4) {};
      \draw[photon big](a.center)
      .. controls ($(a)+(2,0)$) and ($(b)+(2,0)$)..
      node[pos=1/8,below right]{\(Q_1\)}
      node[pos=1/4] (v1){}
      node[pos=3/8,right]{\(Q_2\)}
      node[pos=1/2] (v2){}
      node[pos=5/8,above right]{\(Q_3\)}
      node[pos=3/4] (v3){}
      node[pos=7/8,above right]{\(Q_4\)}
      (b.center) ;
      \draw[potentialf] (m1) --node[pos=1/2,above=-2pt]{\(Q_5\)} (v1);
      \draw[potentialf] (m2) -- node[pos=1/2,above=-2pt]{\(Q_6\)} (v2);
      \draw[potentialf] (m3) -- node[pos=1/2,above=-2pt]{\(Q_7\)} (v3);
    \end{tikzpicture}
  \end{subfigure}
  \quad
    \begin{subfigure}[t]{0.22\textwidth}
    \centering
   \begin{tikzpicture}[smooth]
      \draw[double] (0,-0.5) -- (0,4.5);
      \node[quad src={3}{\(I^{ij}\)}] (a) at (0,0){};
      \node[mass src={2}] (m1) at (0,1){};
      \node[mass src={2}] (m2) at (0,2){};
      \node[mass src={2}] (m3) at (0,3){};
      \node[quad src={3}{\(I^{mn}\)}] (b) at (0,4) {};
      \draw[photon big](a.center)
      .. controls ($(a)+(2,0)$) and ($(b)+(2,0)$)..
      node[pos=1/4,below right]{\(Q_1\)}
      node[pos=27/64](v1){}
      node[pos=9/16,right]{\(Q_3\)}
      node[pos=3/4,above right]{\(Q_4\)}
      node[pos=11/16](v2){}
      (b.center) ;
      \coordinate (vc) at ($(v1) + (180:1)$);
      \draw[potentialf] (m1) --node[pos=1/3,below right=-5pt]{\(Q_5\)} (vc);
      \draw[potentialf] (m2) -- node[pos=1/6,above right=-5pt]{\(Q_6\)}(vc);
      \draw[potentialf](vc)--node[pos=1/2,above]{\(Q_8\)} (v1);
      \draw[potentialf](m3)--node[pos=1/2,below]{\(Q_7\)} (v2);
    \end{tikzpicture}
      \end{subfigure}
    \begin{subfigure}[t]{0.22\textwidth}
    \centering
   \begin{tikzpicture}[smooth]
      \draw[double] (0,-0.5) -- (0,4.5);
      \node[quad src={3}{\(I^{ij}\)}] (a) at (0,0){};
      \node[mass src={2}] (m1) at (0,1){};
      \node[mass src={2}] (m2) at (0,2){};
      \node[mass src={2}] (m3) at (0,3){};
      \node[quad src={3}{\(I^{mn}\)}] (b) at (0,4) {};
      \draw[photon big](a.center)
      .. controls ($(a)+(2,0)$) and ($(b)+(2,0)$)..
      node[pos=1/4,below right]{\(Q_1\)}
      node[pos=5/16](v1){}
      node[pos=7/16,right]{\(Q_2\)}
      node[pos=3/4,above right]{\(Q_4\)}
      node[pos=37/64](v2){}
      (b.center) ;
      \coordinate (vc) at ($(v2) + (180:1)$);
      \draw[potentialf] (m3) --node[pos=1/6,above right=-5pt]{\(Q_7\)} (vc);
      \draw[potentialf] (m2) -- node[pos=1/3,below right=-5pt]{\(Q_6\)}(vc);
      \draw[potentialf](vc)--node[pos=1/2,above]{\(Q_9\)} (v2);
      \draw[potentialf](m1)--node[pos=1/2,below]{\(Q_5\)} (v1);
    \end{tikzpicture}
  \end{subfigure}
  \quad
    \begin{subfigure}[t]{0.22\textwidth}
    \centering
   \begin{tikzpicture}[smooth]
      \draw[double] (0,-0.5) -- (0,4.5);
      \node[quad src={3}{\(I^{ij}\)}] (a) at (0,0){};
      \node[mass src={2}] (m1) at (0,1){};
      \node[mass src={2}] (m2) at (0,2){};
      \node[mass src={2}] (m3) at (0,3){};
      \node[quad src={3}{\(I^{mn}\)}] (b) at (0,4) {};
      \draw[photon big](a.center)
      .. controls ($(a)+(2.5,0)$) and ($(b)+(2.5,0)$)..
      node[pos=1/4,below right]{\(Q_1\)}
      node[pos=1/2] (v1) {}
      node[pos=3/4,above right]{\(Q_4\)}
      (b.center) ;
      \draw[potentialf](m2) --
      node[pos=1/8,above]{\(Q_6\)}
      node[pos=1/3](vc1){}
      node[pos=1/2,below]{\(Q_8\)}
      node[pos=9/12](vc2){}
      node[pos=10/12,above] {\(Q_{10}\)}
      (v1);
      \draw[potentialf] (m1) --node[pos=1/3,below right=-5pt]{\(Q_5\)} (vc1);
      \draw[potentialf](m3)--node[pos=1/3,above]{\(Q_7\)} (vc2);
    \end{tikzpicture}
  \end{subfigure}
  \caption{The cubic TTT diagrams that are used to select a basis of
    momentum invariants}
  \label{fig:ttt-cubic}
\end{figure}
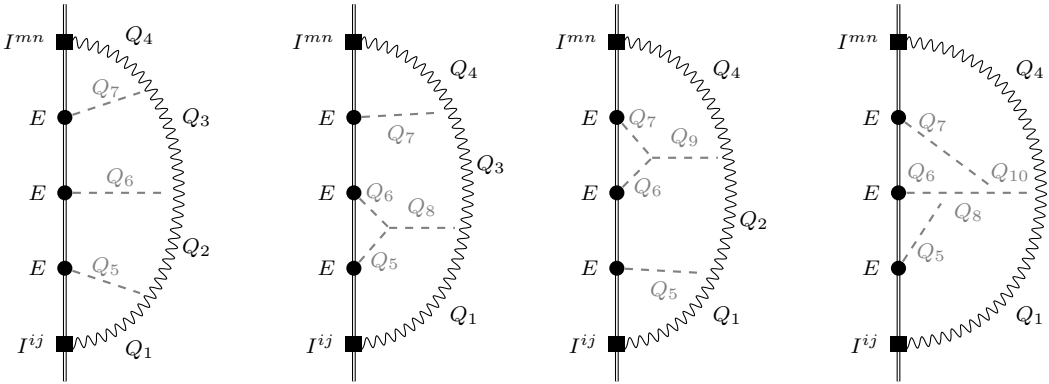

\begin{figure}[h]
  \centering
  \begin{subfigure}[b]{0.2\textwidth}
    \centering
    \begin{tikzpicture}
      \draw pic {hills={3}};
    \end{tikzpicture}
    \caption{}
    \label{fig:ttt-1}
  \end{subfigure}
  \begin{subfigure}[b]{0.2\textwidth}
    \centering
    \pgfmathsetmacro{\mampPushback}{1.5}
    \begin{tikzpicture}[smooth]
      \node[quad src={3}{\(I^{ij}\)}] (a) at (0,0){};
      \node[quad src={3}{\(I^{mn}\)}] (b) at (0,4) {};
      \draw pic {mamp={a}{b}{3}};
    \end{tikzpicture}
    \pgfmathsetmacro{\mampPushback}{1.2}
        \caption{}
    \label{fig:ttt-2}
  \end{subfigure}
  \begin{subfigure}[b]{0.2\textwidth}
    \centering
    \begin{tikzpicture}
      \node[quad src={3}{\(I^{ij}\)}] (a) at (0,0){};
      \node[quad src={3}{\(I^{mn}\)}] (b) at (0,4) {};
      \node[mass src={2}](m1) at (0,1){};
      \draw pic {mamp={m1}{b}{2}};
      \draw pic {cut hill={a}{m1}};
    \end{tikzpicture}
        \caption{}
    \label{fig:ttt-3}
  \end{subfigure}
  \begin{subfigure}[b]{0.2\textwidth}
    \centering
    \begin{tikzpicture}
      \node[quad src={3}{\(I^{ij}\)}] (a) at (0,0){};
      \node[quad src={3}{\(I^{mn}\)}] (b) at (0,4) {};
      \node[mass src={2}](m3) at (0,3){};
      \draw pic {mamp={a}{m3}{2}};
      \draw pic {cut hill={m3}{b}};
    \end{tikzpicture}
        \caption{}
    \label{fig:ttt-4}
  \end{subfigure}
  \caption{The TTT unitarity cuts.}
  \label{fig:ttt-cuts}
\end{figure}
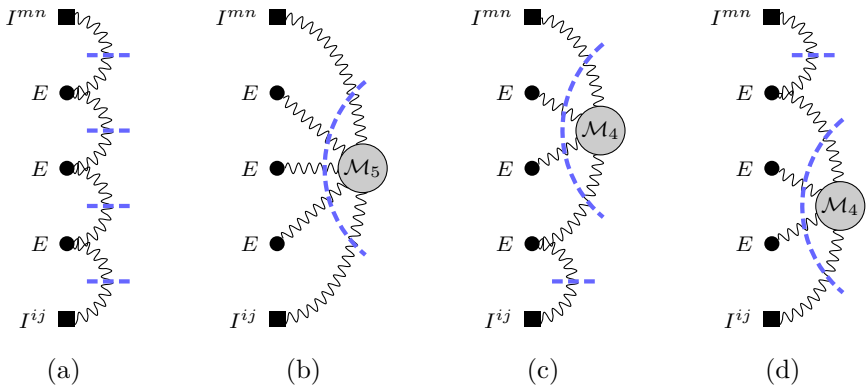

There are few new features to the process of calculating TTT.  It is
primarily a proliferation of all features.  First, we have a 
momentum-invariants basis with $\binom{4}{2} + 4 = 10$ elements that 
we need to choose.
We use the cubic diagrams shown in \cref{fig:ttt-cubic} to define the
set of propagators we select as a basis.  There are significantly more
``uncrossing'' relabelings that can be used to return non-planar pole
structures back to the basis, all of which we will need to employ when
expanding higher-point graviton amplitudes.  Thus we still only need a
single integral family, given by
\begin{equation}
  F^{(4)}(\lambda_1,\dots, \lambda_{10}) 
  = \int \left(\frac{\Diff{d}\ell}{(2 \pi)^d}\right)^4 
  \frac{1}{\prod_{i=1}^{10}Q_i^{\lambda_i}} \,.
\end{equation}
The relevant basis integrals are
\begin{align}
  \mathcal{I}^{(4)}
  &= \{ F^{(4)}(1,1,1,1,0,0,0,0,0,0),
    F^{(4)}(1,0,0,1,1,1,1,0,0,0), \notag \\
  & \qquad  F^{(4)}(1,0,1,1,1,1,0,0,0,0),
    F^{(4)}(1,1,0,1,0,1,1,0,0,0) \} \,,
\end{align}
which correspond to the topologies in \cref{fig:ttt-cuts}.  The first,
third, and fourth basis integrals are all factorizable, while the
third is evaluable via bubble iteration.  Notably, the last two integrals are
actually equal: the integrals both factorize in exactly the same way
so it does not matter whether the one-propagator piece occurs first or
last in the evaluation.  Thus, at the level of the CTP sum, the
$- +$ orientation of one diagram exactly matches the $+ -$ of
the other.  We could use this symmetry to remove one of the two
integrals and corresponding cut from the basis, at which point it
would \emph{no longer carry the reflection symmetry factor}.  Instead
we keep both contributions and the reflection symmetry factor will
remove the over-count of keeping both.  This allows an explicit
verification that both diagrams have identical contributions so either
approach would produce the same result.

The cuts are still one-to-one with the basis coefficients, so we
simply need the symmetry factors of each cut, which come from the
number of equivalent rearrangements of the $E$-connected lines along
with the reflection symmetry
\begin{align}
  |\graph{\cref{fig:ttt-1}}| = 2 \ , \qquad
  |\graph{\cref{fig:ttt-2}}| = 2 \times 3! = 12 \ , \qquad
  |\graph{\cref{fig:ttt-3}}| = |\graph{\cref{fig:ttt-4}}| = 2 \times 2! = 4 \ ,
\end{align}
along with each of the reduced cuts.  The corresponding cuts are all
thousands of terms long prior to reduction, but after reduction each
one produces a basis coefficient that is a relatively simple rational
function of $d$ and $\omega$:
\begin{subequations}
\begin{align}
  a_{\text{TTT,\cref{fig:ttt-1}}}
    &= -\frac{\tttcoup \omega^4(d-2)\mathcal{P}_4^3}{8 (d-3)^3 (d-1)^4 d^3 (d+1)^2(d+2)}\,, \label{eq:ttt-a1}\\
  a_{\text{TTT,\cref{fig:ttt-2}}}
  &=  \frac{\tttcoup \omega^6(d-2)(3d-5) \mathcal{P}_{11}}
    {12(d-3)^2 (d-1)^4d (d+1)^2 (d+2)(2d-3)(3d-4)(3d-2)}\,, \label{eq:ttt-a2}\\
  a_{\text{TTT,\cref{fig:ttt-3}}}
  &=  -\frac{\tttcoup \omega^6(2d-3)\mathcal{P}_4 \mathcal{P}_8}
    {12 (d-3)^2 (d-1)^4 d^2 (d+1)^2 (d+2) (3d-4) (3d-2)}\,, \label{eq:ttt-a3}\\
    a_{\text{TTT,\cref{fig:ttt-4}}}
    &=  -\frac{\tttcoup \omega^6(2d-3)\mathcal{P}_4 \mathcal{P}_8}
  {12 (d-3)^2 (d-1)^4 d^2 (d+1)^2 (d+2) (3d-4) (3d-2)}\,, \label{eq:ttt-a4}
\end{align}
\label{eq:ttt-coefs}
\end{subequations}
with $\mathcal{P}_4$ and $\mathcal{P}_8$ from \cref{eq:p4} and \cref{eq:p8} respectively, and
\begin{align}
  \mathcal{P}_{11} &= -3024 + 3720d + 2980 d^2 + 996 d^3 - 2426 d^4 - 737 d^5 + 799 d^6 \notag\\
                   &- 284 d^7 - 36 d^8 + 223 d^9 - 117 d^{10} + 18 d^{11} \,.
\end{align}
Assembling the effective action by evaluating the integrals (see
\cref{sec:ints-analytic}), combining with the coefficients, and expanding
in $d=3+\dimreg$ results in
\begin{align}
  \label{eq:ttt-raw}
  S_{\text{TTT}}
  &= - \frac{214}{525} G_N^4 E^3 \intw \omega^8 \kappa_{-+}(\omega) \Bigg\{
    \frac{1}{\dimreg^2}
    +\frac{1}{\dimreg}\left[2
    \log\left(\frac{\omega^2 e^{\gamma_E}}{\mu^2 \pi}\right)
    -2 i \pi  \text{sgn}(\omega )-\frac{252583}{29960}\right] \notag \\
  &  +\Bigg[\log\left(\frac{\omega^2 e^{\gamma_E}}{\mu^2 \pi}\right)\left(2 \log\left(\frac{\omega^2 e^{\gamma_E}}{\mu^2 \pi}\right)
    -\frac{252583}{14980}\right)
    + \left(\frac{252583 }{14980}- 4\log\left(\frac{\omega^2 e^{\gamma_E}}{\mu^2 \pi}\right)\right) i \pi \text{sgn}(\omega )\notag \\
  & \quad
    -\frac{29 }{2}\zeta_{2}
    -\frac{840}{107}\zeta_{3}
    +\frac{1583459537}{37749600}\Bigg] \notag \\
  &+\dimreg \Bigg[\log\left(\frac{\omega^2 e^{\gamma_E}}{\mu^2 \pi}\right) \bigg\{
    \log\left(\frac{\omega^2 e^{\gamma_E}}{\mu^2 \pi}\right) \Bigg[
    \frac{4}{3} \log\left(\frac{\omega^2 e^{\gamma_E}}{\mu^2 \pi}\right)
    -\frac{252583 }{14980} \Bigg] \notag \\
  & \qquad + \left(-29 \zeta_{2}-\frac{1680 \zeta_{3}}{107}+\frac{1583459537}{18874800}\right) \bigg\} \notag \\
  & \quad +\frac{7324907 \zeta_{2}}{59920}
    +\frac{14309 \zeta_{3}}{642}+\frac{420\zeta_{4}}{107}
    -\frac{104414536729}{634193280} \notag \\
  &\quad - i \pi \sgn(\omega) \bigg\{
    2 \log\left(\frac{\omega^2 e^{\gamma_E}}{\mu^2 \pi}\right)
    \left(2  \log\left(\frac{\omega^2 e^{\gamma_E}}{\mu^2 \pi}\right)-\frac{252583 }{14980} \right) \notag \\
  & \qquad -13\zeta_{2}-\frac{1680}{107}\zeta_{3}+\frac{1583459537}{18874800} \bigg\}\Bigg] + \odimreg{2} \Bigg\}\,.
\end{align}
We first reported this result in Ref.~\cite{Edison:2022cdu}, which was
the first time the tail-of-tail-of-tail was tackled outside of traditional GR.

\subsection{Tail-of-Tail-of-Tail-of-Tail}
\label{sec:tttt}
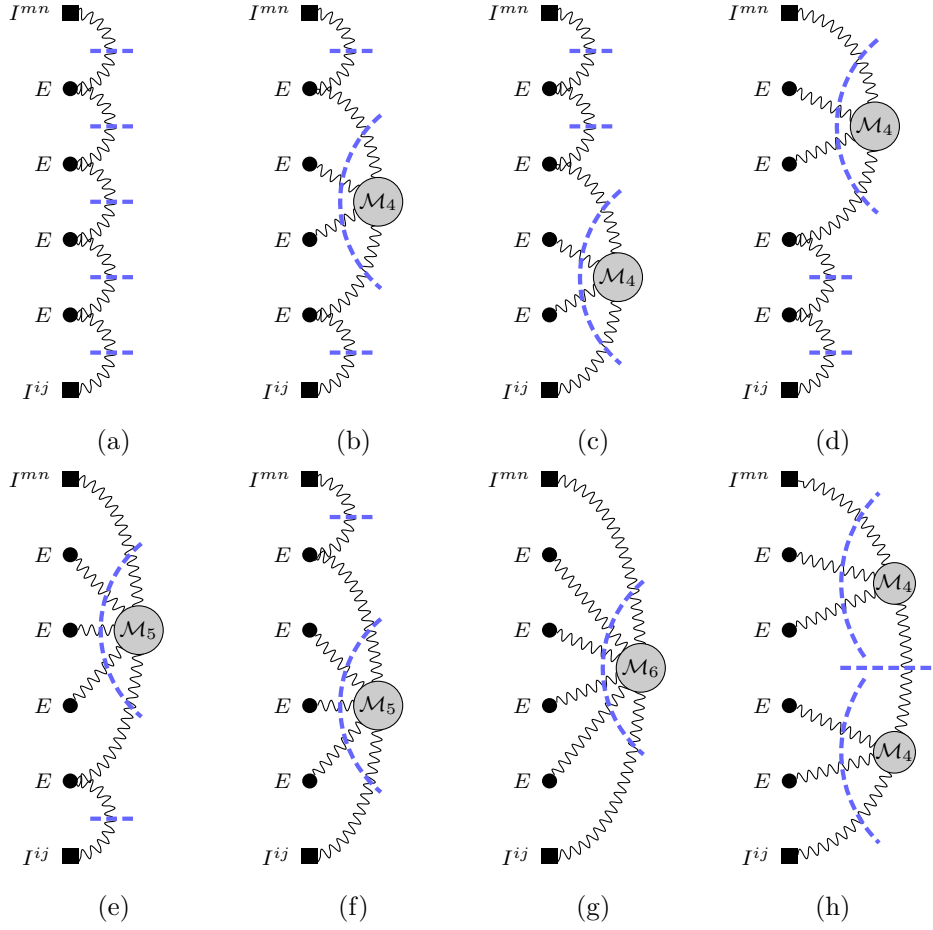
\begin{figure}[h]
  \centering
  \begin{subfigure}[b]{0.2\textwidth}
    \begin{tikzpicture}
      \draw pic {hills={4}};
    \end{tikzpicture}
    \caption{}
    \label{fig:tttt-1}
  \end{subfigure}
    \begin{subfigure}[b]{0.2\textwidth}
    \begin{tikzpicture}
      \node[quad src={3}{\(I^{ij}\)}] (a) at (0,0){};
      \node[quad src={3}{\(I^{mn}\)}] (b) at (0,5) {};
      \node[mass src={2}](m1) at (0,1){};
      \node[mass src={2}](m4) at (0,4){};
      \draw pic {mamp={m1}{m4}{2}};
      \draw pic {cut hill={a}{m1}};
      \draw pic {cut hill={m4}{b}};
    \end{tikzpicture}
        \caption{}
    \label{fig:tttt-2}
  \end{subfigure}
  \begin{subfigure}[b]{0.2\textwidth}
    \begin{tikzpicture}
            \node[quad src={3}{\(I^{ij}\)}] (a) at (0,0){};
      \node[quad src={3}{\(I^{mn}\)}] (b) at (0,5) {};
      \node[mass src={2}](m3) at (0,3){};
      \node[mass src={2}](m4) at (0,4){};
      \draw pic {mamp={a}{m3}{2}};
      \draw pic {cut hill={m3}{m4}};
      \draw pic {cut hill={m4}{b}};
    \end{tikzpicture}
        \caption{}
    \label{fig:tttt-3}
  \end{subfigure}
  \begin{subfigure}[b]{0.2\textwidth}
    \pgfmathsetmacro{\mampPushback}{1.5}
    \begin{tikzpicture}
            \node[quad src={3}{\(I^{ij}\)}] (a) at (0,0){};
      \node[quad src={3}{\(I^{mn}\)}] (b) at (0,5) {};
      \node[mass src={2}](m1) at (0,1){};
      \node[mass src={2}](m2) at (0,2){};
      \draw pic {mamp={m2}{b}{2}};
      \draw pic {cut hill={a}{m1}};
      \draw pic {cut hill={m1}{m2}};
    \end{tikzpicture}
        \caption{}
    \label{fig:tttt-4}
  \end{subfigure}
  \begin{subfigure}[b]{0.2\textwidth}
    \begin{tikzpicture}
      \node[quad src={3}{\(I^{ij}\)}] (a) at (0,0){};
      \node[quad src={3}{\(I^{mn}\)}] (b) at (0,5) {};
      \node[mass src={2}](m1) at (0,1){};
      \draw pic {mamp={m1}{b}{3}};
      \draw pic {cut hill={a}{m1}};
    \end{tikzpicture}
        \caption{}
    \label{fig:tttt-5}
  \end{subfigure}
  \begin{subfigure}[b]{0.2\textwidth}
    \begin{tikzpicture}
      \node[quad src={3}{\(I^{ij}\)}] (a) at (0,0){};
      \node[quad src={3}{\(I^{mn}\)}] (b) at (0,5) {};
      \node[mass src={2}](m3) at (0,4){};
      \draw pic {mamp={a}{m3}{3}};
      \draw pic {cut hill={m3}{b}};
    \end{tikzpicture}
        \caption{}
    \label{fig:tttt-6}
  \end{subfigure}
  \begin{subfigure}[b]{0.2\textwidth}
    \pgfmathsetmacro{\mampPushback}{1.6}
    \begin{tikzpicture}
      \node[quad src={3}{\(I^{ij}\)}] (a) at (0,0){};
      \node[quad src={3}{\(I^{mn}\)}] (b) at (0,5) {};
      \draw pic {mamp={a}{b}{4}};
    \end{tikzpicture}
        \caption{}
    \label{fig:tttt-7}
    \pgfmathsetmacro{\mampPushback}{1.2}
  \end{subfigure}
  \begin{subfigure}[b]{0.2\textwidth}
    \begin{tikzpicture}
      \node[quad src={3}{\(I^{ij}\)}] (a) at (0,0){};
      \node[quad src={3}{\(I^{mn}\)}] (b) at (0,5) {};
      \foreach \ml in {1,...,4}
      \node[mass src={2}] (\ml) at (0,\ml){};
      \node (ci) at (2,2.5) {};
      \node (cf) at (0.5,2.5) {};
      \draw[photon big](a) to[out=10,in=-10] node[circle,fill=blobcolor,draw=black,inner sep=0pt,pos=1/3] (v1){$\mathcal{M}_4$} node[circle,fill=blobcolor,draw=black,inner sep=0pt,pos=2/3] (v2){$\mathcal{M}_4$} (b) ;
      \begin{pgfonlayer}{background}
      \draw[potential] (1) -- (v1.center);
      \draw[potential] (2) -- (v1.center);
      \draw[potential] (3) -- (v2.center);
      \draw[potential] (4) -- (v2.center);
    \end{pgfonlayer}
    \draw[cuts] (ci)--(cf);
      \draw[cuts] ($(v1) +(1,0)+(145:1.7)$) arc (145:360-135:1.7);
      \draw[cuts] ($(v2) +(1,0)+(360-145:1.7)$) arc (360-145:135:1.7);
    \end{tikzpicture}
        \caption{}
    \label{fig:tttt-8}
  \end{subfigure}
  
  \caption{All of the unitarity cut topologies needed for $T^4$.}
  \label{fig:t4-all}
\end{figure}

Analyzing all of the patterns between the previous tails, we are able
to make predictions about both what cuts we need to evaluate, and what
their value should be.  For the cut basis, we expect diagrams
involving no bulk graviton contacts, as well as all diagrams involving
at least four-point bulk contacts.  Thus a reasonable initial guess
for the needed integrals and cuts would be those shown in
\cref{fig:tttt-1,fig:tttt-2,fig:tttt-3,fig:tttt-4,fig:tttt-5,fig:tttt-6,fig:tttt-7}.
However, it turns out that this set is slightly incomplete.  We use
these initial seven diagrams to help define the integral family, with
9 of the $\binom{5}{2} + 5=15$ momentum invariants accounted for by
the explicit propagators in
\cref{fig:tttt-1,fig:tttt-2,fig:tttt-3,fig:tttt-4,fig:tttt-5,fig:tttt-6,fig:tttt-7} via
\begin{subequations}
\begin{align}
  F^{(5)}(\text{\cref{fig:tttt-1}})
  &= F^{(5)}(1, 1, 1, 1, 1, 0, 0, 0, 0, 0, 0, 0, 0, 0, 0) \ ,\\
  F^{(5)}(\text{\cref{fig:tttt-2}})
  &=F^{(5)}(1, 1, 0, 1, 1, 0, 1, 1, 0, 0, 0, 0, 0, 0, 0) \ ,\\
  F^{(5)}(\text{\cref{fig:tttt-3}})
  &= F^{(5)}(1, 0, 1, 1, 1, 1, 1, 0, 0, 0, 0, 0, 0, 0, 0) \ ,\\
  F^{(5)}(\text{\cref{fig:tttt-4}})
  &= F^{(5)}(1, 1, 1, 0, 1, 0, 0, 1, 1, 0, 0, 0, 0, 0, 0) \ ,\\
  F^{(5)}(\text{\cref{fig:tttt-5}})
  &= F^{(5)}(1, 1, 0, 0, 1, 0, 1, 1, 1, 0, 0, 0, 0, 0, 0) \ ,\\
  F^{(5)}(\text{\cref{fig:tttt-6}})
  &= F^{(5)}(1, 0, 0, 1, 1, 1, 1, 1, 0, 0, 0, 0, 0, 0, 0) \ ,\\
  F^{(5)}(\text{\cref{fig:tttt-7}})
  &= F^{(5)}(1, 0, 0, 0, 1, 1, 1, 1, 1, 0, 0, 0, 0, 0, 0) \ ,
\end{align}
\end{subequations}
and the remaining 6 from the internal ``planar'' potential-mode
propagators of $\mathcal{M}_6$\footnote{Since they do not appear as
  part of the integral basis, the choice of these propagators is for
  convenience when handling the intermediate steps during reduction.}.
Note that again we are including contributions that can be identified
with each other,
\begin{equation}
  F^{(5)}(\text{\cref{fig:tttt-2}}) = F^{(5)}(\text{\cref{fig:tttt-3}})=F^{(5)}(\text{\cref{fig:tttt-4}}) \ ,
  \qquad
  F^{(5)}(\text{\cref{fig:tttt-5}}) = F^{(5)}(\text{\cref{fig:tttt-6}}) \,,
\end{equation}
so the reflection symmetry considerations mentioned in \cref{sec:ttt}
continue to apply.  Checking the basis integrals of this family, we
find an eighth relevant basis member, shown in \cref{fig:tttt-8}
\begin{equation}
  F^{(5)}(\text{\cref{fig:tttt-8}})
  = F^{(5)}(1, 0, 1, 0, 1, 1, 1, 1, 1, 0, 0, 0, 0, 0, 0) \,.
\end{equation}
This new basis element spoils the one-to-one correspondence between
basis integrals and cuts as its propagator structure is a superset of
the propagators of \cref{fig:tttt-7}.  However, it does so in a very
mild way: as a cut, \cref{fig:tttt-8} is completely contained within
$\cut_{\text{\cref{fig:tttt-7}}}$ due to factorization.  Only the
underlying scalar integrals are unrelated.  Thus calculating and
matching the cut for \cref{fig:tttt-7} determines \emph{both} basis
coefficients (after taking into account relative symmetry factors).

The explicit rational functions of propagators for each of the cuts is
tens of thousands of terms long.  The cuts for
\cref{fig:tttt-1,fig:tttt-2,fig:tttt-3,fig:tttt-4,fig:tttt-5,fig:tttt-6,fig:tttt-8}
were all just barely computable using a personal computer.
\Cref{fig:tttt-7} required the use of the Quest computing cluster at
Northwestern University.  For all of the cuts except for
\cref{fig:tttt-7}, we can observe patterns occurring in the
reduced cuts and basis coefficients.  Starting with the
``all-radiation'' cuts,
\cref{fig:rr-u,fig:tail-cut,fig:tt-hills,fig:ttt-1}, we see that they
follow a simple iteration which suggests that
\begin{equation}
  \rcut{\text{\cref{fig:tttt-1}}}
  = (2 \pi G_N) \left(\frac{(d+1)(d-2)}{(d+2)(d-1)} \omega^4 \right)
  \left( \frac{(-8 \pi G_N E) \mathcal{P}_4}{(d-3)(d-1)d(d+1)}\right)^4 \prod_{i=1}^5 \delta(Q_i)\,.
\end{equation}
We have verified this prediction through direct calculation of the
cut.  Similarly, comparing \cref{eq:ttt-a3,eq:ttt-a4} to
\cref{eq:tt-bulk-rcut}, we are led to
\begin{align}
  \rcut{\text{\cref{fig:tttt-2}}}
  &= \rcut{\text{\cref{fig:tttt-3}}}
    = \rcut{\text{\cref{fig:tttt-4}}} \notag \\
  &= \frac{\ttcoup \omega^6(2d-3) \mathcal{P}_8 \left( \frac{(-8 \pi G_N E) \mathcal{P}_4}{(d-3)(d-1)d(d+1)}\right)^2 }{3(d-3)(d-1)^3d(d+1)(d+2)(3d-4)(3d-2)}\,,
\end{align}
where from now on the $\delta(Q_i)$ are implicit. These equalities are
borne out through explicit computation.  Moving on to
\cref{fig:tttt-5,fig:tttt-6}, we find that their reduced cuts are
\begin{align}
  \rcut{\text{\cref{fig:tttt-5}}}
  &= \rcut{\text{\cref{fig:tttt-6}}} = \frac{-\tttcoup \omega^6 (d-2)(3d-5) \mathcal{P}_{11} \left( \frac{(-8 \pi G_N E) \mathcal{P}_4}{2(d-3)(d-1)d(d+1)}\right)}{(d-3)^2 (d-1)^4d (d+1)^2 (d+2)(2d-3)(3d-4)(3d-2)}\,,
\end{align}
yet again in line with an iteration, this time between the tail-type
factor and a factor from $\mathcal{M}_5$.  \Cref{fig:tttt-8} is the
last of the easier cuts.  While we mentioned that kinematically it
is contained within \cref{fig:tttt-7}, we explicitly include it with
the other iterated cuts because it also turns out to be iterative.
Specifically, we find that the direct calculation of the cut yields
\begin{align}
  \rcut{\text{\cref{fig:tttt-8}}}
  &= \frac{\ttttcoup (2d-3)^2 \omega^8 \mathcal{P}_8^2}{
    9(d-3)^2 (d-2)(d-1)^5 d^2 (d+1)^3(d+2)(3d-4)^2(3d-2)^2}  \\
  &= (2 \pi G_N) \left(\frac{(d+1)(d-2)}{(d+2)(d-1)} \omega^4 \right)
    \left( \frac{(256 G_N^2 E^2 \pi^2) (2d-3) \omega^2 \mathcal{P}_8}{
    3(d{-}3)(d{-}2)(d{-}1)^2 d (d{+}1)^2 (3d{-}4)(3d{-}2)} \right)^2\,, \notag
\end{align}
which is the square of \cref{eq:tt-bulk-rcut} with an additional
prefactor matching the radiation-reaction term \cref{eq:rr-cut}.

Last, we turn our attention to \cref{fig:tttt-7}.  Since we have
already determined \cref{fig:tttt-8}, we report only the part of
$\rcut{\text{\cref{fig:tttt-7}}}$ that dresses
$F^{(5)}(\text{\cref{fig:tttt-7}})$
\begin{align}
  \rcut{\text{\cref{fig:tttt-7}}}&\big|_{F^{(5)}(\text{\cref{fig:tttt-7}})}
 = \frac{\ttttcoup}{30} \omega^2 \mathcal{P}_{22}
 \left(\frac{(d+1)(d-2)}{(d+2)(d-1)}\omega^4\right) \\
     & \times \frac{1}{(d-3)^3 (d-2)^2 (d-1)^4 d^2 (d+1)^4 (2 d-3) (3 d-4)^2 (3 d-2)^2 (5 d-8) (5 d-6)}  \ , \notag
\end{align}
with
\begin{align}
  \mathcal{P}_{22} &= 291600 d^{22}-5364630 d^{21}+43610967 d^{20}-198558153 d^{19}+478961469 d^{18} \notag \\
  &\quad +14340403 d^{17}-4908635433 d^{16}+21042703665 d^{15}-53485147433
    d^{14} \notag \\
                   &\quad +101573518519 d^{13}-174143104426 d^{12}+309327324244 d^{11}-555238832200 d^{10} \notag \\
  &\quad +944837229872 d^9-1503589473248 d^8+2203112130496 d^7-2911608239232
    d^6 \notag \\
                   &\quad +3432661203456 d^5-3454174264320 d^4+2688616654848 d^3-1430856658944 d^2 \notag \\
  &\quad +447389982720 d-60914073600 \,.
\end{align}

The final piece remaining to assemble the effective action is to
enumerate the symmetry factors.  Since we are using the over-complete
integral basis, all diagrams come with the reflection symmetry factor
in addition to the permutation factor for the $E$ coupling
\begin{align}
  |\graph{\cref{fig:tttt-1}}| &= 2 \ , \qquad
  |\graph{\cref{fig:tttt-2}}|
  = |\graph{\cref{fig:tttt-3}}|
  = |\graph{\cref{fig:tttt-4}}| = 2 \times 2! \ , \notag\\
  |\graph{\cref{fig:tttt-5}}|
  &= |\graph{\cref{fig:tttt-6}}| = 2 \times 3! \ ,
 \qquad  |\graph{\cref{fig:tttt-7}}| = 2 \times 4! \ , \notag\\
  |\graph{\cref{fig:tttt-8}}| &= 2 \times (2! \times 2!) \,.
\end{align}
The integrals for
\cref{fig:tttt-1,fig:tttt-2,fig:tttt-3,fig:tttt-4,fig:tttt-5,%
  fig:tttt-6,fig:tttt-7} are all straightforward to evaluate using the
same techniques as the previous calculations.  Unfortunately, for
\cref{fig:tttt-8} we must resort to numerical evaluation and
reconstruction, discussed in \cref{sec:int-tttt-8}.  Combining all of
the above data, expanding in $d=3+\dimreg$ and evaluating the CTP sums
results in the effective action contribution
\begin{align}
  \label{eq:tttt-raw}
  S_{\text{TTTT}} = - \frac{5}{2}\frac{214^2}{525^2} G_N^5 E^4
  &\intw \omega^9 \kappa_{-+}(\omega) \Bigg\{
    \frac{i}{\dimreg^2}
    +\frac{1}{\dimreg}\Bigg[\frac{5}{2} i \log\left(\frac{\omega^2 e^{\gamma_E}}{\mu^2 \pi}\right)
    +\frac{5 \pi  \sgn(\omega)}{2}  \\
    & \quad -i\left(\frac{840}{107} \zeta_2 +  \frac{5249287}{343470}\right) \Bigg] \notag \\
  &+\Bigg[
    \frac{5}{2}i\log\left(\frac{\omega^2 e^{\gamma_E}}{\mu^2 \pi}\right)
    \left( \frac{5}{4} \log\left(\frac{\omega^2 e^{\gamma_E}}{\mu^2 \pi}\right)
    - \frac{840}{107} \zeta_2
    - \frac{5249287}{343470} \right) \notag \\
  & \qquad +\frac{5}{2} \pi \sgn(\omega) \left(\frac{5}{2}  \log\left(\frac{\omega^2 e^{\gamma_E}}{\mu^2 \pi}\right) 
    -\frac{840}{107}  \zeta_{2}
    -\frac{5249287 \pi  \sgn(\omega )}{343470}\right) \notag \\
  & \qquad  +\frac{5541865 i \zeta_{2}}{91592}
    -\frac{2940}{107} i \zeta_{3}
    -\frac{176400 i \zeta_{4}}{11449}
    +\frac{90973869743 i}{673201200} \Bigg] \notag \\
  & \qquad + \odimreg{1} \Bigg\}\,. \notag
\end{align}
This is the first time that the T$^4$ contribution
has been derived in any approach whatsoever in terms of generic quadrupoles.



\section{Analysis of Dissipative Sector}
\label{sec:renorm}
The effective actions derived in \cref{sec:all-tails,sec:higher-tails}
contain divergences in the dimensional regularization parameter
$\dimreg$.
The divergences in the dissipative sector, namely of the part of the
action that is odd-in-$\omega$, can be handled completely within the tower
of tails. This dissipative analysis takes place purely within the
``binary-as-composite'' EFT at the radiation scale.  As mentioned in
\cref{sec:eft}, the UV (small-scale) theory that defines the
quadrupoles, namely the description of the binary dynamics at the
orbital scale, allows one to directly compute at sufficiently high PN order 
the renormalization of the multipole couplings in terms of the IR
divergences of the small-scale theory.  To highlight that we are
focusing on the purely dissipative contributions, we will work at the
level of the dissipated energy and energy spectra, $\Delta E$ and
$\frac{\diff E}{\diff \omega}$, respectively.  
We will then proceed to demonstrate the renormalization of these 
dissipative contributions up to new subleading orders.  
We take inspiration from Ref.~\cite{Goldberger:2009qd}, which
is essentially textbook QFT renormalization.

\subsection{Energy Loss in the CTP Formalism}
\label{sec:ctp-energy}

The CTP formalism provides a generic extension of Noether's theorem
which accounts for the effects of the non-conservative dynamics on the
total energy of the accessible degrees of freedom.  Specifically, we
have \cite{Galley:2014wla}:
\begin{equation}
	\frac{\diff E^{\text{CTP}}}{\diff t} = - \frac{\partial L}{\partial t} 
	+ \dot{q}^I \left[ \frac{\partial K}{\partial q_-^I(t)} \right]_{\text{PL}} 
	+ \ddot{q}^I \left[ \frac{\partial K}{\partial \dot{q}_-^I(t)} \right]_{\text{PL}} 
	+ \frac{\partial}{\partial \substack{\text{higher}\\\text{time}\\\text{derivs}}}
	\label{eq:ctp-gen-flux}
\end{equation}
for $E^{\text{CTP}}$ the energy of the accessible degrees of freedom
$q$.  To apply this formula in the case of gravitational tails, we
need to analyze the splitting into the conservative and
non-conservative parts of the effective action.  Then we must
transform the time-domain expression of \cref{eq:ctp-gen-flux} into a
frequency-domain.

Splitting the full CTP Lagrangian into the conservative and
non-conservative parts only requires knowledge about the functional
structure of the generalized coordinates.  In the current case of
tails, the quadrupoles themselves, $I_{\pm}(\omega)$ (suppressing
$SO(3)$ tensor indices in the following discussion), are the relevant
generalized coordinates.  We have seen in the computations in
\cref{sec:all-tails,sec:higher-tails} that the CTP effective actions
for the tails take the form:
\begin{equation}
	S_{\text{CTP}}= \int_{-\infty}^{\infty}\diff \omega  f(\omega) \kappa_{-+}(\omega),
\end{equation}
in which $\omega$ is purely real, but the function $f(\omega)$ may be
complex.  The conservative part of the CTP effective Lagrangian is
defined as the piece expressible as the difference between two
distinct histories, which in this case means the part symmetric in the
$(-,+)$ variables:
\begin{equation}
\kappa_{\text{C}}(\omega) \equiv \kappa_{11}(\omega) -
\kappa_{22}(\omega) \sim \kappa_{-+}(\omega) + \kappa_{+-}(\omega).
\end{equation}
We then adopt an orthogonal change of basis to define:
\begin{equation}
\kappa_{\text{NC}}(\omega) \equiv \kappa_{-+}(\omega) -
\kappa_{+-}(\omega),
\end{equation}
so that the CTP effective action becomes:
\begin{equation}
	S_{\text{CTP}}= \int_{-\infty}^{\infty}\diff \omega  \, \big[\frac{1}{2}
	f(\omega)( \kappa_{\text{C}} + \kappa_{\text{NC}})\big] \,.
\end{equation}
Finally, we can use the parity properties of the integral,
and $\kappa_{-+}(-\omega) \to \kappa_{+-}(\omega) $ under
$\omega \leftrightarrow - \omega$ to write:
\begin{equation}
	S_{\text{CTP}} = \frac{1}{2} \int_{-\infty}^{\infty} \diff \omega 
	\Big[f_{\text{even}}(\omega)\kappa_{\text{C}}(\omega)
	+ f_{\text{odd}}(\omega)\kappa_{\text{NC}}(\omega) \Big]
	\,.
\end{equation}
We then identify in \cref{eq:ctp-gen-flux}
$L(\omega) = \frac{1}{2}f_{\text{even}}(\omega) \kappa_{\text{C}}(\omega)$, 
and
$K(\omega) = \frac{1}{2} f_{\text{odd}}(\omega)
\kappa_{\text{NC}}(\omega)$.  We assume that the conservative piece
does not have an explicit time dependence, and so it does not
contribute to \cref{eq:ctp-gen-flux}.

In order to apply \cref{eq:ctp-gen-flux} to the Fourier-space
non-conservative potential, we need to switch the from frequency
dependence in the CTP quadrupole, $I_{-}(\omega)$, back to time
dependence via the Fourier transform
\begin{equation}
	I_a(\omega) = \ftnorm \int \diff t e^{-i \omega t} I_a(t),
	\qquad I_a(t) = \iftnorm \int \diff \omega e^{i \omega t} I_a(\omega)  \,.
	\label{eq:ft-defs}
\end{equation}
This allows to define $K(t)$ as:
\begin{align}
	K(t)
	&= \frac{1}{2}\ftnorm \int \diff \omega f_{\text{odd}}(\omega) 
	\big[I_{+}(\omega)e^{i \omega t} - I_{+}(-\omega)e^{-i \omega t}\big]I_{-}(t)\,,
\end{align}
and in turn the needed pieces for \cref{eq:ctp-gen-flux} read:
\begin{align}
	\left[\frac{\partial K}{\partial I_{-}(t)}\right]_{\text{PL}}
	&= \frac{1}{2} \ftnorm \int \diff \omega 
	f_{\text{odd}}(\omega) \left(I(\omega)e^{i \omega t} - I(-\omega)e^{-i \omega t}\right), \\
	\frac{\diff I(t)}{\diff t}
	&= \iftnorm \frac{\diff}{\diff t} \int \diff \omega' e^{i \omega' t} I(\omega') 
	= \iftnorm \int \diff \omega' ( i \omega') e^{ i \omega' t} I(\omega') \,.
\end{align}
Importantly, this step allows us to ignore terms that contain
$I_{+}(-\omega) I_{+}(\omega)$ and $I_{-}(-\omega) I_{-}(\omega)$,
since the first carries no dependence on $I_{-}$, and the second
vanishes in the physical limit $[\dots]_{\text{PL}}$ after taking the
derivative.

We then assemble
\begin{align}
	\Delta E
	&=\int \diff t \frac{\diff E^{\text{CTP}}}{\diff t} \notag \\
	&= \frac{1}{2} \ftnorm \iftnorm \int \diff t \diff \omega \diff \omega'
	( i \omega') e^{i \omega' t} I(\omega') 
	f_{\text{odd}}(\omega)\big[I(\omega)e^{i \omega t} - I(-\omega)e^{-i \omega t}\big] \,.
\end{align}
Resolving the Fourier transforms of the delta functions that arise:
\begin{equation}
	\int \diff t e^{i t (\omega + \omega')} = 2 \pi \delta(\omega + \omega')\, ,
	\quad
	\int \diff t e^{i t (\omega - \omega')} = 2 \pi \delta(\omega - \omega')\, ,
\end{equation}
leads to
\begin{equation}
	\Delta E = \int \diff \omega \Big [(- i \omega)f_{\text{odd}}(\omega) \kappa(\omega) \Big]\,.
	\label{eq:extract-energy}
\end{equation}
We reiterate that this analysis accounts for the energy change of the
binary system, and that the \emph{radiated energy} carried by
the gravitational field must, by conservation of energy, be opposite.

\subsection{Radiated Energy from Tails}

We begin by applying energy loss formula, \cref{eq:extract-energy}, to
the CTP effective actions derived in
\cref{sec:all-tails,sec:higher-tails}.  Note that since we have not
performed the renormalization at the level of the effective action,
these initial energy contributions will still carry dimensional
regularization divergences.  We will perform the renormalization at
the level of the energy loss in the following subsections.

We begin with the radiation reaction term, \cref{eq:rr-raw}.  Applying
\cref{eq:extract-energy}, we can easily extract the energy loss of the
quadrupoles into the gravitational field
\begin{equation}
  \label{eq:rr-e-raw}
  \deltaE{}{RR} = - \frac{G_N}{5 \pi}\int_{-\infty}^{\infty} \diff \omega \kappa(\omega) \omega^6 \left[ 1 - \frac{\dimreg}{2}\left(\frac{9}{10}- \log\left( \frac{\omega^2 e^{\gamma_E}}{\mu^2 \pi} \right) \right) + \odimreg{2}\right] \,.
\end{equation}
By energy balance, this must be opposite to the energy carried away
by the gravitational field in the form of gravitational waves.  The
finite part of this energy loss is the Fourier transform of the
well-known Einstein quadrupole radiation formula.  We present the
$\odimreg{1}$ term for later use in renormalization.
Similarly, from the tail effective action, \cref{eq:t-raw}, we compute
the energy loss contribution
\begin{align}
  \label{eq:t-e-raw}
  \deltaE{}{T} &= - \frac{2}{5} G_N^2 E \inte{7} \left[1+\dimreg \left(\log\left(\frac{\omega^2 e^{\gamma_E}}{\mu^2 \pi}\right)-\frac{41}{30}\right) + \odimreg{2} \right] \,.
\end{align}
This correction is in agreement with previous derivations
\cite{Blanchet:1987wq,Blanchet:1992br,Blanchet:1993ng,%
  Goldberger:2009qd,Foffa:2011np,Galley:2015kus}, again up to sign
conventions.

The energy loss induced from the tail-of-tail comes from
\cref{eq:tt-raw} , giving
\begin{align}
  \label{eq:tt-e-raw}
  \deltaE{}{TT} = \frac{214 G_N^3 E^2  }{525 \pi }
  &\inte{8} \Bigg\{
    \frac{1}{\dimreg}+
    \left[\frac{3}{2} \log\left(\frac{\omega^2 e^{\gamma_E}}{\mu^2 \pi}\right)-\frac{420 \zeta_{2}}{107}-\frac{675359}{89880}\right] \notag \\
  &+\dimreg\Bigg[\log\left(\frac{\omega^2 e^{\gamma_E}}{\mu^2 \pi}\right)
    \left(\frac{9}{8} \log\left(\frac{\omega^2 e^{\gamma_E}}{\mu^2 \pi}\right)-\frac{ (352800\zeta_{2}+675359)}{59920} \right) \notag \\
  &\qquad    +\frac{4569 \zeta_{2}}{856}-\frac{1050 \zeta_{3}}{107}+\frac{1259125247}{37749600}\Bigg]
    +\odimreg{2} \Bigg\}\,.
\end{align}
From T$^3$, \cref{eq:ttt-raw}, we find
\begin{align}
  \label{eq:ttt-e-raw}
  \deltaE{}{TTT} = \frac{428}{525} G_N^4 E^3
  &\inte{9} \Bigg\{
    \frac{1}{\dimreg}
    +\left[2 \log\left(\frac{\omega^2 e^{\gamma_E}}{\mu^2 \pi}\right)-\frac{252583}{29960}\right] \notag \\
  &\quad+ \dimreg \Bigg[\log\left(\frac{\omega^2 e^{\gamma_E}}{\mu^2 \pi}\right)\left(
    2\log\left(\frac{\omega^2 e^{\gamma_E}}{\mu^2 \pi}\right)-\frac{252583 }{14980}\right) \notag \\
  & \qquad   -\frac{13 }{2}\zeta_{2}
    -\frac{840 }{107}\zeta_{3}+\frac{1583459537}{37749600}
    \Bigg] + \odimreg{2}
    \Bigg\} \,.
\end{align}
Finally, the T$^4$ contribution to the energy
loss is computed from \cref{eq:tttt-raw} as
\begin{align}
  \label{eq:tttt-e-raw}
  \deltaE{}{TTTT} = -& \frac{5}{2}\frac{214^2}{525^2 \pi} G_N^5 E^4
  \inte{10} \Bigg\{
    \frac{1}{\dimreg^2}
    +\frac{1}{\dimreg}\Bigg[\frac{5}{2} \log\left(\frac{\omega^2 e^{\gamma_E}}{\mu^2 \pi}\right)
    -\frac{840\zeta_{2}}{107}
    -\frac{5249287}{343470}\Bigg] \notag \\
  & +\Bigg[
    \frac{5}{2} \log\left(\frac{\omega^2 e^{\gamma_E}}{\mu^2 \pi}\right) \left(
    \frac{5}{4} \log\left(\frac{\omega^2 e^{\gamma_E}}{\mu^2 \pi}\right)
    -\frac{840}{107} \zeta_2  - \frac{5249287}{343470} \right) \notag \\
   &\quad  +\frac{5541865 \zeta_{2}}{91592}
    -\frac{2940 \zeta_{3}}{107}
    -\frac{176400 \zeta_{4}}{11449}+\frac{90973869743}{673201200}\Bigg] + \odimreg{1} \Bigg\}\,.
\end{align}
With the emitted energy contributions computed, we can begin
renormalization analysis.

\subsection{Going to Subleading RG Flow}
\label{sec:tosub-rg}
 
The first appearance of a $\dimreg$ divergence in the dissipated energy occurs in
the tail-of-tail as a simple pole
\begin{equation}
  \deltaE{}{TT}\Big|_{\dimreg^{-1}} = \frac{2}{3 \pi \dimreg} \times \frac{107}{175} G_N^3 E^2 
  \int_{-\infty}^{\infty} \diff\omega \, \omega^8 \kappa(\omega) \,.
\end{equation}
Since no pole appears in the dissipative sector at the tail level, any counterterms and the
renormalization must carry a factor of $(G_NE)^2$ to skip tail orders.
With this in mind, we introduce a renormalized coupling
to the quadrupoles via:
\begin{equation}
  \kappa(\omega) \to \kappa'(\omega) \equiv \kappa(\omega,\mu)
  \left( 1 + \frac{(G_N E)^2 X(\omega)}{\epsilon} + \dots \right) \,,
  \label{eq:kappa-ren}
\end{equation}
where $X$ is an unknown, independent of $(G_NE)^2$ and $\dimreg$, $\mu$ is the
renormalization scale of the logs, and the ellipsis 
indicate higher-order terms in $G_NE$.  
To find $X$, we substitute
\cref{eq:kappa-ren} into \cref{eq:rr-e-raw}, and demand that the
\emph{total} energy dissipation:
\begin{equation}
  \delEInc{TT} \equiv \big[\deltaE{}{RR} + \deltaE{}{T} + \deltaE{}{TT} \big] \Big|_{\kappa \to \kappa'}
\end{equation}
is free from $\dimreg$ poles up to the TT at order $G_N^3$.
Notably, since the pole in $\kappa'$ carries a factor of $G_N^2$, the
$\dimreg^{-1}$ part will only contribute to $\delEInc{TT}$ at the
appropriate order in $G_N$ via $\deltaE{}{RR}$, as
$G_N^2 (\deltaE{}{T} + \deltaE{}{TT})$ is beyond $\mathcal{O}(G_N^3)$.
This pole cancellation requires:
\begin{equation}
  X(\omega) = \frac{214}{105} \omega^2 \Rightarrow
  \kappa'(\omega) \equiv \kappa(\omega,\mu)\left( 1 + \frac{214\omega^2(G_N E)^2}{105\epsilon} 
  + \dots \right)\,.
  \label{eq:ren-e1}
\end{equation}

Moving on to the tail-of-tail-of-tail (T$^3$) we might suspect a new term in $\kappa'$
carrying $G_N^3$.  However, performing the explicit calculation using only \cref{eq:ren-e1}, 
we find:
\begin{align}
  \label{eq:del-e-ttt-inc}
  \delEInc{TTT}
  &=\int^{\infty}_{-\infty} \diff \omega \kappa(\omega,\mu) \Bigg[-\frac{\omega^6 G_N }{5 \pi }
    -\frac{2}{5} \omega ^7 G_N^2E \notag \\
  & + \frac{1}{\pi} G_N^3E^2 \omega^8 \left(
    \frac{214 }{525 } \log\left(\frac{\omega^2 e^{\gamma_E}}{\mu^2 \pi}\right)
    -\frac{634913}{220500 }
    -\frac{8  \zeta_{2}}{5 }\right) \notag \\
  & + G_N^4 E^3\omega^9  \left(
    \frac{428}{525}  \log\left(\frac{\omega^2 e^{\gamma_E}}{\mu^2 \pi}\right)
    -\frac{634913}{110250}\right) + \mathcal{O}(G_N^5) \Bigg]\,,
\end{align}
which is also completely free of $\dimreg$ poles.  This means there is
no term in the renormalized coupling, \cref{eq:kappa-ren}, of
$\mathcal{O}(G_N^3)$.  Further terms in \cref{eq:kappa-ren} must only
contribute then at $G_N^4$, and thus enter via T$^4$.

With a renormalized coupling comes a renormalization-group (RG) flow.
Using counterterm analysis to determine RG flow equations in gravity
is ambiguous due to the presence of topological operators like the
Gauss-Bonnet term \cite{Bern:2017puu}.  Instead, we will study the
renormalization-scale dependence of the observable $\delEInc{TTT}$
directly.  All of the logarithms in $\delEInc{TTT}$ carry a
renormalization scale $\mu$, that comes from compensating for the
misalignment between the mass dimension of $G_N$, and the required
mass dimension of the coupling constant in a dimensionally-regulated
action.  We also introduced a scale dependence in $\kappa$ via the
renormalization of the source coupling for similar reasons.  The RG
flow then follows from demanding that $\delEInc{TTT}$, a perturbative
observable, must be invariant under shifts of the scale:
\begin{equation}
  \frac{\diff}{\diff \mu} \delEInc{TTT} = 0+\mathcal{O}(G_N^5) \,,
\end{equation}
which leads to the RG equation:
\begin{align}
  \frac{\diff}{\diff \log \mu} \kappa(\omega,\mu) = - \frac{428}{105}(G_N E \omega )^2 \kappa(\omega,\mu) 
  + \mathcal{O}(G_N^4)\,,
  \label{eq:rg-flow-ttt}\\
  \Rightarrow \kappa(\omega,\mu) = \left( \frac{\mu}{\mu_0}\right)^{- \frac{428}{105}(G_N E \omega)^2} 
  \kappa(\omega,\mu_0) \,,
  \label{eq:rg-kappa-ttt}
\end{align}
where $\mu_0$ is an arbitrary but fixed reference scale at which
$\kappa$ is measured (or otherwise known, \eg through a matching
calculation with the small-scale theory).  This is in exact
agreement with the RG flow originally found by Goldberger and Ross
\cite{Goldberger:2009qd}, which can be seen by substituting in
$\kappa(\omega,\mu) = I_{ij}(-\omega,\mu)I^{ij}(\omega,\mu)$, that
introduces a factor of $2$ on the LHS of \cref{eq:rg-flow-ttt} but not
on the RHS.

With the leading dissipative renormalization dealt with, we now turn
to the subleading corrections induced by the TTTT terms.  As we saw in
\cref{eq:tttt-e-raw}, the unrenormalized $\deltaE{}{TTTT}$ has both a
double pole, $\dimreg^{-2}$, as well as a single pole.  Since we
successfully removed the divergence in the construction of
$\delEInc{TTT}$ with only a $G_N^2$ counterterm, we know that one of
the new terms in $\kappa'$ must be of the form
$Y (G_N E \omega)^4 \dimreg^{-2}$.  This term will bring the
$\mathcal{O}(G_N)$ term from \cref{eq:rr-e-raw,eq:del-e-ttt-inc} up to
$G_N^5$ while shifting the $\dimreg^0$ term into a double pole.  We
find that after adding the new term to $\kappa'$, the coefficient of
the double pole of $\delEInc{TTTT}$ is given by:
\begin{align}
  \delEInc{TTTT} \Big|_{\dimreg^{-2}} &= \frac{G_N^5 E^4}{5\pi} \intem{10} 
  \left[  \underbrace{-Y}_{\text{RR}} + \underbrace{\frac{45796}{11025}}_{\text{TT}} 
  -\underbrace{ \frac{22898}{11025}}_{\text{TTTT}} \right] \,.
\end{align}
Cancellation of this pole then requires: 
\begin{equation}
  Y = \frac{22898}{11025} = 2\frac{107^2}{105^2}\,,
\end{equation}
in accordance with the expected iteration of the previous counterterm.

However, the iterated counterterm is not the only required correction
to $\kappa'$.  The single pole has been altered, but not
completely removed:
\begin{equation}
  \delEInc{TTTT} \Big|_{\dimreg^{-1}} = 
  \frac{G_N^5 E^4}{5 \pi} \times \frac{1695233}{105^3} \intem{10} \,.
\end{equation}
Removing this pole necessitates a second new term in $\kappa'$ of the
form $Z (G_N E \omega)^4 \dimreg^{-1}$.  Incorporating this new
correction to $\delEInc{TTTT}$ will allow the finite piece of RR,
\cref{eq:rr-e-raw}, to also contribute a $\dimreg^{-1}$
pole at $G_N^5$.  We then find that $Z = \frac{1695233}{105^3}$.
Thus, with two new terms, $\kappa'$ becomes:
\begin{equation}
  \kappa'(\omega) \equiv \kappa(\omega,\mu)\left( 1 + 2 \left( \frac{107}{105} 
  \frac{(G_N E \omega)^2}{\dimreg} + \frac{107^2}{105^2} \frac{(G_N E \omega)^4}{\dimreg^2} \right) 
  +  \frac{1695233}{105^3} \frac{(G_N E \omega)^4}{\dimreg} + \dots \right) \,,
  \label{eq:kappa-ren-2}
\end{equation}
and the inclusive energy loss is:
\begin{align}
  \delEInc{TTTT} &= \delEInc{TTT} + \frac{G_N^5 E^4}{5 \pi} \intem{10} \Bigg\{
  32 \zeta_4 + 2^4 \frac{107}{105} \zeta_3 - \frac{1132438}{105^2} \zeta_2  \notag\\
  & \quad - \frac{275977249}{1944810}
    - 2\frac{107^2}{105^2} \log^2\left(\frac{\omega^2 e^{\gamma_E}}{\mu^2 \pi}\right)
    \notag \\
  &\quad    
+ \log\left(\frac{\omega^2 e^{\gamma_E}}{\mu^2 \pi}\right)\left[ \frac{8301847}{257250} 
+ 2^4 \frac{107}{105} \zeta_2 \right]
    \Bigg\} \,.
    \label{eq:del-e-t4-inc}
\end{align}

Since the TTTT energy loss required a new counterterm (and has a
subleading log), there will be a new term in the RG flow associated to
it.  We again simply demand that:
\begin{equation}
  \frac{\diff}{\diff \mu} \delEInc{TTTT} = 0+\mathcal{O}(G_N^6)\,,
\end{equation}
which leads to the RG equation:
\begin{equation}
  \frac{\diff}{\diff \log \mu} \kappa(\omega,\mu) = -(2G_N E \omega)^2  \kappa(\omega,\mu) 
  \left( \frac{107}{105} + \frac{1695233}{105^3} (G_N E \omega)^2 \right) + \mathcal{O}(G_N^5) \,.
  \label{eq:new-rg}
\end{equation}
This new RG equation necessarily includes the leading RG flow, and
now allows for prediction of subleading logs at all higher-order tails.


\section{Post-Newtonian and Self-Force Results}
\label{sec:compare}
In this section, we present comparisons of our energy loss with those
derived via traditional GR results in PN and self-force theory for 
further checks that go beyond the tail.  Observable results from these
GR approaches are presented in a PN expansion, eventually
specified to a quasi-circular orbit.  Our results are also in the PN
regime due to the multipole expansion of the inspiral.  Thus, matching
against the known PN results primarily entails inserting a PN-expanded
quadrupole expression, and then aligning related scheme choices.

\subsection{PN Mapping to the Binary Inspiral}

We will focus on the leading PN expansion, which for a binary system
is simply a circular orbit.  WLOG, we take the circular orbit to be in
the $x$-$y$ plane, in which the frequency-space quadrupole components
can be chosen as
\begin{align}
  I_{xx}(\omega)&= I_{yy}(\omega) 
  = \nu E r^2 \pi/2 ( \delta(\omega - 2 \Omega) +\delta(\omega + 2 \Omega))\,,\notag\\
  I_{xy}(\omega) &= I_{yx}(\omega) 
  = -i \nu E r^2 \frac{\pi}{2}( \delta(\omega - 2 \Omega) -\delta(\omega + 2 \Omega))  \,,\\
  I_{zj}(\omega) &= 0 \notag\,,
\end{align}
with $\omega$ the radiation frequency, $\Omega>0$ the orbital
frequency, $r$ the radius of the circular orbit, $\nu$ the symmetric
mass ratio of the binary, and $E$ the energy of the binary.  From
these expressions, we easily arrive at the quadrupole-quadrupole ``two-point'' 
contraction:
\begin{equation}
  \kappa(\omega) = E^2 \pi^2 r^4 \nu^2 
  \left( \delta(\omega - 2 \Omega)^2 + \delta(\omega + 2 \Omega)^2 \right)\,.
  \label{eq:pn-kappa-start}
\end{equation}
Upon integration in $\omega$, the $\delta(\omega \pm 2 \Omega)^2$ will
leave behind a $\delta(0)$.  These are resolved by noting that the PN
energy loss for the binary is actually computed as the time-averaged
energy loss of the system over a sufficiently long period of time
\cite{Blanchet:2013haa}.  Invoking one of the definitions of
the frequency space $\delta(0)$,
\begin{equation}
  \delta(0) \equiv \lim_{T \to \infty}\frac{1}{2 \pi} \int_{-T/2}^{T/2} e^{- i t 0} \diff t 
  = \lim_{T \to \infty} \frac{T}{2 \pi} \,,
  \label{eq:delta-zero}
\end{equation}
we can formally align the time-averaging interval with the $T$ in
\cref{eq:delta-zero}, canceling said $T$ dependence from the final
result.  The net result is that we can effectively use
\begin{equation}
 \kappa(\omega) =\frac{E^2 \pi r^4 \nu^2}{2} \left( \delta(\omega - 2 \Omega)
    + \delta(\omega + 2 \Omega)\right)  
  \label{eq:pn-kappa}
\end{equation}
as the quadrupole contraction in order to match against the PN
results.

We will also eventually need Kepler's law, $GE/r = (r \Omega)^2$, to
rewrite expressions in terms of the PN parameter
$x \equiv (G E \Omega)^{2/3}$. All of the above are leading-order
expressions in the PN expansion, which have subleading corrections
that we ignore here. These leading expressions are sufficient for us
to verify critical features of our results: the leading logs and
leading transcendental numbers.

\subsection{Direct Comparisons}
We begin by looking at the radiation-reaction and the tail.  
Since these actions contain no divergences or logarithms in their dissipative part,
there is no subtlety about aligning choices of logarithm scales.  Thus, we simply insert 
\cref{eq:pn-kappa} into
\cref{eq:rr-e-raw,eq:t-e-raw} which gives:
\begin{align}
  \delEInc{$G_N,G_N^2$}^{\text{LO PN}} &= 
  - \frac{32 x^5 \nu^2}{5 G_N}  - \frac{256 \pi x^{13/2} \nu^2}{5 G_N} + \dots
\end{align}
in agreement (up to the energy balance sign) with the long-known
results of Blanchet and Damour
\cite{Blanchet:1987wq,Blanchet:1993ng,Blanchet:2013haa}.

For higher-order tails, we need to carefully process the
total PN energies, including the renormalization of the
quadrupoles, and align subtraction schemes. 
The RG flows from \cref{sec:tosub-rg} trade out dependence on
$\log(\mu)$, the renormalization flow parameter, for $\log(\mu_0)$, the
scale at which we perform EFT matching to the short-scale theory, 
the orbital separation $r$ in our case. Thus we take $\mu_0^{-1} \to r$.  
A proportionality constant is left undetermined, and amounts to different 
choices of subtraction scheme, which we consider next.

We must align subtraction schemes between our results and traditional 
GR literature.  In \cref{sec:renorm}, the counterterms we introduced
only absorbed the divergences, and not any additional constants.  Thus
we are technically working in a pure minimal subtraction scheme.
However, we have explicitly packaged the logarithms into
$\log \frac{\omega^2 \exp(\gamma_E)}{\mu^2 \pi}$, which makes it easy
to switch to $\overline{\text{MS}}$ subtraction by sending
$\mu^2 \to \frac{\exp(\gamma_E)}{4 \pi} \mu^2$, or to other
nonstandard schemes via similar replacements.  For the TT and TTT, 
we compare against the work of Blanchet et al
\cite{Blanchet:1997ji,Blanchet:1997jj,Blanchet:2001aw,Blanchet:2013haa,Marchand:2016vox},
whose results include $\gamma_E$ but not $\log \pi$, suggesting that
their implicit renormalization scheme is not equivalent to either MS
or $\overline{\text{MS}}$.  We will thus adopt a generic subtraction
via $\mu^2 \to A \pi^{-1} \mu^2$, and determine the proper choice of 
$A$ to align schemes.

Pushing our $\mathcal{O}(G_N^3)$ term from \cref{eq:del-e-ttt-inc}
through the transformation to PN variables, including the generic
subtraction and renormalization considerations, we arrive at:
\begin{equation}
  \delEInc{$G_N^3$}^{\text{LO PN}} \to - \frac{32}{5 G_N} \nu^2 x^8 \left[ \frac{634913}{11025} 
  +32 \zeta_2 -  \frac{856}{105}\left( \log x + 2 \log 2 + \gamma_E - \log A \right) \right] \,.
\end{equation}
Comparing with known PN results
\cite{Blanchet:1997ji,Blanchet:1997jj,Blanchet:2001aw,Blanchet:2013haa,Marchand:2016vox}
\begin{equation}
  \mathcal{F}_{\nu^2 x^8} = \frac{32}{5 G_N} \nu^2 x^8 \left[\frac{6643739519}{69854400} + 32 \zeta_2 - 
  \frac{856}{105} \left(\log x + 4 \log 2 + 2 \gamma_E\right) \right] \,.
\end{equation}
The $\log x$ terms match exactly, up to the energy balance sign.
Since all of the transcendental numbers are newly-appearing at this
order, we expect to also match them exactly up to choice of
subtraction scheme and the energy balance sign.  We see that choosing a
subtraction scheme with $A = (4 \exp(\gamma_E))^{-1}$ leads to the
desired matching.  Note that this subtraction scheme is equivalent to
using a dimensionally-regulated gravitational constant:
\begin{equation}
	\label{eq:msbar}
  \mu^2 \to \frac{\mu^2}{4 \pi e^{\gamma_E}} 
  \Rightarrow G_N \to G_d \equiv G_N \left( \frac{\sqrt{4 \pi e^{\gamma_E}}}{\mu}\right)^{d-3} \,,
\end{equation}
which is also motivated by PN calculations in the conservative sector,
see for instance Refs.~\cite{Levi:2020kvb,Henry:2023sdy}.  Matching
the rational number would require inserting the higher-order PN terms
of $\kappa(\omega)$ (as well as $E$) into the RR contribution, which
would allow the rational terms at $\mathcal{O}(G_N)$ to contribute at
$\mathcal{O}(x^8)$.

Similarly, we extract the
PN terms coming from our $\mathcal{O}(G_N^4)$ contribution to
\cref{eq:del-e-ttt-inc} using the above fixed subtraction scheme:
\begin{equation}
  \delEInc{$G_N^4$}^{\text{LO PN}} \to  - \frac{32}{5 G_N} \nu^2 x^{19/2} 4 \pi 
  \left[ \frac{634913}{11025} - \frac{856}{105} \left( \log x + 4 \log 2 + 2\gamma_E \right) \right] \,,
\end{equation}
and compare against the known PN result \cite{Marchand:2016vox}:
\begin{equation}
  \mathcal{F}_{\nu^2 x^{19/2}} = \frac{32}{5 G_N} \nu^2 x^{19/2} 4 \pi 
  \left[ \frac{265978667519}{2980454400} - \frac{856}{105} 
  \left( \log x + 4 \log 2 + 2 \gamma_E \right) \right] \,,
\end{equation}
with which we again find agreement between all terms, except for the 
rational contribution, where the piece from TTT is partial to the full PN correction.

For T$^4$, the only available results are from self-force theory.
Results at the appropriate order we need here are available in
Refs.~\cite{Fujita:2011zk,Fujita:2012cm}, written in terms of the
orbit velocity $ v \sim x^{1/2}$.  With our leading quadrupole in the
PN expansion we should exactly match terms like $\zeta_4$ and
$\log^2 x$.  The $\log 3$ and $\log 5$ come from higher-order
multipoles, while the other terms receive contributions from
higher-order PN terms in the quadrupole bringing terms from RR and TT
up to $x^{11} \sim v^{22}$.  We organize the results so that the terms
we should match appear first, the ones that we cannot appear later,
and we ignore completely the $\log 3$ and $\log 5$ terms from
Refs.~\cite{Fujita:2011zk,Fujita:2012cm}.  From our results in
\cref{eq:del-e-t4-inc}, we obtain:
\begin{align}
  \delEInc{$G_N^5$}^{\text{LO PN}}&= -\frac{32}{5 G_N} \nu^2 v^{22} \bigg[- 512 \zeta_4 - 
  \frac{27392}{105} \zeta_3  + \frac{1465472}{11025}(\log(v) + \gamma_E + 2 \log 2)^2 \notag \\
  & \quad -\frac{54784}{105} \zeta_2 \left(\log v + \gamma_E + 2 \log 2\right) \notag \\
  & \quad +\frac{2207817992}{972405} - \frac{132829552}{128625} \gamma_E - \frac{265659104}{128625} 
  \log 2 - \frac{132829552}{128625} \log v \notag \\
  & \quad+ \frac{18119008}{11025} \zeta_2 \bigg] \,,
\end{align}
which we compare against the expressions from Refs.~\cite{Fujita:2011zk,Fujita:2012cm}:
\begin{align}
  \frac{\diff E^{(12)}}{\diff t}
  &= \left( \frac{\diff E}{\diff t}\right)_{\text{N}}\bigg[
    - 512 \zeta_4 - \frac{27392}{105} \zeta_3 + \frac{1465472}{11025}(\log(v) + \gamma_E + 2 \log 2)^2 
    \notag \\
  &\qquad  - \frac{54784}{105} \zeta_2 (\log v  + \gamma_E + 2 \log 2) \notag \\
  & \qquad + \frac{2067586193789233570693}{602387400044430000} - \frac{246137536815857}{157329572400} 
  \gamma_E \notag \\
  & \qquad - \frac{271272899815409}{157329572400} \log 2
    - \frac{246137536815857}{157329572400} \log v \notag \\
  & \qquad + \frac{3803225263}{1746360} \zeta_2 \bigg] \,.
\end{align}
We find that all of the terms match as expected, namely the first
two lines in both expressions.


\section{Conclusions}
\label{sec:conc}

In this paper we presented in detail a novel methodology to treat
higher-order non-linear effects of gravitational radiation that is
scattered from binary inspirals, where conservative and dissipative
dynamics are inevitably intertwined. The primary new idea that was
first introduced in \cite{Edison:2022cdu}, and enabled our uniquely
distinct approach to these type of effects, is that we make our
analysis directly at level of the whole binary taken as a single
composite particle interacting with gravity. This distinguishes our
current line of study from all the many amplitudes-driven works, which
study the unbound 2-to-2 scattering problem rather than the actual
bound two-body problem which is the primary focus of present and
planned GW experiments.

We treat the $l$-th multipole moments of the whole radiating binary
coupled to gravity in the EFT of the composite particle in analogy to
massive elementary particles of spin $l/2$ and their gravitational
scattering amplitudes. In this paper we go one step forward in
grounding our approach, where in \cref{sec:tail-bb} we start from pure
tree amplitudes as our analogous building blocks from which we
construct the necessary unitarity cuts, rather than using the EFT
vertices as a given. We verified that these pure amplitude
replacements work through to the highest orders reached in the present
work.  In \cref{sec:all-tails}, we spelled out our new method for the
well-studied lower-order effects of radiation-reaction and tail, where
the CTP formalism adopted to our problem is layered on top of our
integral basis and generalized-unitarity inspired procedure.

In view of the pressing need to push such predictions to high PN
orders, we proceeded in \cref{sec:higher-tails} to study higher-order
tails all through to the third subleading tail effect: This is the
$5$-loop tail-of-tail-of-tail-of-tail, or T$^4$, at order $G^5$
corresponding to $8.5$PN.  One interesting benefit of our method is
that it naturally organizes the results at each tail level according
to an iterative pattern.  We pointed out explicit examples for the cut
coefficients in \cref{sec:tttt}.  However, there are other interesting
hints at iterative and recursive structure.  For instance, the number
of \emph{actually distinct} cut diagrams and contributions at each
tail order so far tracks the Fibonacci sequence:
\begin{equation}
	\text{RR} \to 1 \quad \text{T} \to 1 \quad \text{TT} \to 2 \quad
	\text{TTT} \to 3 \quad \text{T}^4 \to 5.
\end{equation}
It would be interesting to explore these patterns and attempt to
identify a structure which directly produces the various cut
coefficients.

Let us highlight that also in contrast with all other
amplitudes-driven related studies, we only deal with classical
propagating gravitons, and land directly on causal effective actions,
which encompass the full conservative and dissipative dynamics of
these effects. For the lower-order results in \cref{sec:all-tails},
these could be checked against previous EFT results
\cite{Galley:2009px,Galley:2015kus}. However, as of the TT level the
causal effective actions we obtained in our approach in
\cref{sec:higher-tails} have never been previously derived.  Yet, in
\cref{sec:renorm} after we formulate the consequent energy loss of the
tails, we could verify through a renormalization analysis that the
related leading RG flow of the quadrupole coupling is in perfect
agreement with that of \cite{Goldberger:2009qd}, where only the TT
level was reached.

Additionally, the new T$^4$ corrections we obtained to the effective
action, \cref{eq:tttt-raw}, and its associated correction to the
emitted energy, \cref{eq:tttt-e-raw}, led us to identify a novel
counterterm in the quadrupole coupling, \cref{eq:kappa-ren-2}, and an
associated new term in the RG flow of the renormalized quadrupoles,
\cref{eq:new-rg}. Because the new effects continue the pattern of
skipping loop orders, it would be interesting to check our results by
computing the T$^5$ contributions, which should produce the same
counterterms and RG flows. Beyond that we could only establish that
our energy emissions are consistent with those derived via traditional
PN-theory results \cite{Blanchet:2013haa,Marchand:2016vox}, available
only up through $T^3$, as well as specific pieces of the results from
self-force theory \cite{Fujita:2011zk,Fujita:2012cm}.


\begin{acknowledgments}
  We thank John Joseph Carrasco, Sasank Chava, and Radu Roiban for
  feedback on the manuscript.  AE is supported by the USDOE under
  contract DE-SC0015910 and by Northwestern University via the
  Amplitudes and Insight Group, Department of Physics and Astronomy,
  and Weinberg College of Arts and Sciences.
ML has been supported by the Science and Technology Facilities Council 
(STFC) Rutherford Grant ST/V003895 \textit{``Harnessing QFT for Gravity''}, 
and by the Mathematical Institute University of Oxford.

This research was supported in part through the computational
resources and staff contributions provided for the Quest high
performance computing facility at Northwestern University which is
jointly supported by the Office of the Provost, the Office for
Research, and Northwestern University Information Technology.

fill\TeX was used as part of writing the bibliography
\cite{2017JOSS....2..222G}.

\end{acknowledgments}

\appendix

\section{Handling Tensor Reductions}
\label{sec:tens-red}
Throughout our tail calculations, we encounter cuts which are
functions of the loop momenta $\ell_x$, the radiation frequency
$\omega$ and the Euclidean metric $\delta^{op}$ with four Euclidean
indices contracted against the quadrupoles,
\begin{equation}
  \mathcal{N}^{ij;mn}(\{\ell_x,\omega\}) I^{ij}(\omega) I^{mn}(-\omega) \ , 
\end{equation}
that will require tensor reduction to reach the final state factor
$\kappa(\omega) = I^{ij}(-\omega) I^{ij}(\omega)$.  While we could
perform this reduction term-by-term at the level of the individual
numerator and propagator combinations that appear, this order of
processing delays the introduction of $\ell^2$ that should be zeroed
by on-shell conditions in the construction of the cut.  We will
instead introduce a generic tensor reduction scheme that can easily be
applied during the process of cut assembly by constructing a tensor
$\mathbf{U}^{ij;mn}$ such that
\begin{equation}
  \mathcal{N}^{ij;mn} I^{ij}I^{mn} = \mathcal{N}^{ij;mn}\mathbf{U}^{ij;mn} \kappa(\omega) \ ,
  \label{eq:u-red}
\end{equation}
subject to the symmetry and trace constraints
\begin{align}
  \mathbf{U}^{ij;mn} &= \mathbf{U}^{ji;mn} = \mathbf{U}^{ij;nm} \ ,\\
  \mathbf{U}^{ii;mn} &= \mathbf{U}^{ij;mm} = 0 \,.
\end{align}
With the spatial Euclidean metric as the only available (even-parity)
tensor to construct $\mathbf{U}$ from, we find a unique object
\begin{equation}
  \mathbf{U}^{ij;mn}  = - 2 \frac{\delta^{ij}\delta^{mn}}{(d + 2)d(d-1)} 
  + \frac{\delta^{im}\delta^{jn} + \delta^{in}\delta^{jm}}{(d+2)(d-1)} \,.
\end{equation}
We can then insert \cref{eq:u-red} during the evaluation of a cut,
after splitting the four-momenta into frequencies and spatial momenta.

We have explicitly checked that this method reproduces the
term-by-term reduction method for a number of integrals relevant to
the tails.


\section{Evaluating Basis Integrals}
\label{sec:eval-ints}
\subsection{Analytic Evaluation and Bubble Iteration}
\label{sec:ints-analytic}

The two most important integrals we need for evaluating all of the
basis integrals appear in Chapter 10 of Smirnov's \emph{Analytic Tools
  for Feynman Integrals} \cite{Smirnov:2012gma}.  Important to note is
that Ref.~\cite{Smirnov:2012gma} works in mostly-minus Minkowski
signature, whereas the integrals we need to evaluate are in Euclidean
signature.  To compensate for this, we need to Wick rotate the
Minkowski integrals to Euclidean signature via $\ell_0\to i \ell_E$
which takes $\ell^2 \to - \ell_E^2$ and
$\Diff{d}\ell \to i \Diff{d}\ell_E$.  The $i$ from the change in
measure cancels against the $i$ as part of the $i \pi^{d/2}$
normalization, and the change-in-sign of the propagators will induce
an extra phase $(-1)^{\lambda}$ for each propagator, as well as
\emph{change the relative sign} between $\ell^2$ and $m^2$ for the
massive propagators.  For example, the massive one-propagator
integral, the ``tadpole'', in Minkowski signature is
\begin{equation}
\int \frac{\Diff{d}k}{(-k^2 + m^2)^\lambda} = 
i \pi^{d/2} \frac{\Gamma(\lambda - d/2)}{\Gamma(\lambda)} \frac{1}{(m^2)^{\lambda -d/2}} \,.
\end{equation}
Switching to Euclidean signature, we get
\begin{equation}
\int_E \frac{\Diff{d}k_E}{(-k_E^2 - m^2)^\lambda} = (-1)^{\lambda} \pi^{d/2} 
\frac{\Gamma(\lambda - d/2)}{\Gamma(\lambda)} \frac{1}{(m^2)^{\lambda -d/2}} \,.
\label{eq:euc-tad}
\end{equation}
From here on, we drop the explicit $E$ label on the integrated
momenta. The scale of the tadpole integrals that actually occurs
throughout the basis integrals used in \cref{sec:all-tails} is really
the graviton frequency $\omega^2$, and always appears with the opposite
sign compared with an actual mass as in \cref{eq:euc-tad}.  In
addition, we need to switch to the physical $\pi$ normalization. Thus,
the integral we need for evaluation is
\begin{equation}
  F^{(1)}(\lambda;\omega^2) = G^{(1)}(\lambda) = \int_E \frac{\Diff{d}k}{(2 \pi)^d}\frac{1}{(-k^2 -(-\omega^2))^\lambda} = 
   \frac{\Gamma(\lambda - d/2)}{\Gamma(\lambda)} \frac{(-1)^\lambda(4\pi)^{-d/2} }{(-\omega^2)^{\lambda -d/2}} \,,
\end{equation}
where we have supressed the explicit $\ctpeps$ component of $\omega$
because \emph{we are not integrating over it as part of $\Diff{d}k$.}

Similarly, the $m_1=m_2=m,m_3=0$ two-loop three-propagator integral in
Euclidean signature is given by
\begin{align}
\int_E &\frac{\Diff{d}k \, \Diff{d} l}
{(-k^2 - m^2)^{\lambda_1}(-l^2-m^2)^{\lambda_2}[-(k+l)^2]^{\lambda_3}} \notag\\ 
&= \big( \pi^{d/2}\big)^2 (-1)^{\lambda_1+\lambda_2+\lambda_3} 
\frac{\Gamma(\lambda_1 + \lambda_3 - d/2)\Gamma(\lambda_2 + \lambda_3 
	- d/2)\Gamma(d/2-\lambda_3)}{\Gamma(\lambda_1)\Gamma(\lambda_2)} \notag\\
&~\times \frac{\Gamma(\lambda_1+\lambda_2+\lambda_3 - d)}{\Gamma(\lambda_1 + \lambda_2 
	+ 2 \lambda_3 -d)\Gamma(d/2) (m^2)^{\lambda_1 + \lambda_2 + \lambda_3 - d}} \notag\\
&= ( \pi^{d/2})^2 (-1)^{\lambda_1+\lambda_2+\lambda_3} B_{\lambda_1,\lambda_2,\lambda_3;d} 
(m^2)^{-\lambda_1-\lambda_2-\lambda_3+d} \,.
\label{eq:banana}
\end{align}
As in the case of the tadpole, the basis integrals we encounter have
the opposite relative sign between $k^2$ and $\omega^2$, meaning
instead we are interested in
\begin{equation}
  G^{(2)}(\lambda_1,\lambda_2,\lambda_3) =(4 \pi)^{-d} (-1)^{\lambda_1+\lambda_2+\lambda_3} 
  B_{\lambda_1,\lambda_2,\lambda_3;d} (-\omega^2)^{-\lambda_1-\lambda_2-\lambda_3+d}  \,.
\end{equation}
The generic bubble integral is also useful as an intermediate step, namely
\begin{align}
\int_E \frac{\Diff{d}k}{(2 \pi)^{d}}\frac{1}{(-k^2)^{\lambda_1}[-(q-k)^2]^{\lambda_2}} 
&=  \frac{(-1)^{d/2}}{(4\pi)^{d/2}} \frac{\Gamma(d/2-\lambda_1)\Gamma(d/2-\lambda_2)\Gamma(\lambda_1+\lambda_2 
	- d/2)}{\Gamma(\lambda_1)\Gamma(\lambda_2)\Gamma(d-\lambda_1-\lambda_2) (-q_E^2)^{\lambda_1 
		+ \lambda_2 - d/2}} \notag\\
&= (-1)^{d/2} (4\pi)^{-d/2} A_{\lambda_1,\lambda_2;d} (-q_E^2)^{-\lambda_1-\lambda_2+d/2} \,.
\label{eq:bub-int}
\end{align}
Importantly, the Euclidean bubble produces a new Euclidean propagator,
and matching the sign choice for this new propagator to the one used
for integration absorbs the ubiquitous phase
$(-1)^{\lambda_1 + \lambda_2}$.  Note that, as expected from the
topology, the expression is symmetric in $\lambda_1$ and $\lambda_2$.

With these ingredients, we can begin evaluating the higher-loop basis
integrals recursively.  The first non-trivial integral we need to
evaluate is the TT bulk contact integral, $F^{(3)}(1,1,1,1,0,0)$ from
\cref{eq:tail-raw-s}.  The two potential mode propagators can be
integrated together as a bubble using \cref{eq:bub-int} with
$q = \ell_1+\ell_3$, resulting in a single new potential mode
propagator.  The remaining integral is now of the form
\cref{eq:banana}, with a shifted index on the scaleless propagator.
Putting everything together, we have
\begin{align}
F^{(3)}(1,1,1,1,0,0) &=  (-1)^{d/2}(4\pi)^{-d/2} A_{1,1;d}G^{(2)}(1,1,2-d/2) \notag\\
&=( 4\pi)^{-3d/2} A_{1,1;d}B_{1,1,2-d/2;d} (-\omega^2)^{-4+3d/2}\,.
\end{align}
This expression is readily expandable near $d=3$.  Evaluating the TTT
and T$^4$ bulk contacts proceeds in a similar manner, just with more
levels of bubble iteration, yielding
\begin{align}
  F^{(4)}(\text{\cref{fig:ttt-2}})
  &=(-1)^{d} (4 \pi)^{-d} A_{1,1;d}A_{1,2-d/2;d}G^{(2)}(1,1,3-d) \notag\\
&=(4 \pi)^{-2d} A_{1,1;d}A_{1,2-d/2;d}B_{1,1,3-d;d} (-\omega^2)^{-5+2d}  \ , \\
  F^{(5)}(\text{\cref{fig:tttt-7}})
  &= (4 \pi)^{-5d/2} A_{1,1;d}A_{1,2-d/2;d}A_{1,3-d;d}B_{1,1,4-3d/2;d} (-\omega^2)^{-6+5d/2} \,.
\end{align}

\subsection{Evaluation of \Cref{fig:tttt-8}}
\label{sec:int-tttt-8}

We do not know of a way to analyitcally evaluate the T$^4$ integral
corresponding to the $\mathcal{M}_4 \otimes \mathcal{M}_4$ topology,
\cref{fig:tttt-8}.  However, we can exploit the fact that, like all
the other integrals we consider, the $\omega^2$ scale of the integral
is completely factorizable
\begin{equation}
  F^{(5)}(\text{\cref{fig:tttt-8}}) \equiv (-\omega^2)^{5d/2-7} \mathcal{I}_{4 \otimes 4}(1) \,,
\end{equation}
so that we just need to numerically determine $\mathcal{I}_{4 \otimes 4}(1)$ as
an expansion in the dimensional regularization parameter.  We use the
program AMFlow \cite{Liu:2022chg} in conjunction with Kira
\cite{Klappert:2020nbg} to evaluate $\mathcal{I}_{4 \otimes 4}(1)$ up
to $\mathcal{O}(\dimreg^2)$ with 500 digits of precision at each
order.  We can then use the PSLQ algorithm \cite{Bailey:1999nv,pslq}
to reconstruct the transcendental numbers, using the transcendental
numbers appearing in the other T$^4$ integrals as a guide for guessing
the basis.  We find, using AMFlow's definition of
$d = 3 - 2 \dimreg$ instead of the $d=3+\dimreg$ that we use in the
rest of the paper,
\begin{align}
  \mathcal{I}_{4 \otimes 4}(1) = \frac{1}{8(4 \pi)^5}\Bigg[
  & \frac{1}{ \dimreg^2}
    + \frac{16 +5(\log(\pi) - \gamma_E)}{\dimreg} \notag \\
  &+ \big( 184 + 80(\log(\pi) -  \gamma_E) + \frac{25}{2}(\log(\pi) - \gamma_E)^2
    + \frac{47}{2} \zeta_2 \big) \notag \\
  &+\dimreg\Big( 1888 + 920(\log(\pi) - \gamma_E) + 200(\log(\pi) - \gamma_E)^2
    + \frac{125}{6}(\log(\pi) - \gamma_E)^3 \notag \\
  &\ + 408 \zeta_2 - \frac{611}{3} \zeta_3
    + \frac{235}{2} \zeta_2 (\log(\pi) - \gamma_E)\Big)
     \notag \\
  & \dimreg^2\Big( 18544 + 9440 (\log(\pi) - \gamma_E)
    + 2300 (\log(\pi) - \gamma_E)^2 \notag\\
  &\ + \frac{1000}{3}(\log(\pi) - \gamma_E)^3
    + \frac{625}{24}(\log(\pi) - \gamma_E)^4 \notag \\
  &\ + 5092 \zeta_2 + 2040\zeta_2 (\log(\pi) - \gamma_E)
    + \frac{1175}{4} \zeta_2 (\log(\pi) - \gamma_E)^2 \notag \\
  &+ \frac{42193}{40} \zeta_2^2 - \frac{9872}{3} \zeta_3 - \frac{3055}{3} \zeta_3(\log(\pi) - \gamma_E) \Big)
  + \mathcal{O}(\dimreg^3)\Bigg] \,.
\end{align}
We have verified the reconstruction using an additional numerical
evaluation in AMFlow to 1000 digits.  This depth in the $\dimreg$
expansion is more than sufficient, after matching $\dimreg$
conventions, to obtain up through the $\mathcal{O}(\dimreg^0)$ part of
the T$^4$ effective action.


\bibliographystyle{JHEP}
\bibliography{gwbibtex}

\end{document}